\documentclass[a4paper, usenatbib]{mnras}

\usepackage[T1]{fontenc}
\usepackage{ae,aecompl}

\usepackage{graphicx,amsmath,amsfonts,amssymb}
\usepackage{epstopdf}
\usepackage{epsf, color}
\usepackage{url}
\usepackage{verbatim}
\usepackage{graphicx}

\bibpunct{(}{)}{;}{a}{}{,} 

\def\setsymbol#1#2{\expandafter\def\csname #1\endcsname{#2}}
\def\getsymbol#1{\csname #1\endcsname}

\def\Planck{\textit{Planck}}

\newbox\tablebox    \newdimen\tablewidth
\def\leaderfil{\leaders\hbox to 5pt{\hss.\hss}\hfil}
\def\endPlancktablewide{\tablewidth=\textwidth 
    $$\hss\copy\tablebox\hss$$
    \vskip-\lastskip\vskip -2pt}
\def\tablenote#1 #2\par{\begingroup \parindent=0.8em
    \abovedisplayshortskip=0pt\belowdisplayshortskip=0pt
    \noindent
    $$\hss\vbox{\hsize\tablewidth \hangindent=\parindent \hangafter=1 \noindent
    \hbox to \parindent{$^#1$\hss}\strut#2\strut\par}\hss$$
    \endgroup}
\def\doubleline{\vskip 3pt\hrule \vskip 1.5pt \hrule \vskip 5pt}

\def\L2{\ifmmode L_2\else $L_2$\fi}

\def\DeltaT{\ifmmode \Delta T\else $\Delta T$\fi}
\def\deltat{\ifmmode \Delta t\else $\Delta t$\fi}
\def\fknee{\ifmmode f_{\rm knee}\else $f_{\rm knee}$\fi}
\def\Fmax{\ifmmode F_{\rm max}\else $F_{\rm max}$\fi}
\def\solar{\ifmmode{\rm M}_{\mathord\odot}\else${\rm M}_{\mathord\odot}$\fi}
\def\Msolar{\ifmmode{\rm M}_{\mathord\odot}\else${\rm M}_{\mathord\odot}$\fi}
\def\Lsolar{\ifmmode{\rm L}_{\mathord\odot}\else${\rm L}_{\mathord\odot}$\fi}
\def\inv{\ifmmode^{-1}\else$^{-1}$\fi}
\def\mo{\ifmmode^{-1}\else$^{-1}$\fi}
\def\sup#1{\ifmmode ^{\rm #1}\else $^{\rm #1}$\fi}
\def\expo#1{\ifmmode \times 10^{#1}\else $\times 10^{#1}$\fi}
\def\,{\thinspace}
\def\lsim{\mathrel{\raise .4ex\hbox{\rlap{$<$}\lower 1.2ex\hbox{$\sim$}}}}
\def\gsim{\mathrel{\raise .4ex\hbox{\rlap{$>$}\lower 1.2ex\hbox{$\sim$}}}}
\let\lea=\lsim
\let\gea=\gsim
\def\simprop{\mathrel{\raise .4ex\hbox{\rlap{$\propto$}\lower 1.2ex\hbox{$\sim$}}}}
\def\deg{\ifmmode^\circ\else$^\circ$\fi}
\def\pdeg{\ifmmode $\setbox0=\hbox{$^{\circ}$}\rlap{\hskip.11\wd0 .}$^{\circ}
          \else \setbox0=\hbox{$^{\circ}$}\rlap{\hskip.11\wd0 .}$^{\circ}$\fi}
\def\arcs{\ifmmode {^{\scriptstyle\prime\prime}}
          \else $^{\scriptstyle\prime\prime}$\fi}
\def\arcm{\ifmmode {^{\scriptstyle\prime}}
          \else $^{\scriptstyle\prime}$\fi}
\newdimen\sa  \newdimen\sb
\def\parcs{\sa=.07em \sb=.03em
     \ifmmode \hbox{\rlap{.}}^{\scriptstyle\prime\kern -\sb\prime}\hbox{\kern -\sa}
     \else \rlap{.}$^{\scriptstyle\prime\kern -\sb\prime}$\kern -\sa\fi}
\def\parcm{\sa=.08em \sb=.03em
     \ifmmode \hbox{\rlap{.}\kern\sa}^{\scriptstyle\prime}\hbox{\kern-\sb}
     \else \rlap{.}\kern\sa$^{\scriptstyle\prime}$\kern-\sb\fi}
\def\ra[#1 #2 #3.#4]{#1\sup{h}#2\sup{m}#3\sup{s}\llap.#4}
\def\dec[#1 #2 #3.#4]{#1\deg#2\arcm#3\arcs\llap.#4}
\def\deco[#1 #2 #3]{#1\deg#2\arcm#3\arcs}
\def\rra[#1 #2]{#1\sup{h}#2\sup{m}}

\def\dots{\relax\ifmmode \ldots\else $\ldots$\fi}
\def\WHzsr{\ifmmode $W\,Hz\mo\,sr\mo$\else W\,Hz\mo\,sr\mo\fi}
\def\mHz{\ifmmode $\,mHz$\else \,mHz\fi}
\def\GHz{\ifmmode $\,GHz$\else \,GHz\fi}
\def\mKs{\ifmmode $\,mK\,s$^{1/2}\else \,mK\,s$^{1/2}$\fi}
\def\muKs{\ifmmode \,\mu$K\,s$^{1/2}\else \,$\mu$K\,s$^{1/2}$\fi}
\def\muKRJs{\ifmmode \,\mu$K$_{\rm RJ}$\,s$^{1/2}\else \,$\mu$K$_{\rm RJ}$\,s$^{1/2}$\fi}
\def\muKHz{\ifmmode \,\mu$K\,Hz$^{-1/2}\else \,$\mu$K\,Hz$^{-1/2}$\fi}
\def\MJysr{\ifmmode \,$MJy\,sr\mo$\else \,MJy\,sr\mo\fi}
\def\MJysrmK{\ifmmode \,$MJy\,sr\mo$\,mK$_{\rm CMB}\mo\else \,MJy\,sr\mo\,mK$_{\rm CMB}\mo$\fi}
\def\microns{\ifmmode \,\mu$m$\else \,$\mu$m\fi}
\def\micron{\microns}
\def\muK{\ifmmode \,\mu$K$\else \,$\mu$\hbox{K}\fi}
\def\microK{\ifmmode \,\mu$K$\else \,$\mu$\hbox{K}\fi}
\def\muW{\ifmmode \,\mu$W$\else \,$\mu$\hbox{W}\fi}
\def\kms{\ifmmode $\,km\,s$^{-1}\else \,km\,s$^{-1}$\fi}
\def\kmsMpc{\ifmmode $\,\kms\,Mpc\mo$\else \,\kms\,Mpc\mo\fi}
\providecommand{\sorthelp}[1]{}

\usepackage{ifthen}

\providecommand{\simlt}{\lea}
\providecommand{\simgt}{\gea}

\newcommand{\healpix}{{\tt HEALPix\ }}
\newcommand{\camspec}{{\tt CamSpec\ }}

\newcommand{\Herschel}{{\it Herschel}}
\newcommand{\hi}{\ion{H}{i}}

\newcommand{\reffig}[1]{Fig.~\ref{fig:#1}}          
\newcommand{\refFig}[1]{Figure~\ref{fig:#1}}
\newcommand{\reftab}[1]{Table~\ref{t:#1}}
       
\newcommand{\refsec}[1]{Sec.~\ref{sec:#1}}  
\newcommand{\refapp}[1]{Appendix~\ref{app:#1}}   

\defcitealias{planck2011-6.6}{PEP}
\defcitealias{planck2013-pip56}{P13}
\defcitealias{Bethermin2012}{B12}
\defcitealias{Viero2013}{V13}

\newcommand{\PEP}{\citetalias{planck2011-6.6}}
\newcommand{\cppcib}{\citetalias{planck2013-pip56}}
\newcommand{\bcount}{\citetalias{Bethermin2012}}
\newcommand{\viero}{\citetalias{Viero2013}}
 
\begin{document}

\title[\Planck\ CIB anisotropies]{Measurement of CIB power spectra over large sky areas from \Planck\ HFI maps}

\author[D.~S.~Y. Mak et al.]{Suet Ying, Daisy, Mak$^1$\thanks{dmak@ast.cam.ac.uk}, Anthony Challinor$^{1,2}$, George Efstathiou$^1$, Guilaine Lagache$^3$\\
$^1$ Institute of Astronomy and Kavli Institute for Cosmology
Cambridge, Madingley Road, Cambridge CB3 0HA, UK \\
$^2$ DAMTP, Centre for Mathematical Sciences,
Wilberforce Road, Cambridge CB3 0WA, UK \\
$^3$ Aix Marseille Universit\'e, CNRS, LAM (Laboratoire d'Astrophysique de Marseille) UMR 7326, 13388, Marseille, France}

\date{Accepted XXX. Received YYY; in original form ZZZ}
\pubyear{2016}

\maketitle
\begin{abstract}
We present new measurements of the power spectra of the cosmic
infrared background (CIB) anisotropies using the \Planck\ 2015
full-mission HFI data at 353, 545, and 857\,GHz over 20\,000 square
degrees. We use techniques similar to those applied for the cosmological analysis of \Planck,
subtracting dust emission at the power spectrum level.  Our analysis gives stable solutions for the 
CIB power spectra with increasing sky coverage up to about 50\% of the sky. These spectra agree well with \ion{H}{i} cleaned
spectra from \Planck\ measured on much smaller areas of sky with low Galactic dust emission. 
At 545 and 857\,GHz our CIB spectra agree well with those measured from \textit{Herschel} data. 
We find that the CIB spectra at $\ell \simgt 500$ are well fitted by a power-law model for the
clustered CIB, with a shallow index $\gamma^{\rm cib} =0.53 \pm 0.02$. This is consistent with the
CIB results at 217\,GHz from the cosmological parameter analysis of \Planck. We show that a linear combination of the 545 and 857\,GHz
\Planck\  maps  is dominated by CIB fluctuations at multipoles $\ell
\simgt 300$.
\end{abstract}


\section{Introduction}
\label{sec:intro}


\vspace{2cm}

The CIB arises as the integrated emission from dust heated by
starlight in star-forming galaxies. The CIB carries a wealth of
information about the growth of galaxies and hence the process of star
formation. The most informative method to study the CIB is to resolve
the individual sources and then perform analyses of clustering and
counts on the resolved sources. However, the high density of faint,
distant galaxies makes this very challenging, particularly at the
lower frequencies seen by \Planck\ which are expected to probe the
highest redshift sources. In fact, only $10\,\%$ of the CIB has been
resolved into galaxies by {\it Herschel}\ at
857\GHz~\citep{Bethermin2010,Oliver2010} and negligibly so for \Planck\
with its poorer angular resolution~\citep{planck2012-VII}. An
alternative approach, and the only one feasible at lower frequencies
in the foreseeable future, is to study the statistical properties of
the unresolved background. Correlated anisotropies in the background
reflect the clustering properties of the unresolved galaxies and so
are a powerful probe of large-scale
structure~\citep[e.g.,][]{Haiman2000}. On large scales, the angular
power spectrum of galaxy clustering should reflect the underlying
distribution of dark matter halos (the two-halo term) and hence halo
bias; on smaller non-linear scales pairs of galaxies within the same
parent halo become an important contribution to the signal (the
one-halo term). The CIB also depends on the mean emissivity per
comoving volume and so probes models of galaxy evolution, star
formation and the initial stellar mass function.

CIB fluctuations have been measured at 3\,300\GHz\ (AKARI;
\citealt{Matsuura2011}) and 3\,000\GHz\ (IRAS/IRIS;
\citealt{Penin2012, planck2013-pip56}), 1\,875\GHz\ (Spitzer;
\citealt{Grossan2007,Lagache2007}), 600\GHz\ with
BLAST~\citep{Viero2009,Hajian2012} and
\textit{Herschel}/SPIRE~\citep{Amblard2010,Viero2013}, 220\GHz\ with
ACT~\citep{Dunkley2011} and SPT~\citep{Hall2010} and at 217, 353, 545,
and 857\GHz\ by \Planck\ (\citealt{planck2011-6.6}, hereafter \PEP,
and~\citealt{planck2013-pip56}, hereafter \cppcib). The results in
\PEP\ are based on power spectrum measurements over a total of
$140\,{\rm deg^2}$. \cppcib\ extends this to $2\,200\, {\rm deg^2}$,
in the cleanest regions of the sky for which \hi\ data, used as a
tracer to subtract Galactic dust emission from the CIB, is
available. The mean redshift of the CIB is expected to increase with
decreasing frequency, with models suggesting that over 90\,\% of the
clustering power at scales of $5\arcm$ in the 353 and
217\GHz\ \Planck\ bands comes from $z >
2$~\citep{planck2011-6.6}. \Planck\ data is therefore able to place
strong constraints on models of large-scale structure, galaxy
evolution and star formation at high redshift via CIB measurements.

Accurate removal of Galactic dust emission is critical for
measurements of CIB clustering. At present almost all measurements of
CIB power spectra either assume that the power spectrum of Galactic
dust emission can be described by a simple power-law template (e.g.,
$C_\ell\propto \ell^{-2.8}$) or use relatively low-resolution (around $10\arcm$ FWHM) \hi\ maps as a tracer of Galactic dust emission in
regions of low \hi\ column density to subtract dust in the map domain
at low multipoles, together with a power-law template in the power
spectrum domain (fitted to the \hi\ power-spectrum on larger angular scales) to subtract 
dust emission at higher multipoles.
However, comparison of \Planck\ maps with \hi\ observations reveals
excess dust emission at \hi\ column densities greater than $2.5 \times
10^{20} {\rm cm}^2$, probably caused by the formation of molecular
hydrogen in dusty regions \citep[see][]{planck2011-7.12}. \hi\ cleaning
is therefore possible only in the cleanest regions of the sky.

Here, we take a different approach to separate Galactic dust
emission from the CIB over large areas of the sky. We do this by
exploiting the statistical isotropy of the extragalactic signals to
construct dust power spectrum templates from differences of spectra
measured over different areas of the sky. This approach is very similar to that
adopted for the CMB likelihood analysis by the
\Planck\ team~\citep{planck2013-p08,planck2014-a13}. The amplitudes of
the Galactic dust templates at the three frequencies that we consider
in this paper (353, 545, and 857\,GHz) are fitted as part of the
likelihood analysis of the six frequency (cross-)spectra. Similarly,
we model the cosmic microwave background (CMB) anisotropies at the
power spectrum level rather than using component separation
techniques to produce a cleaned CMB map.
Our approach to cleaning the CIB spectra through model fitting
therefore differs from previous analyses of the \Planck\ data in
\PEP\ and \cppcib, which used map-based removal of Galactic dust
and the CMB. 

Our model fitting approach can be applied to large areas of the sky and
can therefore be used to test the stability of the recovered CIB power spectra as a function of
sky coverage. This is the primary aim of this paper.
The downside is that the dust and CMB lead to a high sampling variance compared
to subtraction of these components at the map level.
Removing an estimate of our realisation of the CMB at the
map level on large  scales would improve the CIB
constraints at 353\GHz\ (and at lower frequencies). This is the approach taken
in~\cppcib, where 100\,GHz \Planck\ maps were used as a CMB template. 
However, this requires modelling 
of additional components at 100\,GHz, particularly  thermal Sunyaev--Zel'dovich (SZ) emission, 
and also modelling the mismatch of the beam profiles at 100\,GHz and higher
frequencies. Removing Galactic dust emission at the map level requires a
tracer, e.g. \hi\ as in~\cppcib. However, as 
 discussed above, using \hi\ data limits the sky coverage and
angular resolution.  An alternative is to use internal estimates of dust
emission (see~\citealt{planck2016-XLVIII} for recent progress with the
\Planck\ data). The difficulty with this approach  lies in 
separating CIB and dust with the limited frequency coverage of the
\Planck\ data and quantifying  accurately any residuals.

This paper is organized as follows.  Section \ref{sec:data} describes the
\Planck\ data and  sky masks used in this paper. The parametric models
that we adopt for the CIB and our modelling of foreground components
are discussed in~\refsec{model}.  Details of the power spectra
measurements and construction of the likelihood are given
in~\refsec{likelihood}. In that section we also present tests to validate
the likelihood framework and discuss the expected degeneracies between
parameters.\footnote{Throughout this paper, we use the COSMOMC package
  (\cite{Lewis:2002}, http://cosmologist.info/cosmomc) to perform Markov chain Monte Carlo (MCMC) exploration of the likelihood.}
In~\refsec{cs}, we show the main results from our likelihood
analysis, presenting parameter constraints for the CIB and dust
foregrounds and comparing  our results with those of previous
measurements. Section \ref{sec:gal} develops a modification to the dust
contribution to the covariance matrix of the measured power spectra and explores whether Galactic
dust emission can be removed by forming a linear combination of the $545$ and  857\,GHz \Planck\ maps.
Our conclusions are summarized in 
~\refsec{conclusion}. The
appendices give further details about the structure of the covariance
matrix of the measured power spectra and our likelihood validation
(\refapp{cov}), the datasets and modelling used to make predictions
for the Poisson power from source counts (\refapp{ps}), and our
attempts to account for the statistically-anisotropic nature of
Galactic dust in the covariance matrix of the power spectra
(\refapp{dust}). \refapp{con} presents additional tests of the stability of
our results.

Throughout the paper, we adopt the standard $\Lambda$CDM cosmological
model as our fiducial cosmology, with parameter values derived from
the CMB power spectra as measured by \Planck. Specifically, we adopt
the \Planck\ TT+lowP values from Table 4 of~\citet{planck2014-a13}.

\section{Data sets and sky masks}
\label{sec:data}
\subsection{Planck maps}

The results in this paper are based on the \Planck\ 2015 full-mission
frequency maps at the three highest frequencies: 353, 545, and
857\GHz. More specifically, we use yearly maps for this analysis;
these are constructed by combining Surveys 1 and 2 to form the Year-1
map, and Surveys 3 and 4 to form the Year-2 map (see Table A.2
in~\citealt{planck2014-a09}). Power spectra are estimated by
cross-correlating Year-1 and Year-2 maps. As the instrument noise is
very nearly uncorrelated between the yearly maps, we make no further
correction for noise bias in the estimated power spectra. We consider
alternative splits of the data in \refapp{con}, including
half-mission maps and those made from detector sets.

The maps are calibrated on the CMB orbital dipole for 353\GHz\ and on
planets for 545 and 857\GHz. For 353\GHz, the calibration is accurate
to about $0.78\,\%$. At 545\GHz\ and 857\GHz, the
errors on the absolute calibration is about $6\,\%$, and the relative
calibration between the two channels is better than $3\,\%$. We
include map-level calibration parameters in our
likelihood analysis with Gaussian priors reflecting the errors on the
absolute calibration\footnote{Note, however, that recent work on the relative cailbration between the 545\,GHz channel and the CMB channels of \Planck, based on the Solar dipole, show that this relative calibration is, in fact, correct to within $1.5\,\%$ (see~\citealt{plancklowl2016}).
The accuracy of the absolute calibration of the 545\,GHz maps is therefore also accurate to this level, given the precise absolute calibration of the CMB channels off the orbital dipole.}.
Conservatively, we do not impose the more
precise relative calibration of the 353--857\GHz\ channels in our
priors.  Compared to the maps in the 2013 \Planck\ release, the
calibration factors changed by 1.9\,\% and 4.1\,\% at 545 and 857\GHz,
respectively, which is within the planet modelling uncertainty. The
combination of these, and other changes to the data processing (such as
beam characterization and corrections for non-linearities in the on-board analogue-to-digital
conversion), lead to the 2015 frequency maps being less bright by
1.8\,\% at 545\GHz\ and 3.3\,\% at 857\GHz\ compared to those in the 2013
release~\citep{planck2013-p03f}.  The 353\GHz\ maps in the 2015
release are brighter by 2.1\,\% compared to the 2013 release. Maps are given in units either of
MJy\,sr$^{-1}$ (with the photometric convention $\nu
I_{\nu}=\text{const.}$ ) or ${\rm K_{CMB}}$. The conversion between the
two can be  computed exactly given knowledge of the bandpass filters. 
Throughout
this paper, we convert measurements in ${\rm K_{CMB}}$ units to MJy
sr$^{-1}$ units using the mean coefficients: 287.45 (353\GHz); 58.04
(545\GHz); and 2.27 (857\GHz) as given in \cite{planck2013-pip56}.

We estimate the noise properties of the \Planck\ maps from the
differences of maps constructed from the first and second half of each
ring period (i.e., the half-ring half-difference
maps;~\citealt{planck2014-a09}). These maps provide a good statistical
representation of the noise in the actual sky maps. We compute the
noise spectra from these maps, which we use in combination with the
hit-counts to construct the noise model used in the likelihood analysis.

The maps are provided in \healpix format, with resolution parameter
$N_{\rm side}=2048$, corresponding to pixels with a typical width of
$1.7\,\text{arcmin}$. In analysing these maps, we use the fiducial
beam transfer functions described in~\cite{planck2014-a09}. Our
primary goal is to compute the six (cross-)frequency spectra by
cross-correlating the yearly maps at 353, 545, and 857\GHz. In forming
cross-spectra at different frequencies, e.g., $353\times 545$, we
average the Year-1$\times$Year-2 spectra, discarding the
Year-1$\times$Year-1 and Year-2$\times$Year-2 spectra.  We measure
cross-spectra using the \camspec software, described in the context of
CMB power spectrum estimation in~\citet{planck2013-p08}. Briefly,
pseudo-spectra are computed from masked sky maps, and are subsequently
deconvolved for the mode-coupling effect of the mask and beam and
pixel effects. The covariance matrices for these spectra are
calculated with analytic approximations that account for masking and
the statistically-anisotropic nature of the \Planck\ instrument noise.
Our interpretation of the spectra is based on fitting simple
parametrised models of the CIB and foreground components in an MCMC
likelihood analysis, as presented in \refsec{model}. The signal power
in the best-fitting models is used to construct the sample variance in
the covariance matrices of the power spectra.  We also interpret our
results in the context of halo models
using the CIB template spectra adopted in~\cppcib. 
This allows us 
to compare and test the
quality of our fits of the \Planck\ 2015 data directly against the analysis of
the 2013 data carried out by the \Planck\ team in~\cppcib.

\subsection{Sky masks for the Galactic region and bright sources}\
\label{sec:mask}

We mask regions of strong Galactic emission and bright point sources
before estimating power spectra. A key part of our analysis is to
demonstrate that our results are stable as a function of sky coverage.
 We do this by considering a set of Galactic masks as described 
below.

\begin{figure*}
  \begin{center}
\vskip 0.5 truein
        \includegraphics[width=50mm, angle=90, bb=0 0 460 737]{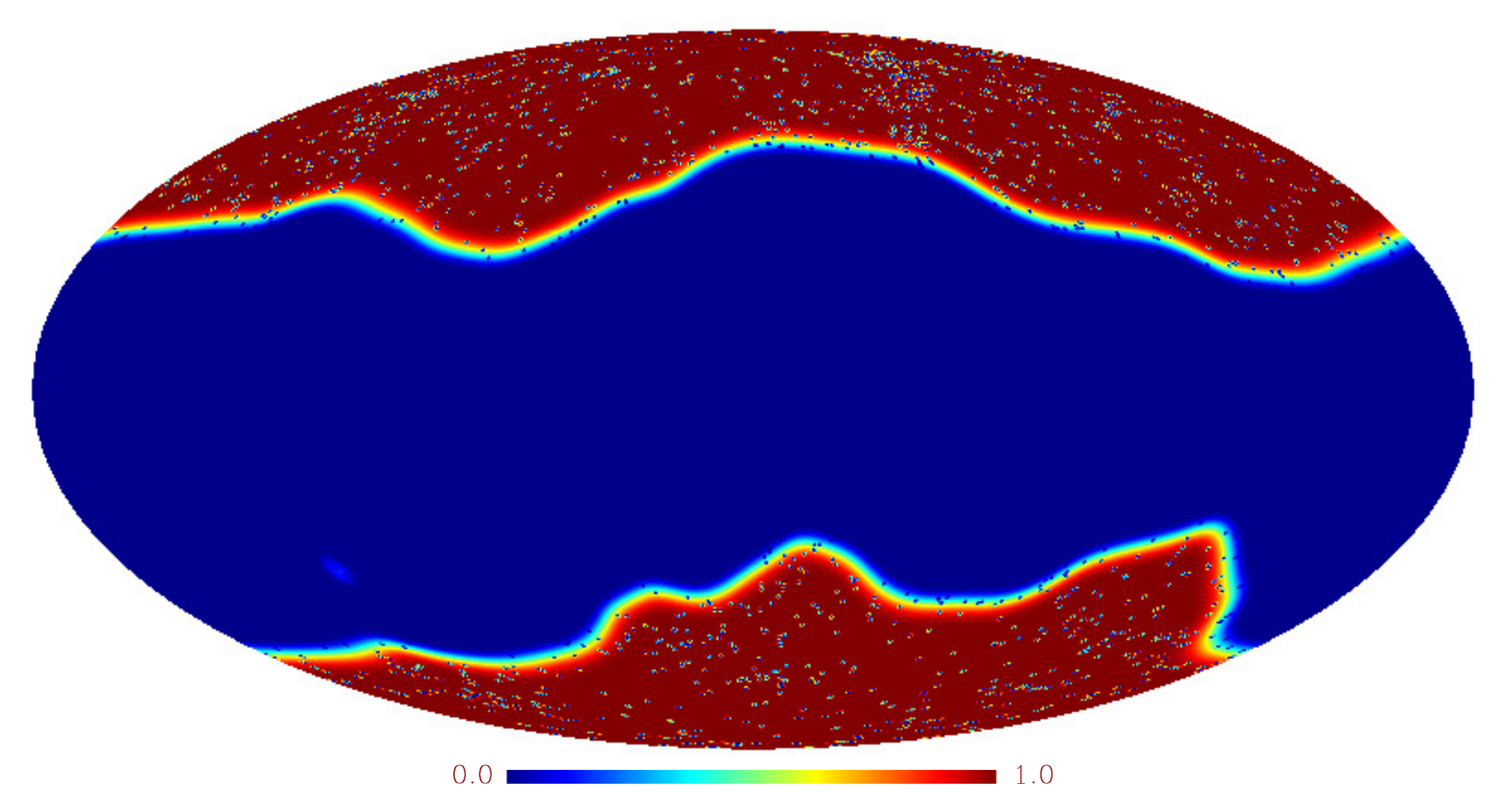}
        \includegraphics[width=45mm, angle=90]{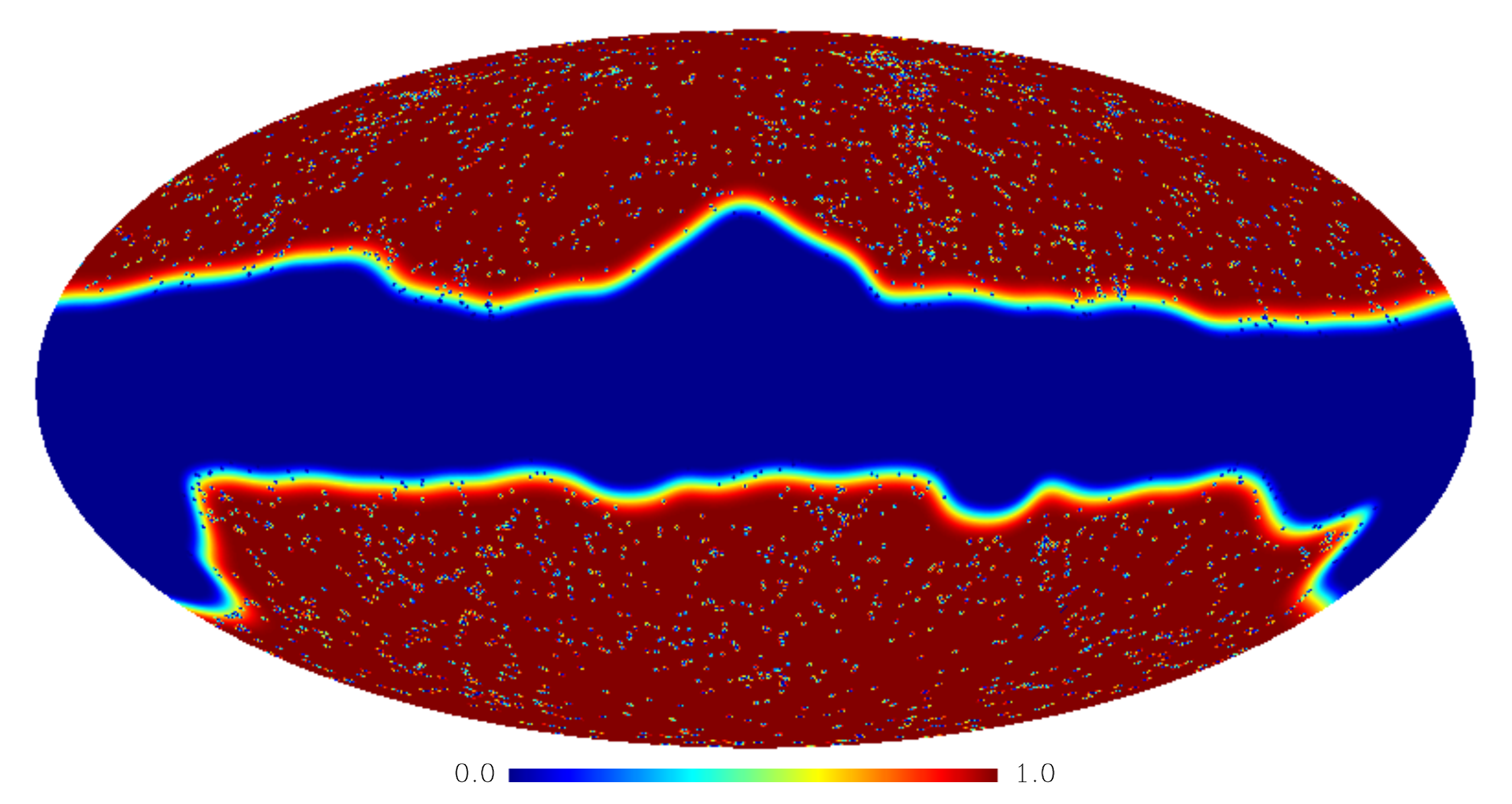} \\
        \includegraphics[width=88mm, angle=00]{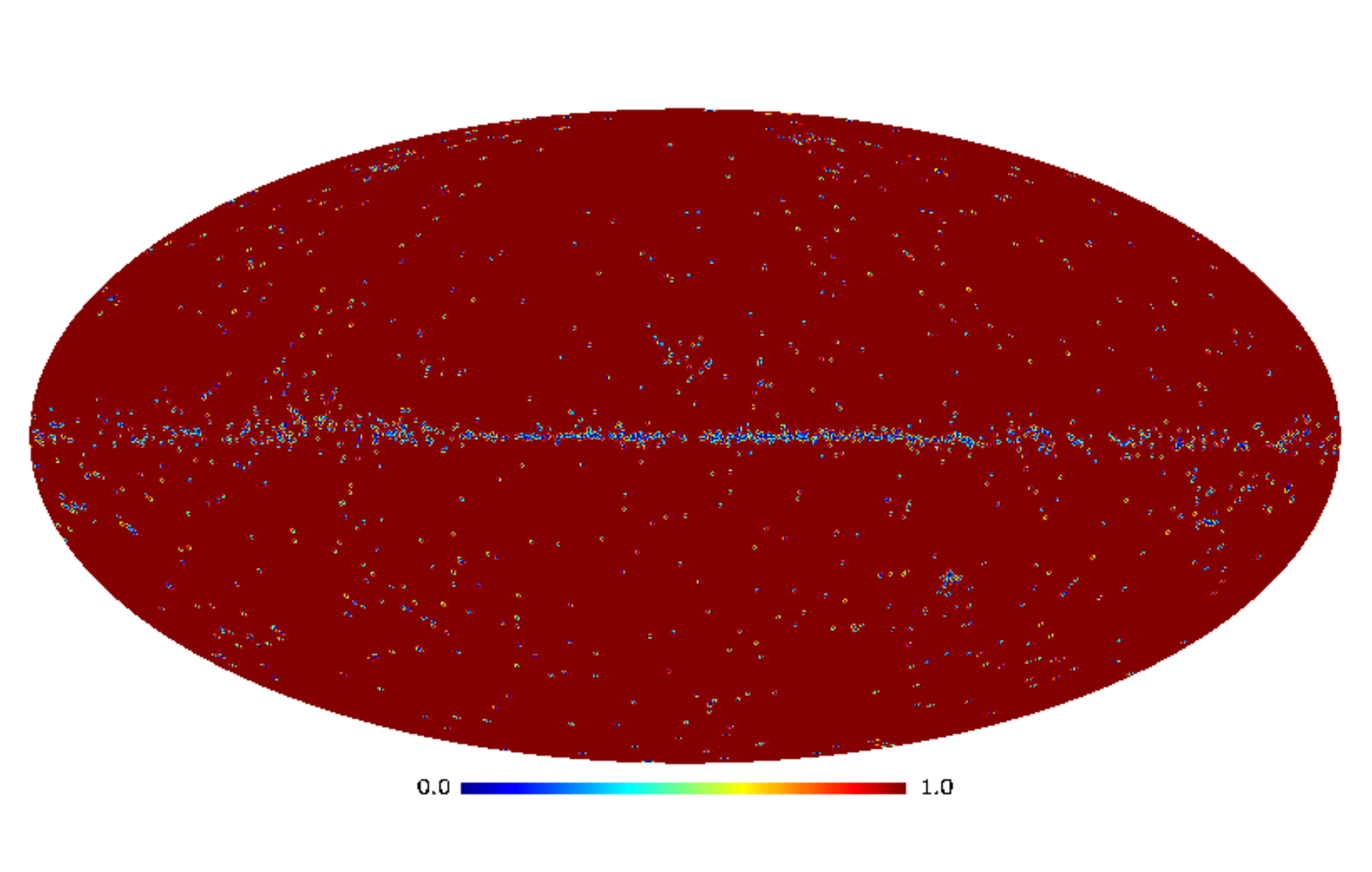}
        \includegraphics[width=88mm, angle=00]{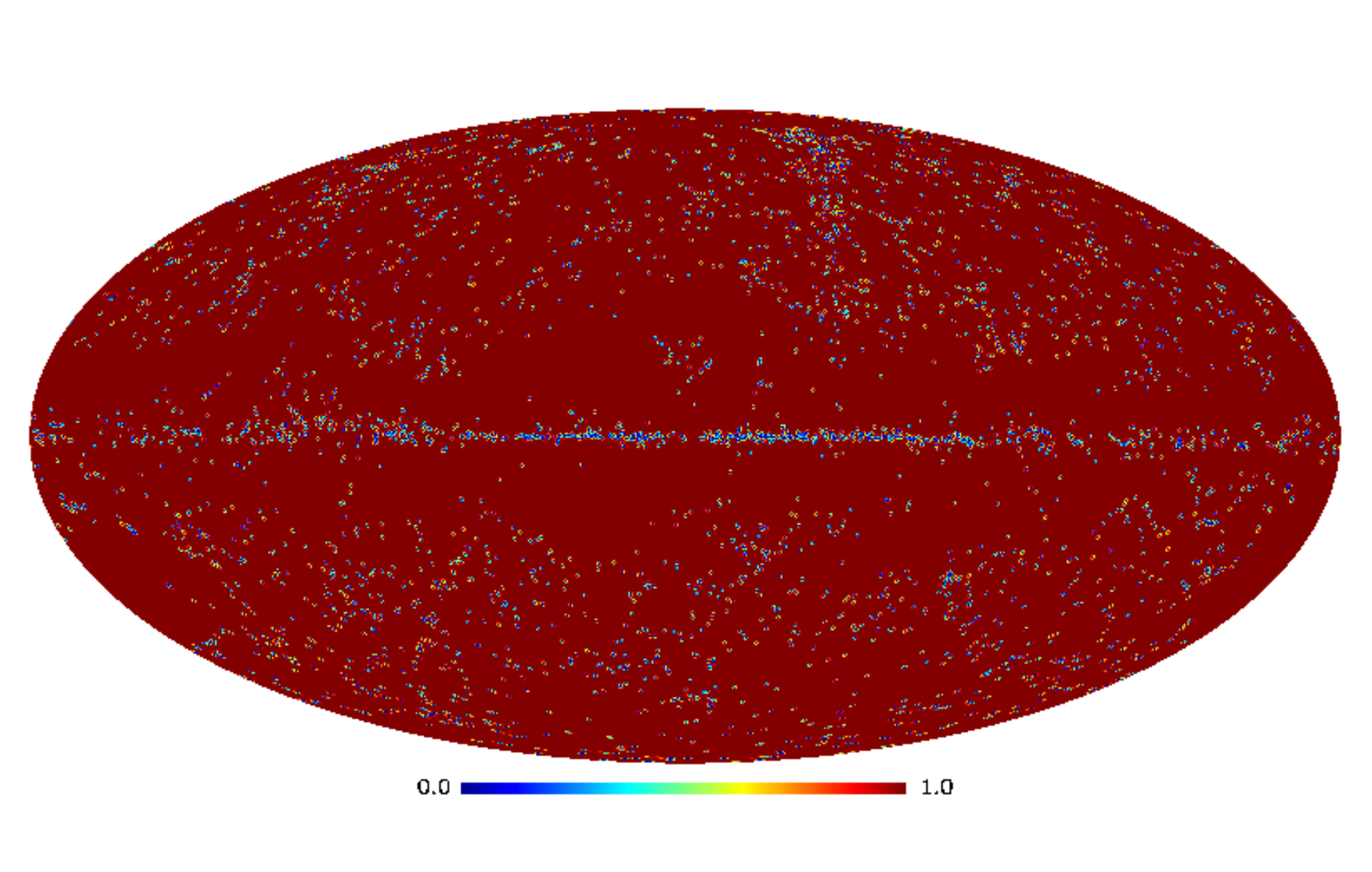} 

       \caption{\emph{Upper}: apodised Galactic mask 40 (left) and mask
         70 (right) combined with the union point source mask adopted in this paper. \emph{Lower}: 353\GHz\ point source mask (left) and 857\GHz\ point source mask (right). }

\vskip 0.5 truein
     \label{fig:mask}
  \end{center}
\end{figure*}

\subsubsection{Galactic masks}

We use a set of Galactic masks that
are obtained by thresholding a smoothed CMB-cleaned 353\GHz\ map at
different levels in order to obtain different desired sky
fractions. All of the Galactic masks are apodised with a $2\deg$ FWHM
Gaussian to reduce mode-coupling caused by sharp mask boundaries.  In
this paper, most of our results are based on unmasked sky fractions of
$30\,\%$, $40\,\%$, and $50\,\%$, corresponding to effective sky
fractions after apodisation of $24.5\,\%$, $33.8\,\%$, and
$42.8\,\%$.\footnote{We use a $25\,\%$ Galactic mask for calculating
  dust template spectra.}
For our baseline results, we use the $40\,\%$ mask.  As we
show in~\refsec{dust}, there are small changes in the shape of the
inferred dust power spectrum as we extend the sky coverage beyond
60\,\% associated with 
the anisotropy of the point source masks (which increases dramatically
as the Galactic plane is approached) and small changes 
in the properties of Galactic dust.  These, combined with the large
increase in the amplitude of the dust power spectrum, limit the
accuracy with which we can model and remove the dust power over larger
sky fractions. It is important to note that as the sky fraction is increased,
the sample variations in the dust power spectrum also increase. The optimal
sky fraction for the CIB analysis presented here therefore involves a trade-off
between improved signal-to-noise at small scales, where Galactic dust is 
sub-dominant compared to the CIB, versus increased sample variance at large scales where Galactic dust dominates.\footnote{This behaviour could be corrected by
  more careful weighting of the data during power spectrum
  estimation. Here, we weight the data uniformly after application of
  the mask.}
Unless otherwise specified, we name the masks by the
percentage of the sky that they retain before apodisation, e.g., mask
40 corresponds to the $40\,\%$ unmasked sky fraction.

\subsubsection{Point source masks}

We remove bright point sources
using a mask constructed from compact sources identified in the Planck
Catalogue of Compact Source (PCCS;~\citealt{planck2013-p05}). In the
\Planck\ analysis of cosmological parameters, point sources detected
in the PCCS with a signal-to-noise $S/N>7$ were masked and the
remaining Poisson point source levels at 100--217\GHz\ were treated
as nuisance parameters in the likelhood analysis.  For the analysis
presented here using the 353--857\GHz\ channels, we have constructed
point source masks with a higher detection threshold and completeness
level.

We use the PCCS as an entry catalogue, but select only those sources
that satisfy a number of criteria to ensure that the majority are
extragalactic (e.g., having counterparts in external catalogues, not
being in regions of bright cirrus or molecular clouds). We validate
the criteria using the observations of large Galactic fields by
\textit{Herschel}/SPIRE. The flux cut is about 400, 600, and 1000\,mJy at 353,
545, and 857\GHz, respectively. We call these the
``frequency-dependent masks''. We also construct a ``union mask'' by
combining the three frequency-dependent masks and use this as our
baseline point source mask. We explore the effects on the CIB
constraints when using these two versions of the point source masks
in~\refapp{con}.

The point source masks are apodised with a Gaussian of FWHM 30\arcmin,
and combined with the apodised Galactic masks to form the final masks
used in the power spectrum analysis. Examples are shown in the upper panels of~\reffig{mask}. Here the left- and right-hand panels show Galactic mask 40 and 70, respectively, 
together with the union point source mask.  Because we apply a dust background criterion, the point
source masks become incomplete for sky fractions greater than around
60\,\%. This is illustrated by the lower panels in~\reffig{mask}, which show the 353\GHz\ (left)  and 857\GHz\ (right)  point source masks.
Note that the union mask is dominated by sources detected at 857\GHz.

\section{Parametric model of the power spectra}
\label{sec:model}

We model the theoretical angular power spectrum for frequencies $\nu$ and $\nu'$ as
\begin{equation}
\mathcal{D}^{\rm th, \nu \times \nu'}_\ell =   \mathcal{D}^{\rm clu, cib, \nu \times \nu'}_\ell+  \mathcal{D}^{\rm dust, \nu \times \nu'}_\ell+\mathcal{D}^{\rm cmb, \nu \times \nu'}_\ell + \mathcal{D}^{\rm ps, \nu \times \nu'}_\ell ,
\label{eqn:model} 
\end{equation}
where e.g., $\mathcal{D}^{\rm th, \nu\times\nu'}_\ell\equiv\ell
(\ell+1)C^{\rm th, \nu \times \nu'}_\ell/2\pi$. The terms in Eq.~(\ref{eqn:model}), in order,
 are the clustered contribution from the  CIB, 
Galactic dust, primordial CMB, and the Poisson power from infrared
sources, respectively. The CMB signal is significant only at 353\GHz, though
we include it at all frequencies in our analysis. The above decomposition separates the
CIB into  clustered and Poisson components. The sum of these components is well constrained
by the \Planck\ data, but the decomposition into clustered and Poisson
components is partially degenerate with extent that depends on the multipole range used and the 
assumed shape of the clustered CIB power spectrum. Most of the 
statistical power of the  \Planck\ spectra comes from multipoles $\ell \simlt 1500$, where
we find clear evidence that the power spectrum of the clustered CIB component is shallower
than the Poisson component. We therefore solve for the Poisson amplitudes in our likelihood
analysis, which can then be compared with the expected Poisson levels computed from source
counts (see \refsec{ps}).

In the following subsections we describe our parametrizations of the
individual contributions in Eq.~\ref{eqn:model}. The model
parameters and prior ranges are summarised in~\reftab{par}.

\begin{table*}
\caption{Summary of the 19 parameters describing the CIB, foregrounds, and calibration that we vary in the likelihood analysis. All amplitude parameters have units of $(\mu{\rm K})^2$. Square brackets denote uniform prior ranges, while parentheses indicate the mean and standard deviation of Gaussian priors.}
\begin{center}
\begin{tabular}{clcl}
\hline\hline
Type & Parameter & Prior range  & Definition \\
\hline
CIB & $A_{353}^{\rm cib}$ & $[10^2,10^5]$ & Clustered CIB power at $\ell=2000$ (in $\mathcal{D}_\ell$) at 353\GHz \\
       &$A_{545}^{\rm cib}$ &$[10^4,10^8]$ & As for $A_{353}^{\rm cib}$ but at 545\GHz \\ 
       &$A_{857}^{\rm cib}$ &$[10^7,10^{10}]$ & As for $A_{353}^{\rm cib}$ but at 857\GHz \\ 
      &  $r_{353\times545}^{\rm  cib}$ & [-0.5,1] & CIB correlation coefficient between 353 and 545\GHz \\
& $r_{353\times857}^{\rm cib}$ & [-0.5,1] & As for $r_{353\times545}^{\rm cib}$ but between 353 and 545\GHz \\
& $r_{545\times857}^{\rm cib}$ & [-0.5,1] & As for $r_{353\times545}^{\rm cib}$ but between 545 and 857\GHz \\
& $\gamma^{\rm cib}$ & [0.0,1.5]&Spectral index of the CIB angular power spectrum\\
\hline
Dust &$A_{353}^{\rm dust}$ & $[10^2,10^5]$ & Dust spectrum amplitude at 353\GHz \\
         &$A_{545}^{\rm dust}$ &$[10^5,10^8]$ & As for $A_{353}^{\rm dust}$ but at 545\GHz \\ 
        &$A_{857}^{\rm dust}$ &$[10^8,10^{11}]$ & As for $A_{353}^{\rm dust}$ but at 857\GHz \\ 

\hline
Poisson power &$A_{353}^{\rm ps}$ & $[10^{2},10^5]$ &Poisson point source power at $\ell=2000$ (in $\mathcal{D}_{\ell}$) at 353\GHz \\
         &$A_{545}^{\rm ps}$ &$[10^{4},10^7]$ & As for $A_{353}^{\rm ps}$ but at 545\GHz \\ 
        &$A_{857}^{\rm ps}$ &$[10^5,10^{10}]$ & As for $A_{353}^{\rm ps}$ but at 857\GHz \\ 

& $r_{353\times545}^{\rm ps}$ & [-0.5,1] & Poisson power correlation coefficient between 353 and 545\GHz \\
& $r_{353\times857}^{\rm ps}$ & [-0.5,1] & As for $r_{353\times545}^{\rm ps}$ but between 353 and 545\GHz \\
& $r_{545\times857}^{\rm ps}$ & [-0.5,1] & As for $r_{353\times545}^{\rm ps}$ but between 545 and 857\GHz \\
\hline
Calibration & $cal_{353}$ & (1, 0.0078)& Calibration factor at 353\GHz \\ 
 & $cal_{545}$ & (1, 0.061) & Calibration factor at 545\GHz \\ 
 & $cal_{857}$ & (1, 0.064)& Calibration factor at 857\GHz \\ 
\hline
\end{tabular}
\end{center}
\label{t:par}
\end{table*}

\subsection{Clustered CIB}
\label{sec:mcib}

\subsubsection{Power-law CIB model}

Following~\cite{planck2013-p08}, we consider a power-law model for the CIB signal
\begin{equation}
 \mathcal{D}^{\rm clu, cib,\nu \times \nu'}_\ell = A^{\rm cib}_{\nu \times \nu'}  \left (\frac{\ell}{2000}\right )^{\gamma^{\rm cib}} \, ,
 \label{eqn:cib}
\end{equation}
where $A^{\rm cib}_{\nu \times \nu'}$ is the amplitude of the
clustered CIB power at multipole $\ell=2000$ in the $\nu\times\nu'$
spectrum and $\gamma^{\rm cib}$ is the power-law index. Our analysis of the \Planck\ data is 
 mostly sensitive to the clustered CIB around $\ell \approx 1000$ and,
 as we shall see, a simple
power-law model provides a reasonably good description of the \Planck\ spectra. 

Theoretical work based on the halo model (see Sec. \ref{sec:halomodel})
suggests that the power-law index $\gamma^{\rm cib}$ should
 steepen on smaller scales ($\ell \simgt 2000$) where non-linear
clustering becomes significant.  Since the inferred value of $\gamma^{\rm cib}$ is expected to  depend on scale, care is needed in comparing our results 
for the power-law model with others in the literature. For high-$\ell$ CMB experiments (e.g.,~\citealt{Story2013, Dunkley2013}), which have sensitivity to the non-linear clustering regime, the power-law index is usually held fixed at $\gamma^{\rm cib}=0.8$, as inferred from measured galaxy correlation functions at optical wavelengths~\citep[e.g.,][]{Zehavi2002}. 

\PEP\ reported  values of $\gamma^{\rm cib} \approx 0.82$--$0.96$ from fits to the \Planck\ auto  power spectra 
in the multipole range $\ell =200$--$2000$. However, in that analysis
the Poisson power levels were fixed to the model predictions
from~\citet{Bethermin2011}, which are lower than the Poisson power levels found in this paper and lower than the more recent model
predictions of \cite{Bethermin2012}. In the 2013 \Planck\ cosmological
parameter analysis~\citep{planck2013-p08}, the clustered CIB was
modelled as a power law with a Gaussian prior $\gamma^{\rm cib} =
0.7\pm 0.2$ together with  weak priors  on the Poisson
power levels. In the frequency range used for the CMB analysis,
100--217\GHz, consistently
low values, $\gamma^{\rm cib} = 0.40\pm 0.15$, were found from the  \Planck\ spectra.

We can anticipate values for the amplitude of the clustered CIB power based on previous measurements. For example, using the measured spectra reported in Table D.2 of~\cppcib, and subtracting their best-fitting Poisson powers, gives the following estimates:
\begin{eqnarray}
A^{\rm cib}_{353} \approx & 2.2\times10^3\,\mu {\rm K}^2 , \nonumber  \\
A^{\rm cib}_{545} \approx& 4.2\times10^5\,\mu {\rm K}^2 , \label{eqn:fcib} \\ 
A^{\rm cib}_{857} \approx & 1.1\times10^9\,\mu {\rm K}^2 .\nonumber
\end{eqnarray}
In our likelihood, we fit for these three amplitudes, plus three correlation coefficients that parametrize the cross-frequency spectra. For example, we take $A^{\rm cib}_{353\times545}=r^{\rm cib}_{353\times545}\sqrt{A^{\rm cib}_{353} A^{\rm cib}_{545}}$.

\subsubsection{Halo model for the CIB}
\label{sec:halomodel}

\begin{figure}
  \begin{center}
        \includegraphics[width=95mm]{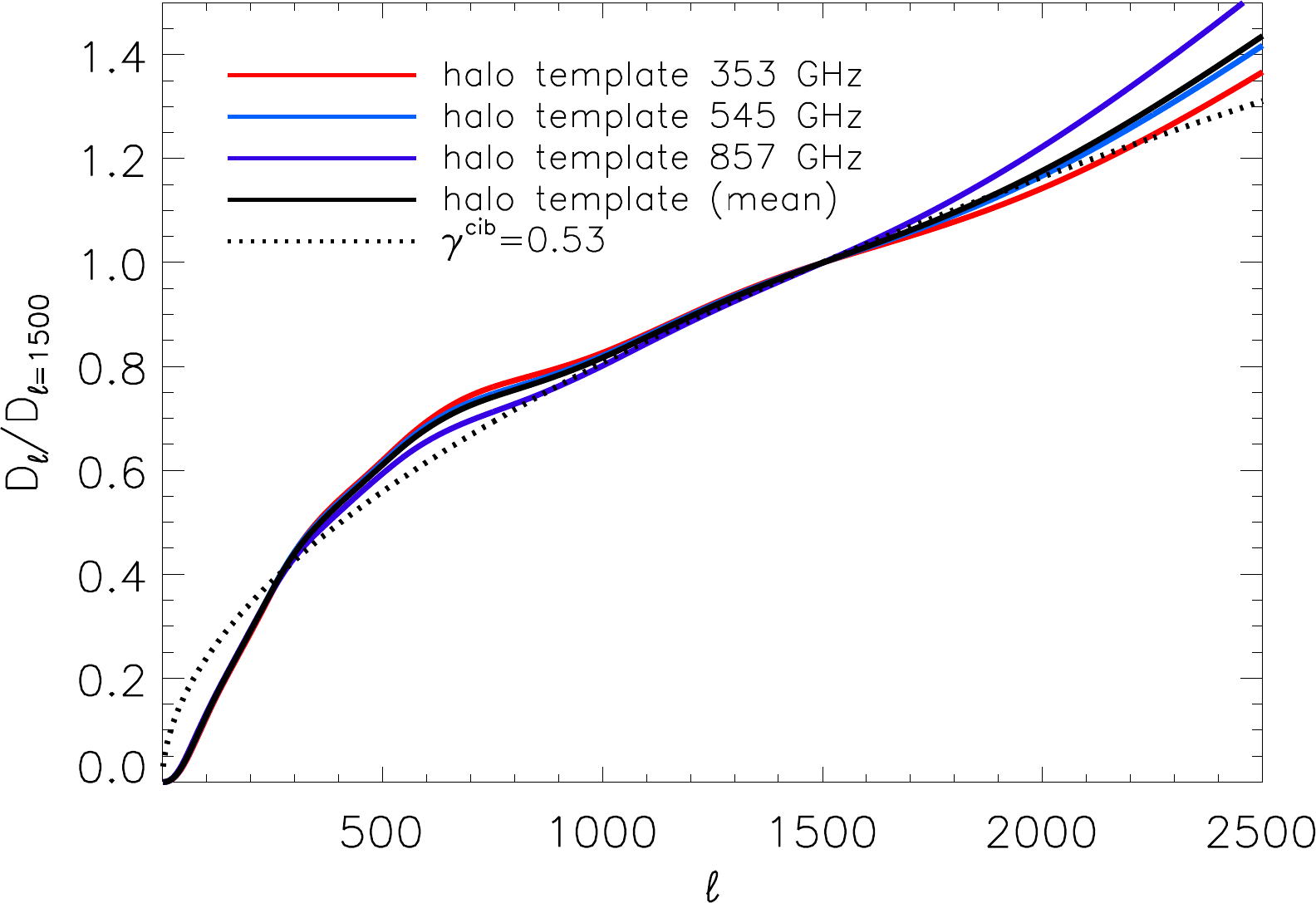}  
       \caption{Halo model templates from ~\cppcib\ for the three auto-frequency spectra at $353$, $545$, and $857$\,GHz, normalized to
unity at $\ell=1500$. The solid black line shows the fiducial halo
model template shape adopted in this paper (which is similar to
the~\cppcib\ template at 545\,GHz. The dotted line shows a power-law CIB spectrum with index  $\gamma^{\rm cib}=0.53$, the best fitting 
index found in our analysis.}
     \label{fig:cibtemp}
  \end{center}
\end{figure}

We also consider an extended halo model, as in ~\cppcib\ which
associates galaxies with dark matter halos and sub-halos, and adopts a
specific parametric relation between galaxy luminosity and mass of the
host sub-halo. In~\cppcib\, such a model is fitted to auto- and
cross-frequency spectra from \Planck\ and IRAS data in the frequency range
217--857\,GHz and 3000\,GHz, respectively. The auto-frequency spectra
of the ~\cppcib\ halo model at $353$, $545$, and $857$\,GHz are shown in
Fig. \ref{fig:cibtemp}.  Here the spectra are normalized, arbitrarily,
at $\ell=1500$. One can clearly see a transition from a two-halo term
dominating at multipoles $\ell \simlt 1000$ to a steeper behavior at
multipoles $\ell \simgt 2000$, where the one-halo term dominates.  The
dotted line in Fig. \ref{fig:cibtemp} shows a power-law with an index
$\gamma^{\rm cib} = 0.54$, close to the best fit spectral index found
in our likelihood analysis. Evidently over most of the multipole range
($500 \simlt \ell \simlt 2000$), the differences between these
template shapes are small. In our analysis, we have therefore
constructed a single halo model template shape from the mean of the three
auto-spectra plotted in Fig. \ref{fig:cibtemp}, which is shown as the
solid black line in the figure. (This is actually quite close to the
545\,GHz halo model template.) We use this mean template in most of our
analysis of the halo model, although in ~\refsec{compare} we
investigate the effect of 
 changing to one of the halo model templates of \cite{Viero2013}.

Our analysis of the halo model assumes the mean template shape plotted in Fig. 
\ref{fig:cibtemp}, normalized to unity at $\ell=2000$. The clustered CIB contribution 
to the \Planck\ spectra is then parameterised by three amplitudes  $A_\nu^{\rm cib}$ 
and three correlation coefficients $r_{\nu\times \nu'}^{\rm cib}$, as in our analysis of
the power-law model.

\subsection{CMB and thermal SZ}
\label{subsec:CMB}
The CMB power is significant at 353\GHz, but sub-dominant compared to
all other components at 545 and 857\GHz\ (see Fig. \ref{fig:cib} below,
which shows the various components deduced from the likelihood
analysis). In this paper, we do not fit for the CMB power spectrum
since it is determined to high precision by the \Planck\ data at lower
frequencies. We therefore fix the CMB spectrum to that of the
fiducial $\Lambda$CDM cosmology  described in~\refsec{intro}.  We do
not include thermal SZ signal in our model, since its amplitude is
negligible at frequencies $\nu \ge 353$\,GHz compared to the CIB
amplitude (see ~\cppcib).

\subsection{Galactic dust}
\begin{figure}
  \begin{center}
        \includegraphics[width=80mm]{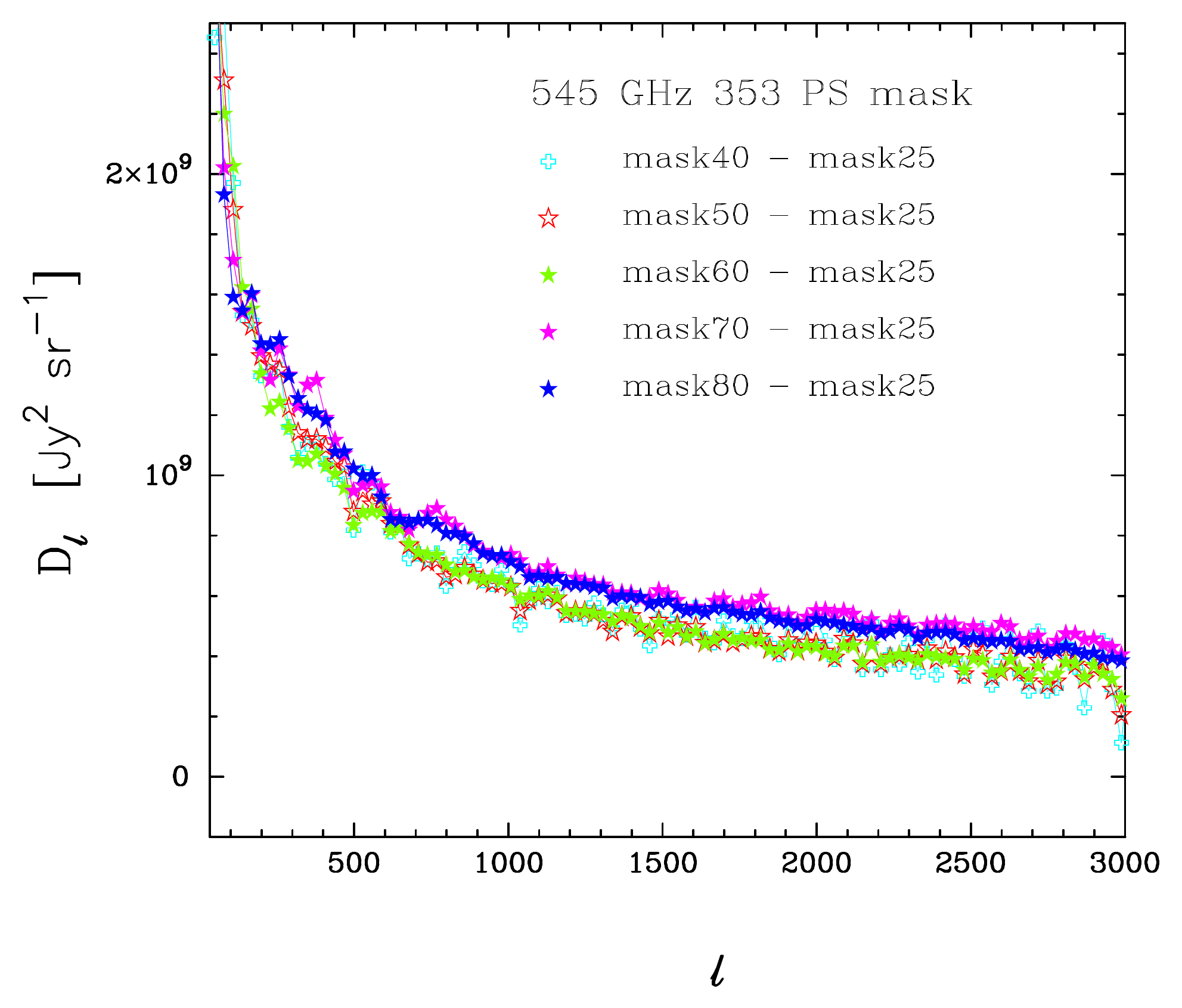}  
        \includegraphics[width=80mm]{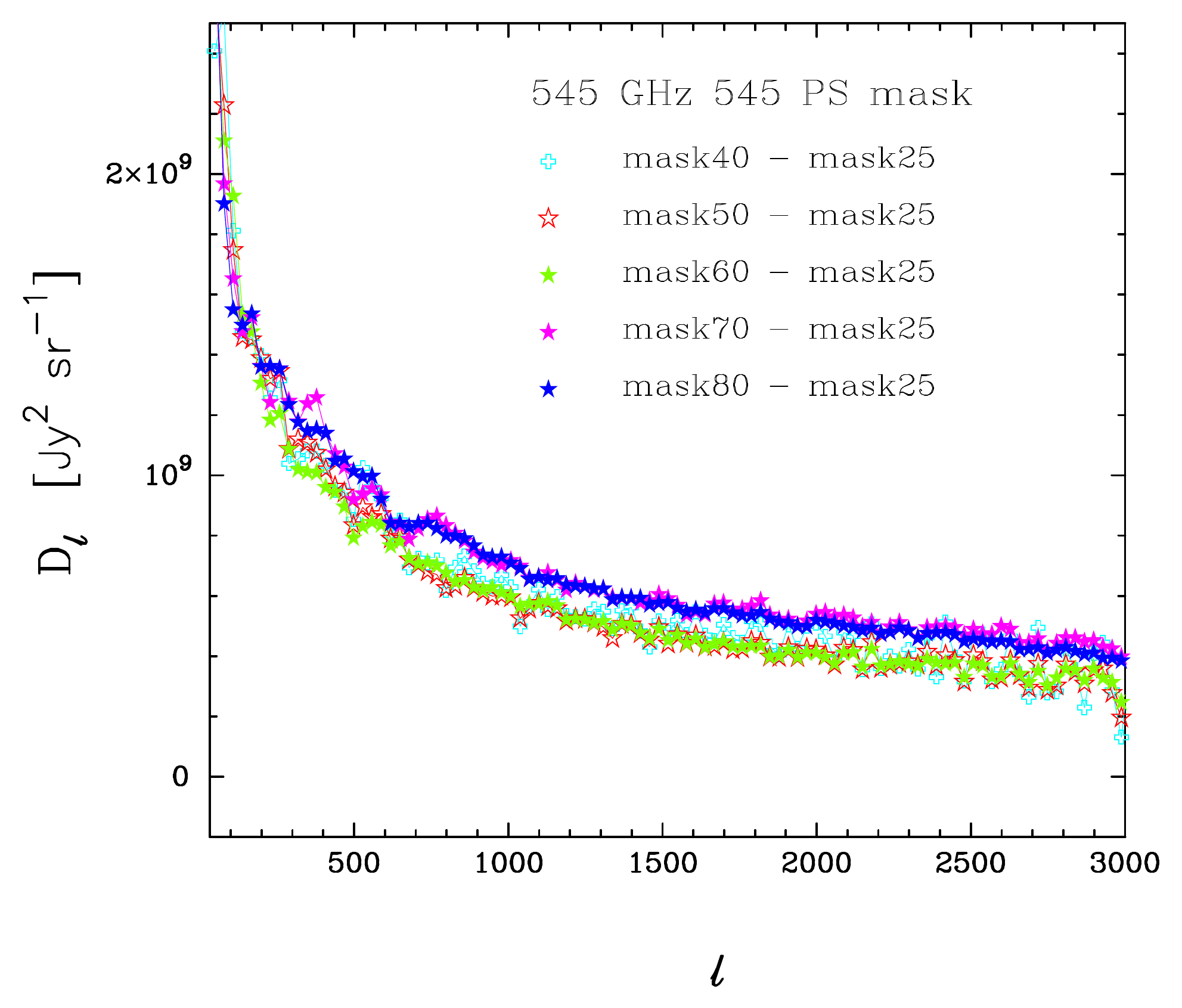}  
        \includegraphics[width=80mm]{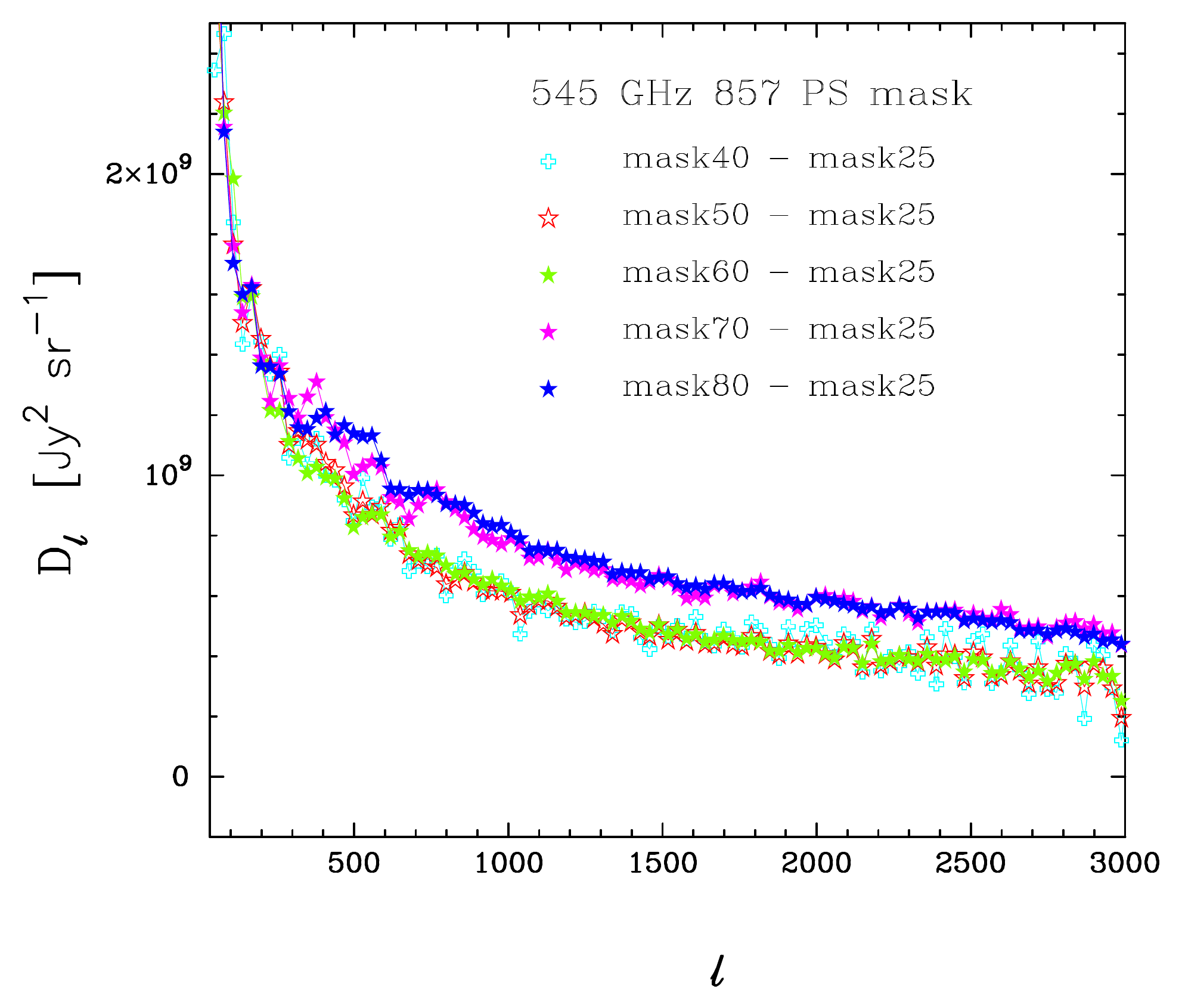}  
       \caption{Mask-differenced power spectra at 545\,GHz for the three point source masks used in this paper.
In each panel, the spectra have been rescaled to match the amplitude of the mask 40$-$mask 25 spectrum over the 
multipole range $100 \le \ell \le 300$.}
     \label{fig:dustspec}
  \end{center}
\end{figure}

\label{sec:dust}

Diffuse Galactic emission dominates over the CIB at low multipoles
($\ell \simlt 500$) even in the cleanest regions of the sky. We subtract
Galactic emission in the power spectrum domain following the approach
adopted in the \Planck\ CMB
analysis~\citep{planck2013-p08,planck2014-a13}.  The idea is to
construct a power spectrum template for  Galactic dust emission
by differencing the spectra computed on different masks. The isotropic
components, including CMB, CIB, and extragalactic sources, should
cancel in forming the difference leaving only a contribution from
Galactic dust emission. Our main assumption is that the
\emph{shape} of the power spectrum of Galactic dust emission outside
of a given mask can be approximated by a smooth fit to the
mask-differenced power spectrum. This provides templates for the dust
power spectra that are introduced into the likelihood analysis 
together with nuisance parameters to fix the amplitudes of the 
dust contribution at each frequency\footnote{We note that even in analyses using \ion{H}{i} as a tracer of 
dust emission (e.g.,~\PEP\ and~\cppcib), some assumptions need to be made to extrapolate the Galactic
dust spectrum to scales smaller than the resolution limit of the \ion{H}{i} data.}.

We begin by investigating the `universality' of Galactic dust
emission, i.e., whether the shape of the dust power spectrum varies
with frequency and/or sky area. Figure \ref{fig:dustspec} tests
variations of the shape of the dust spectrum with sky coverage. We
show the masked differenced 545\,GHz spectra for each of the point
source masks used in this paper. In each case, we have rescaled the
amplitudes of the spectra to match the amplitude of the mask 40-mask 25
spectrum over the multipole range $100 \le \ell \le 300$. Up to
mask 60, the power spectra are almost independent of the point source
mask, and scale accurately independent of the sky coverage. We begin
to see some departures from universality for mask 70 and mask 80, which
are stronger  when we apply the 857\,GHz point source mask compared
to the 353\,GHz point source mask.  The departures from universality
over these large sky areas are caused mainly by the anisotropy of the
point source masks (see Fig. \ref{fig:mask}) rather than by 
changes in the properties of the diffuse Galactic dust
emission. Anisotropy in the point source masks arises, for example,
from not masking sources in regions of strong Galactic cirrus. It
induces statistical anisotropy in the unmasked sources, potentially leaving a
non-zero contribution to the mask-differenced spectra in the mean. However,
up to mask 60
there is no evidence of any change in the shape of the dust power spectrum with
sky area for any of the point source masks used in this paper. We find very similar
behaviour for the 353\,GHz and 857\,GHz mask-differenced spectra.

\begin{figure}
  \begin{center}
        \includegraphics[width=85mm]{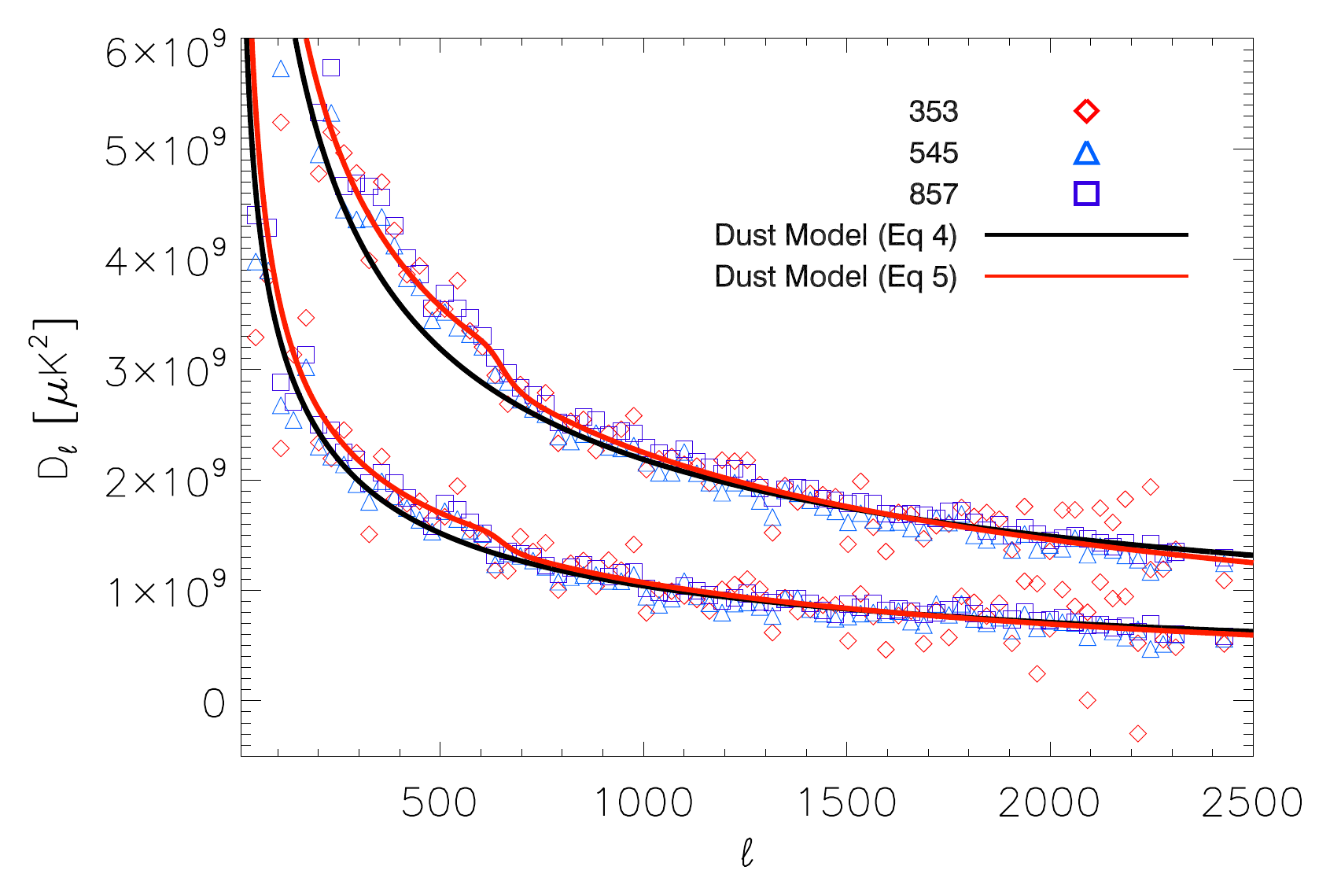}  
       \caption{Mask-differenced power spectra at 353\,GHz (red diamonds),
         545\,GHz (blue triangles), and at 857\,GHz (purple squares). The 353 and 545\GHz\
 spectra have been rescaled to match the amplitudes of the 857\GHz\ spectra. Results are shown for
         the difference of masks 40 and 25 and masks 50 and 25, the
         latter having larger amplitude. The union point source masks are used for all of these estimates. 
         Dust templates based on Eqs.~(\ref{eqn:dust})
         and~(\ref{eqn:dust_mask50}) are shown by the black and red solid lines, respectively.}
     \label{fig:dustf}
  \end{center}
\end{figure}

Next, we test for variations in the shape of the dust power spectrum
with frequency.  Figure~\ref{fig:dustf} shows the 353, 545, and
857\GHz\ mask-differenced spectra for two mask sizes, mask 40$-$mask 25
and mask 50$-$mask 25, adopting the union point source mask. The 353 and
545\GHz\ spectra have been rescaled to match the amplitude of the 857\,GHz spectrum. For each of the two masks shown here, the dust power
spectra at each frequency are closely similar. In fact, the
universality of the dust power spectrum as a function of frequency
extends to larger sky areas, but for our purposes the higher amplitude
of Galactic dust emission and the associated higher cosmic variance at
lower multipoles limits the useful area over which we can separate CIB and
Galactic dust at low multipoles. For this reason, most of our results are 
based on likelihod fits to spectra computed over 50\,\% of the sky or less.
{\it Over this area of sky, the results shown in Figs. \ref{fig:dustspec} and \ref{fig:dustf}
demonstrate  that the Galactic dust power spectrum is accurately universal over sky area
and frequency.}

\begin{figure}
  \begin{center}
        \includegraphics[width=85mm]{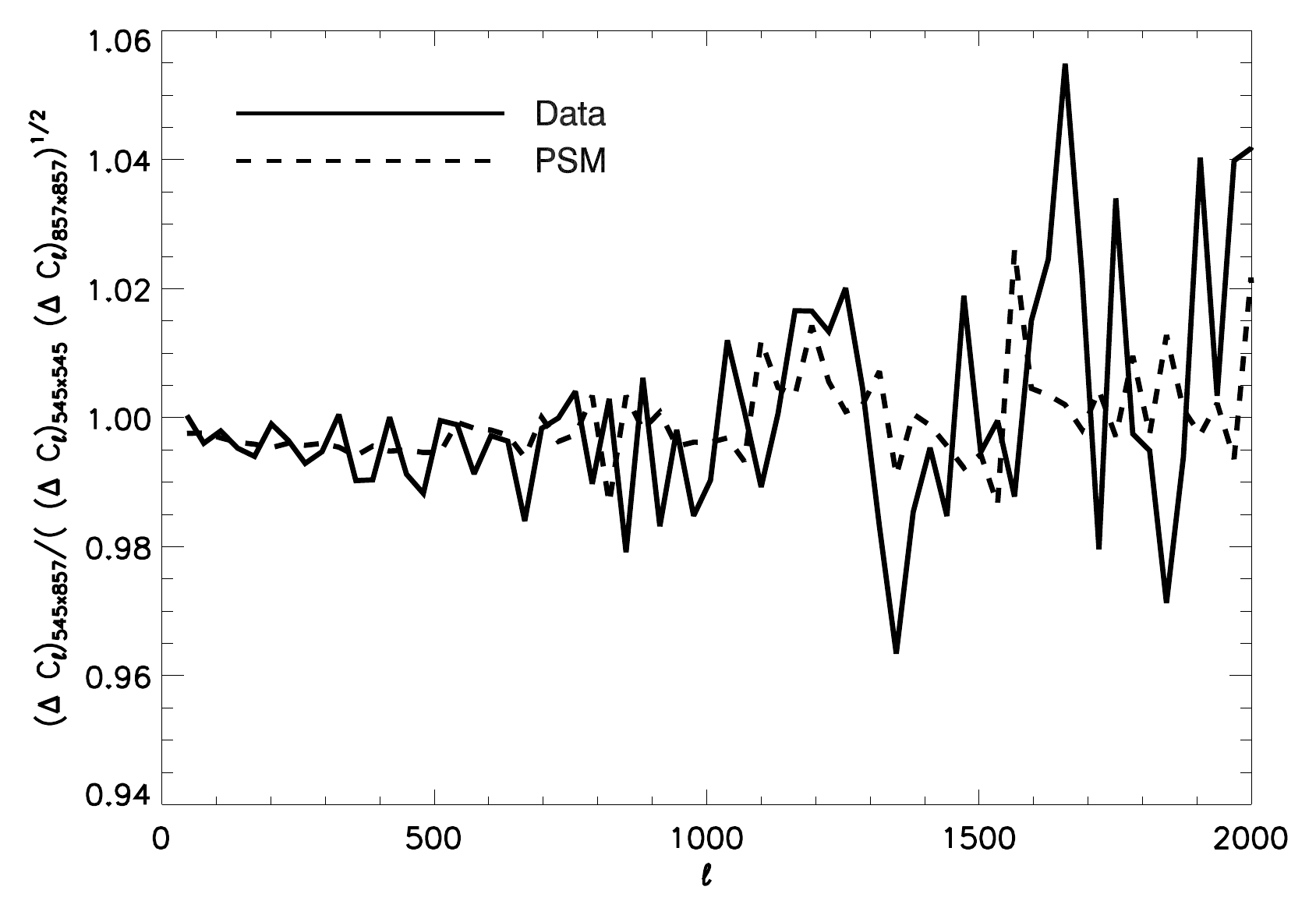}  
       \caption{Measured correlation coefficient of the mask-differenced power spectrum for $545\times 857$ between masks 40 and 25 (solid line). This is in good agreement with the correlation coefficient measured from the 545 and 857\GHz\ dust maps in the Planck Sky Model (dashed lines). Dust is more than 99\,\% correlated between these frequencies.}
     \label{fig:dustcorr}
  \end{center}
\end{figure}

We assume further that the dust emission does not decorrelate across
frequencies, i.e., we set $A^{\rm dust}_{\nu \times \nu'} =
(A^{\rm dust}_{\nu} A^{\rm dust}_{\nu'})^{1/2}$.\footnote{Unlike the
  \Planck\ analysis of cosmological parameters \citep{planck2014-a13},
  in this paper we always use the same Galactic mask for all
  frequencies.}
Some variations in the spectral energy distribution (SED)
of dust across the sky are expected from local variations in the dust
temperature and frequency dependence of the opacity (due to variations
in the dust grain properties). SED variations would 
lead to decorrelation across frequencies, with $C_\ell^{{\rm
    dust},\nu\times \nu'} \neq (C_\ell^{{\rm dust},\nu\times
  \nu}C_\ell^{{\rm dust},\nu'\times \nu'})^{1/2}$. We can test the
degree of correlation across frequencies by forming a mask-differenced
spectrum between different frequencies, e.g., $\hat{C}_\ell^{\nu\times
  \nu'}(\text{mask 40}) - \hat{C}_\ell^{\nu\times \nu'}(\text{mask
  25})$. The resulting correlation coefficient for $\nu=545\,{\rm
  GHz}$ and $\nu'=857\,{\rm GHz}$ is shown in~\reffig{dustcorr}. This
shows that any decorrelation is less than 1\,\%, and so we ignore it
in our dust modelling. The high level of correlation shown in Fig. \ref{fig:dustcorr}
is consistent with that expected given the SED variations inferred from analysis of
\Planck\ maps and the IRAS $100\,\mu{\rm m}$ map
in~\citet{planck2013-p06b}. This can be demonstrated by computing the 
mask-differenced correlation coefficients from the dust maps that form
part of the \Planck\ full focal plane
simulations~\citep{planck2014-a14}. Full details of the construction
of these dust maps are given in~\citet{planck2014-a14}; see
also~\citet{planck2016-XLVIII}.  Briefly, they are based on an
estimate of the dust emission at 353\GHz, constructed with the
Generalized Needlet Internal Linear Combination method
(e.g.,~\citealt{planck2016-XLVIII}), which is scaled in frequency in
each pixel as a modified blackbody with temperature and spectral index
taken from~\citet{planck2013-p06b}. The measured correlation of these
dust maps is shown in~\reffig{dustcorr} and agrees well with our analysis of the 
545 and 857\GHz\ \Planck\ maps.

In the \Planck\ 2013 analysis of cosmological parameters, the
following simple parameteric model for the dust power spectrum was adopted:
\begin{equation}
\mathcal{D}_{\ell}^{{\rm dust},\nu\times\nu'} = A^{\rm dust}_{\nu\times\nu'} \frac{(100/\ell)^\alpha}{[1+(\ell/\ell_c)^2]^{\gamma/2}} .
\label{eqn:dust}
\end{equation}
The parameters $\alpha$, $\ell_c$, and $\gamma$ are fixed by matching
to the mask-differenced dust power spectra. Given the universality of
the dust power spectrum, neither the choice of frequency nor mask sizes
are critical in fixing these parameters.  Using the 857\GHz\
mask 40$-$mask 25 spectrum, we find $\alpha=0.387$, $\ell_c=162.9$, and
$\gamma=0.168$, so that asymptotically $C_\ell^{\rm dust} \propto\ell^{-2.55}$. 
This $\ell$ dependence at high multipoles is compatible with previous
high-resolution measurements of the dust power spectrum in high
Galactic latitude cirrus clouds, $C_\ell \propto \ell^{-2.7 \pm
  0.1}$~\citep{Miville2002,Miville2010}, although our mask-differenced results consistently give
slopes at the shallower end of this range. For a given mask, the dust
power in the $\nu\times\nu'$ spectrum is specified by the parameter
$A^{\rm dust}_{\nu\times\nu'}$. For the values of $\alpha$, $\ell_c$,
and $\gamma$ given above, $A^{\rm dust}_{\nu\times\nu'}$ is the dust
power at multipole $\ell=10$. 

The solid black lines in Fig. \ref{fig:dustf} show the model of
Eq.~(\ref{eqn:dust}) compared to the mask 40$-$mask 25 and mask 50$-$mask 25 spectra for 
the 353, 545, and 857\GHz\ spectra. The model provides a reasonably good fit but 
one can see evidence for an excess at $\ell \approx 500$. We have therefore fit the dust spectra to 
a slightly more complicated fitting function: 
\begin{equation}
\mathcal{D}_{\ell}^{\rm dust, mask50}\propto \left ( \frac{100}{\ell} \right )^{\alpha_2} + B \left ( \frac{\ell}{1000} \right )^{\beta_2} \left[1+\left(\frac{\ell}{\ell_{c2}}\right)^{\gamma_2}\right]^{\delta} ,
\label{eqn:dust_mask50}
\end{equation}
where $B$, $\alpha_2$, $\beta_2$, $\ell_{c2}$, $\gamma_2$, and
$\delta$ are parameters that we fit to the $857\times 857$
mask 50$-$mask 25 spectrum. The best fit parameters are $B= 0.74$, $\alpha_2 = 0.19$, $\beta = -0.58$, $\ell_{c2}=543.1$, 
$\gamma_2=11.6$, and $\delta = 0.07$.   We normalise the template at $\ell=10$, so
that the dust amplitudes $A^{\rm dust}_{\nu\times\nu'}$ derived using the templates
of Eqs.~(\ref{eqn:dust_mask50}) and~(\ref{eqn:dust}) can be compared directly.
The best-fit template of  Eq.~(\ref{eqn:dust_mask50}) is shown by the red lines in Fig. \ref{fig:dustf}.
Equation~(\ref{eqn:dust_mask50}) provides a better fit to both the mask 40$-$mask 25 and mask 50$-$mask 25
spectra than Eq.~(\ref{eqn:dust}), capturing the small excess at $\ell \approx 500$. Unless otherwise
stated, we use the template shape of  Eq.~(\ref{eqn:dust_mask50}) in our baseline analysis.

In summary, using mask-differenced spectra to isolate the contribution
from Galactic dust, we find that the dust power spectrum has a nearly universal
shape independent of frequency and sky fraction up to mask 60. For larger sky fractions
we see relatively small departures from universality that depends on the point source
mask and are most likely associated with the anisotropy of the point source masks at
low Galactic latitudes. The shape of the dust power spectrum is modelled accurately
by the fitting function of Eq.~(\ref{eqn:dust_mask50}), which provides a template shape
used in our likelihood analysis. We have verified that the dust power spectra are almost
perfectly correlated over the frequency range $353$--$857$\GHz. We therefore solve for 
three dust amplitudes, $A^{\rm  dust}_{353}$, $A^{\rm  dust}_{545}$, and $A^{\rm  dust}_{857}$ in our 
likelihood analysis assuming that the dust power spectrum is completely correlated across
frequencies.

\subsection{Poisson power from unmasked point sources}
\label{sec:ps}
The Poisson contribution to the power spectra arises from shot noise
of extragalactic sources. \cppcib\ estimate that the Poisson power
from radio sources is less than $2\,\%$ of that from infrared sources at
$\nu \ge353$\GHz.\footnote{Note that the  effective flux-density cuts for  point source masks
at 353\GHz\  used in~\cppcib\ and in this paper are quite similar
(around $315\, {\rm mJy}$ for~\cppcib\ compared to $400\,{\rm mJy}$ here).}
As a result, we consider here only the Poisson
power from infrared sources.  The Poisson contribution to the power
spectrum is constant, $C^{\rm ps}_\ell=\text{const}$. We model this
contribution  with a single amplitude parameter, giving $\mathcal{D}_\ell^{\rm
  ps}$ at multipole $\ell=2000$, for each auto-frequency spectrum ($A^{\rm
  ps}_{353}$, $A^{\rm ps}_{545}$, and $A^{\rm ps}_{857}$) and a
cross-correlation coefficient for each cross-frequency spectrum (e.g., $A^{\rm
  ps}_{353\times545}=r^{\rm ps}_{353\times545}\sqrt{A^{\rm ps}_{353}
  A^{\rm ps}_{545}}$). In the rest of this section we summarize briefly
the methodology that we have used to calculate the Poisson power levels from
source counts. These calculations are not used directly in the
main likelihood analysis; instead they serve as a rough consistency check of
the Poisson power that we infer from the \Planck\ spectra.

\begin{figure}
  \begin{center}
        \includegraphics[width=77mm]{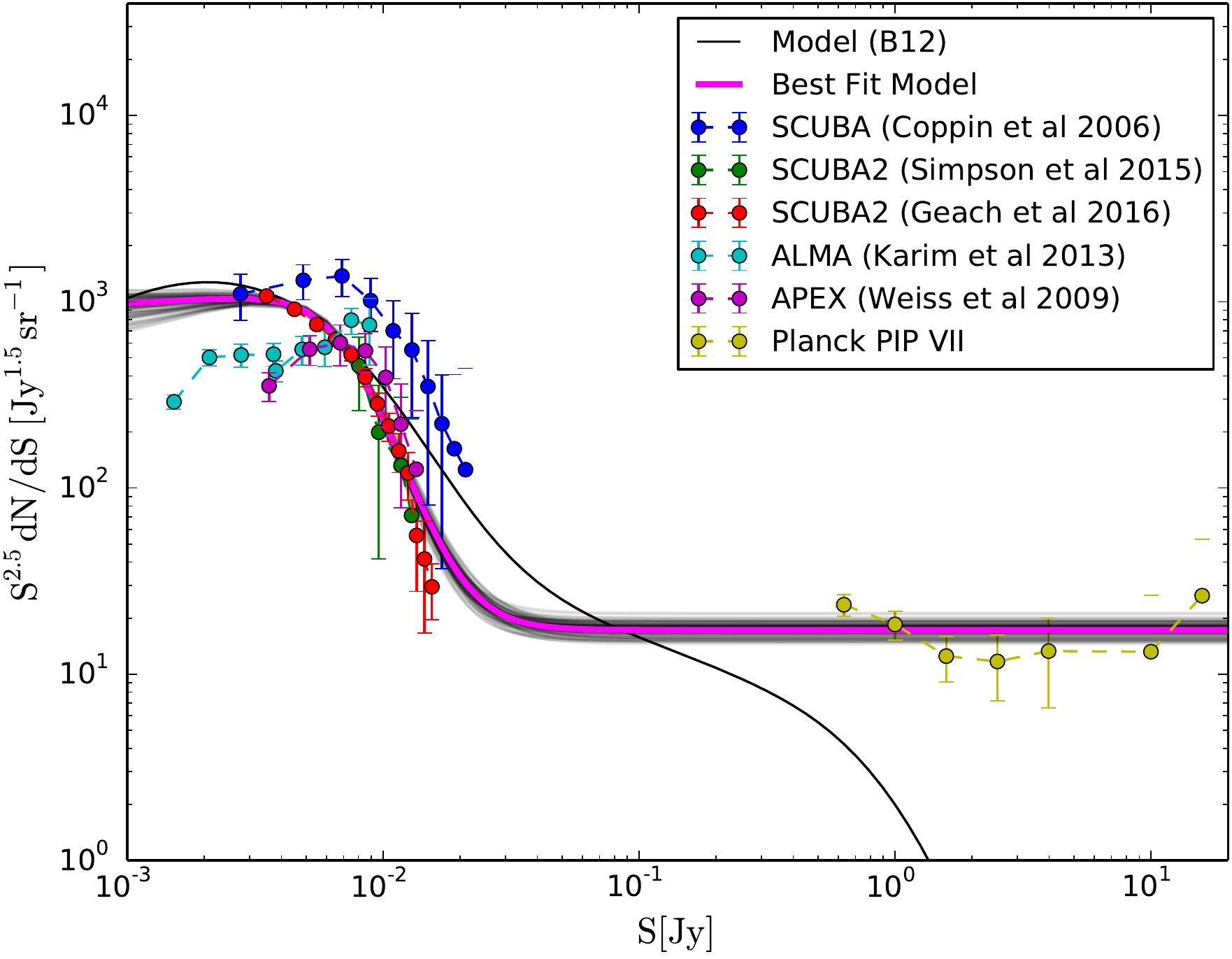}  
        \includegraphics[width=77mm]{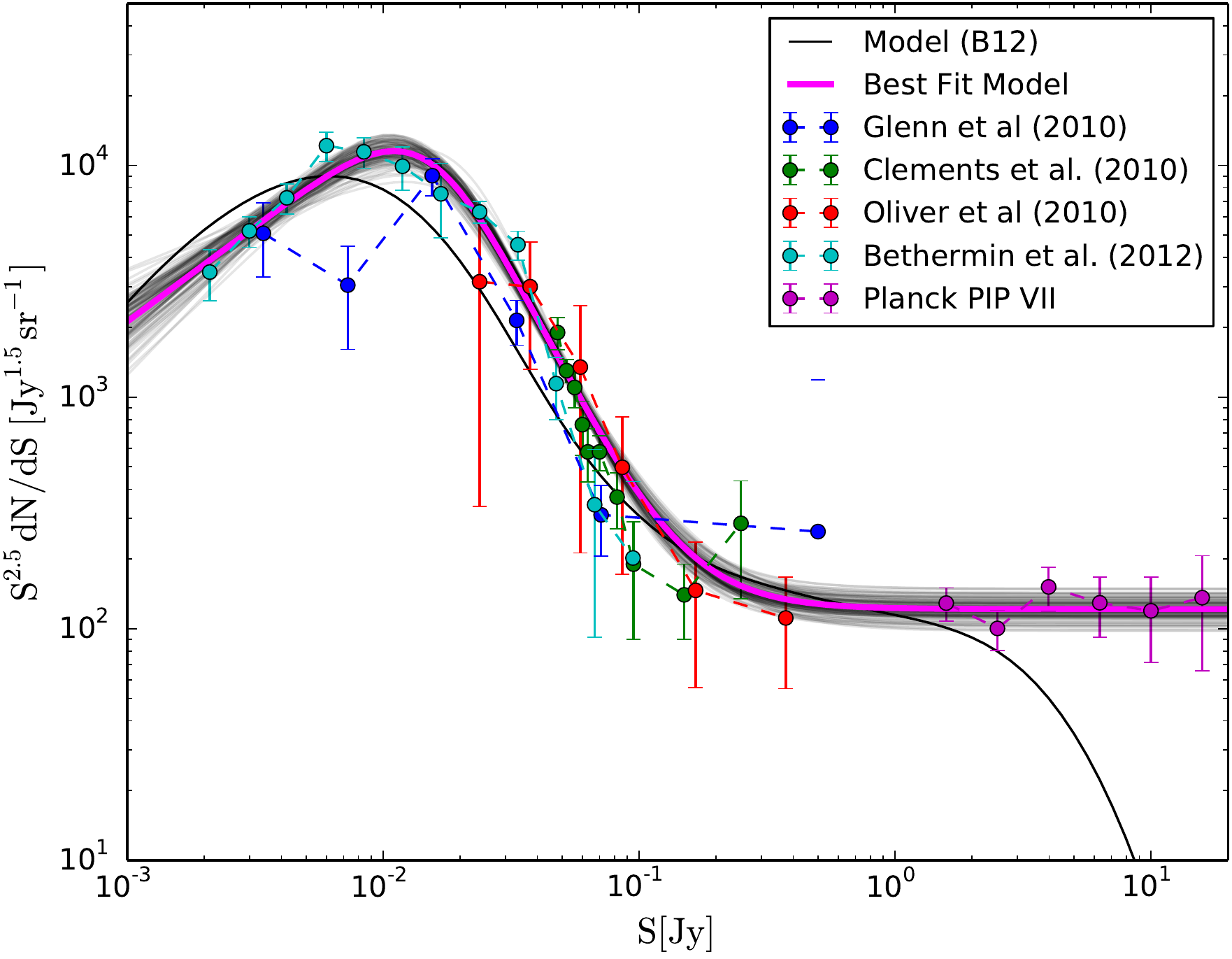}  
        \includegraphics[width=77mm]{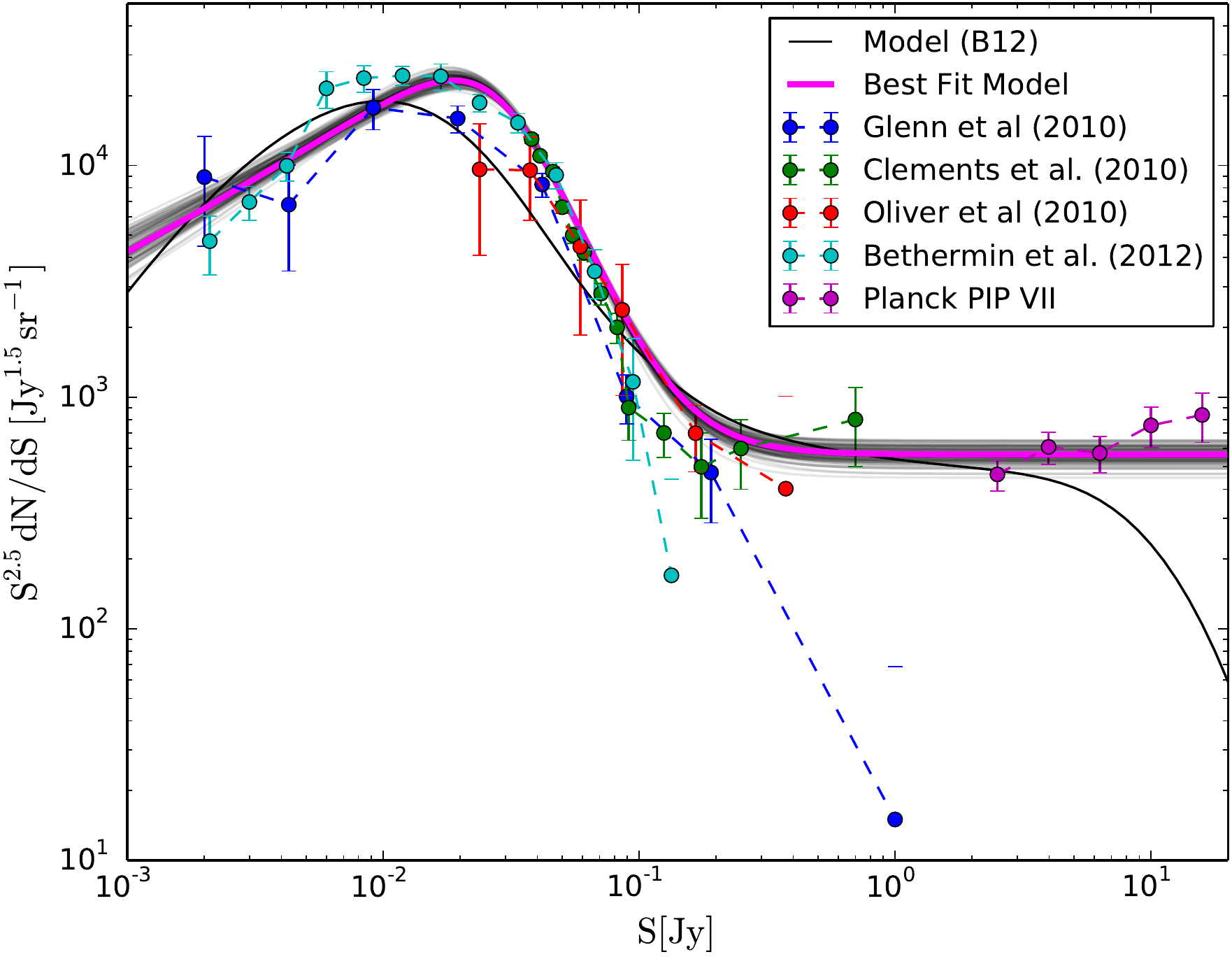}  
\caption{Differential number counts from measurements with
  \Planck~\citep{planck2012-VII},
  \Herschel~\citep{Oliver2010,Clements2010,Glenn2010, Bethermin2012b},
  and APEX, SCUBA, and
  ALMA~\citep{Coppin2006,Weiss2009,Simpson2015,Geach2016} at
  $870\,\mu\text{m}$ or $850\micron$ (345\GHz; top), 
$500\,\mu\text{m}$ (600\GHz; middle), and $350\,\mu\text{m}$ (857\GHz;
bottom). Also shown (thick purple lines) are the best-fit double power-law models (together with the models from individual MCMC chains in grey to give some indication of the statistical uncertainty in the model) and the \bcount\ model prediction (thin black lines). The break in the \bcount\ model at high flux-densities comes from the minimum redshift considered in the model ($z=10^{-3}$). Note that for $870\,\mu\text{m}$
the datasets from~\citet{Coppin2006,Weiss2009,Simpson2015} are shown here for illustration only and  are not used in our fits (see~\refapp{ps}). 
The \Planck\ counts have been colour corrected to match the central
frequencies of the external data.}
     \label{fig:sourcecount}
  \end{center}
\end{figure}

The Poisson power at frequency $\nu$ from sources below a flux-density limit is
\begin{equation}
C^{{\rm ps},\nu\times \nu}_{\ell} = \int_0^{S_{\nu,{\rm cut}}}{S_\nu^2 \frac{dN}{dS_\nu} dS_\nu} \, ,
\label{eqn:ps}
\end{equation}
where $dN/dS_\nu$ is the differential number count measured at
frequency $\nu$, and $S_{\nu,{\rm cut}}$ is an effective flux-density cut
above which sources are masked.  The Poisson power from infrared
sources is dominated by sources well below the \Planck\ flux density
cuts. We must therefore rely on theoretical models for $dN/dS_\nu$, or
measurements of number counts from higher-resolution experiments
capable of detecting sources to  lower flux densities.  Here, we
model the source count distribution by fitting to number counts
determined from experiments at frequencies similar to \Planck. For
this purpose, we use  number counts at 350 and 500\,$\mu$m from
\Herschel~\citep{Oliver2010, Clements2010, Glenn2010,Bethermin2012b},
and at 850\,$\mu$m from the SCUBA2 Cosmology Legacy Survey (S2CLS;~\citealt{Geach2016}). We also plot source count measurements at $870\,\mu\text{m}$:
APEX/LABOCA observations ~\citep{Weiss2009} of the Extended Chandra Deep Field South (ECDF);
SCUBA~\citep{Coppin2006,Simpson2015}; and ALMA pointings of the ECDF LABOCA field~\citep{Karim2013}. Since these are based on smaller areas of sky, and are not in agreement with each other, 
 we use only the larger catalogue from S2CLS (together with \Planck\
 counts) to predict the Poisson power level. We  comment on the differences in the number counts 
 from different surveys around 850\micron\ in~\refsec{psprior}.
At higher flux-density, we use the source counts
from~\citet{planck2012-VII}. The details of the individual data sets, 
(including colour corrections to the \Planck\ bands), source count model
and our fitting procedure are described in~\refapp{ps}. The source counts are
plotted in~\reffig{sourcecount}~\citep[see also  Fig.~10 of][for an earlier version 
of a similar figure]{planck2012-VII}.

 Briefly, we fit a simple model consisting of a double power-law
 term and a constant term to the
 observed counts $S^{2.5} dN/dS$ using MCMC\footnote{We use {{\tt
       emcee}}, an open-source Python
   package~(\citealt{Foreman2013}; \url{http://dan.iel.fm/emcee/current/}) to perform the MCMC.}, propagating the
 uncertainties in the model parameters to our predicted Poisson power
 levels for each frequency band.  Our fits to the counts are shown by
the purple lines and grey bands in~\reffig{sourcecount} 
and provide good fits to the data, except 
at $850\,\mu\text{m}$ where the data are discrepant. This discrepancy is discussed in more
detail in~\refapp{ps}. The solid black lines in~\reffig{sourcecount} show 
number counts predicted from the theoretical models of 
~\citet[hereafter B12]{Bethermin2012}; see
also~\citet{Bethermin2012b}.  Qualitatively, the \bcount\ model
matches the source counts at $500\,\mu\text{m}$ and
$350\,\mu\text{m}$, but not at $870\,\mu\text{m}$, 
where the model predicts more 
faint sources at flux-densities below $10\,{\rm mJy}$ than reported by 
\cite{Weiss2009} and \cite{Karim2013}. The \bcount\ model is in agreement with the new counts of
\citet{Geach2016} at the faint flux-densities, but predicts more
bright sources at flux-densities greater than $10\,{\rm mJy}$. 

From our MCMC fits to the number counts, the integrals of Eq.~(\ref{eqn:ps})
give the following amplitudes and $1\,\sigma$ statistical errors:
\begin{eqnarray}
A^{\rm ps}_{353}&=& (1.4\pm 0.1)\times10^3\,\mu {\rm K}^2 , \nonumber  \\
A^{\rm ps}_{545}&=& (3.7\pm 0.2)\times10^5\,\mu {\rm K}^2 ,   \label{eqn:PSpred} \\
A^{\rm ps}_{857}&=& (9.0 \pm 0.3)\times10^8\,\mu {\rm K}^2 . \nonumber
\end{eqnarray}
These numbers are computed using the nominal flux-density limits appropriate 
to our point source masks at each frequency  as the upper limits in Eq.~(\ref{eqn:ps}).
However, since the point source amplitudes are dominated by faint sources  the
predictions of Eq.~(\ref{eqn:PSpred}) are rather insensitive to the upper limits; for example,
doubling the upper limits increases the predicted amplitude at 353\GHz\ by 14\,\% and
the 545 and 857\GHz\ amplitudes by a few percent. The statistical
errors quoted in Eq.~(\ref{eqn:PSpred}) almost certainly underestimate the
true errors given the poor quality of the fits
at $350\,\mu\text{m}$ and  $850\,\mu\text{m}$. This should be borne in
mind when comparing the predictions in Eq.~(\ref{eqn:PSpred})
with the Poisson power levels deduced from our likelihood analysis.

For reference, the predictions for the \bcount\ model (from Table 6
of~\cppcib) are $2.1\times 10^3$, $3.1\times 10^5$, and $6.9\times
10^8\,\mu{\rm K}^2$ at 353, 545, and 857\GHz, respectively. We note
that the model predictions that we quote here have been
colour-corrected from a CIB SED to the photometric convention $\nu
I_\nu = {\rm constant}.$ Generally, the modelled $C_\ell^{{\rm
    ps},\nu\times \nu'}$ has to be multiplied by a factor $cc_\nu
\times cc_{\nu'}$ in comparing to the data, where for the
colour-corrections $cc_\nu$ we adopt the values in~\cppcib: 1.097,
1.068, and 0.995 at 353, 545, and 857\GHz, respectively.

\section{CIB likelihood}
\label{sec:likelihood}

We construct a likelihood using the \camspec software, following
closely the approach described for the \Planck\ CMB analysis
in~\citet{planck2013-p08}. We form the six beam- and mask-deconvolved
power spectra, $\hat{C}_\ell^{\nu\times \nu'}$, in the multipole range
$50 \le \ell \le 2500$, by cross-correlating the yearly maps.
Although Galactic dust dominates the lowest multipoles for all
spectra, except for $353\times 353$ where the CMB is dominant, we
retain multipoles down to $\ell=50$. As we shall see below, the low amplitude of
the CIB compared to the dust power spectrum at multipoles $\ell \simlt 300$, together
with cosmic variance in the dust, means that it is difficult to separate these
components accurately at low multipoles.

We adopt a Gaussian likelihood for the $\hat{C}_\ell^{\nu\times
  \nu'}$, and include all correlations between the six cross-spectra
in the covariance matrix. The covariance matrix is calculated assuming
a fiducial model, and includes sample variance from the CMB, Galactic
dust, CIB, and also the \Planck\ anisotropic instrument
noise, using the analytic
expressions in~\citet{Efstathiou2004}; see~\refapp{cov} for a summary. 
We compare the measured spectra with the theoretical model predictions as  in 
Eq.~\eqref{eqn:model}. In doing so, the theoretical model is divided by the
product of the appropriate map-level calibration parameters, $cal_\nu
\times cal_{\nu'}$.  The parameters used in our likelihood analysis
and their prior ranges are summarised in~\reftab{par}.

In constructing the covariance matrix for the power spectra, we
approximate the CIB as Gaussian random fields, with total power spectra given by the
sum of the clustered and Poisson powers. We also treat the dust as a
statistically-isotropic, Gaussian random field, i.e., for a given mask
the sample variance is constructed from a power spectrum with
amplitude appropriate to the mask. These assumptions for dust are
clearly incorrect~\citep{Miville2007}. Much of the power comes from
regions close to the edge of the Galactic mask, so the effective
number of modes that contribute to the sample variance for a given
mask is lower than if the fields were statistically isotropic.  This
will increase both the variance of the measured power spectra and also
their covariance. In~\refsec{ngnoise} and~\refapp{dust}, we develop a heuristic model that
describes the dust emission as a Gaussian random field subject to a
large-scale, anisotropic modulation. The sample variance in this model
can be calculated straightforwardly. We show in~\refsec{ngnoise} that the corrections
from statistical anisotropy in this heuristic model have only a minor impact
on our estimation of CIB model parameters, although they do improve
the goodness-of-fit  of the high-frequency spectra
considerably.

Further discussion of the construction and validation of the likelihood can be found in~\refapp{cov}.

\subsection{Expected parameter errors and degeneracies}
\label{sec:fisher}

Before presenting the results of the likelihood analysis, it is
instructive to consider the expected errors on parameters and their
covariances  using a Fisher matrix analysis. We shall see
that our likelihood results using \Planck\ data, presented
in~\refsec{cs}, agree well with the forecasts summarised in this
section.

For our assumed Gaussian likelihood, with a fiducial covariance matrix, the Fisher matrix has elements
\begin{equation}
F_{ij}=\sum_{N,N'}\sum_{\ell ,\ell'}^{\ell_{\rm max}}{\frac{\partial C_{\ell}^{{\rm th},N}}{\partial p_i}{\rm Cov}_{(\ell\,N) (\ell'\,N')}^{-1}\frac{\partial C_{\ell}^{{\rm th},N'}}{\partial p_j}} \, , \label{eqn:fisher}
\end{equation}
where the $p_i$ are the model parameters, $C_\ell^{{\rm th},N}$ is the
theoretical power spectra from Eq.~\eqref{eqn:model}, ${\rm
  Cov}_{(\ell\,N)( \ell'\,N')}$ is the covariance matrix of the
spectra, and the sums over $N$ and $N'$ are over the six distinct
frequency combinations $\nu\times \nu'$. The partial derivatives are
calculated at the fiducial values of the parameters. Since the
parametric model is very simple, the derivatives can all be computed
analytically.  The fiducial values that we adopt for the model
parameters are given in~\reftab{fisher}; they are chosen to be close
to the best-fit values from our likelihood analysis for mask 40.  For
this Fisher analysis, we shall ignore the uncertainty in the
absolute calibrations by fixing the calibration factors to unity. The calibration
factors are totally degenerate with the dust, CIB, and Poisson
amplitudes.

\begin{table*}
\caption{\emph{Forecasted} $1\,\sigma$ errors on the CIB (power-law model), dust, and Poisson power levels for mask 40. The calibration parameters are  fixed to unity in this Fisher analysis. The forecasted errors are shown assuming different prior information on the levels of Poisson power: no prior (4th column); Gaussian priors with a $20\,\%$ relative error in $A^{\rm ps}_\nu$ (5th column); and $A^{\rm ps}_\nu$ fixed at their fiducial values (6th column). All amplitude parameters have units of $\mu{\rm K}^2$.}
\begin{center}
\begin{tabular}{lccccc}
\hline\hline
Types & Parameter & Fiducial value & \multicolumn{3}{c}{$1\,\sigma$ forecasted error}   \\
          &      & & No prior & $20\,\%$ prior & Fixed \\
\hline
CIB & $A_{353}^{\rm cib}$ & $2.2\times10^3$ & $8.7\times10^1$ &$8.3\times10^1$&$3.2\times10^1$ \\
       &$A_{545}^{\rm cib}$ & $4.1\times10^5$ & $1.2\times10^4$&$1.2\times10^4$& $3.4\times10^3$ \\ 
       &$A_{857}^{\rm cib}$ & $8.4\times10^8$ & $3.3\times10^7$  &$3.2\times10^7$&$1.1\times10^7$\\ 
      &  $r_{353\times545}^{\rm cib}$ & 0.95 & 0.01& 0.01&0.01 \\
& $r_{353\times857}^{\rm cib}$ & 0.89 & 0.01& 0.01& 0.01 \\
& $r_{545\times857}^{\rm cib}$ & 0.95& $<0.01$& $<0.01$& $<0.01$\\
& $\gamma^{\rm cib}$ & 0.50 & 0.04 & 0.04& 0.02 \\
\hline
Dust &$A_{353}^{\rm dust}$ & $3.3\times10^3$ & $1.3\times10^2$&$1.3\times10^2$& $1.0\times10^2$ \\
         &$A_{545}^{\rm dust}$ & $8.9\times10^5$ & $2.1\times10^4$&$2.1\times10^4$&$1.6\times10^4$ \\ 
        &$A_{857}^{\rm dust}$ & $5.1\times10^9$ & $7.0\times10^7$ &$7.0\times10^7$&$4.4\times10^7$\\ 

\hline
PS &$A_{353}^{\rm ps}$ & $2.3\times10^3$ & $1.0\times10^2$ &$9.7\times10^1$&--\\
         &$A_{545}^{\rm ps}$ & $3.6\times10^5$ & $1.4\times10^4$&$1.4\times10^4$&--\\ 
        &$A_{857}^{\rm ps}$ & $8.4\times10^8$ & $2.9\times10^7$ &$2.8\times10^7$&--\\ 

& $r_{353\times545}^{\rm ps}$ & 0.96& 0.01&0.01&0.01\\
& $r_{353\times857}^{\rm ps}$ & 0.84&0.02&0.02&0.01\\
& $r_{545\times857}^{\rm ps}$ & 0.95&$<0.01$&$<0.01$& $<0.01$\\
\hline
\end{tabular}
\end{center}
\label{t:fisher}
\end{table*}

\begin{figure}
  \begin{center}
       \includegraphics[width=75mm, bb=0 0 441 391]{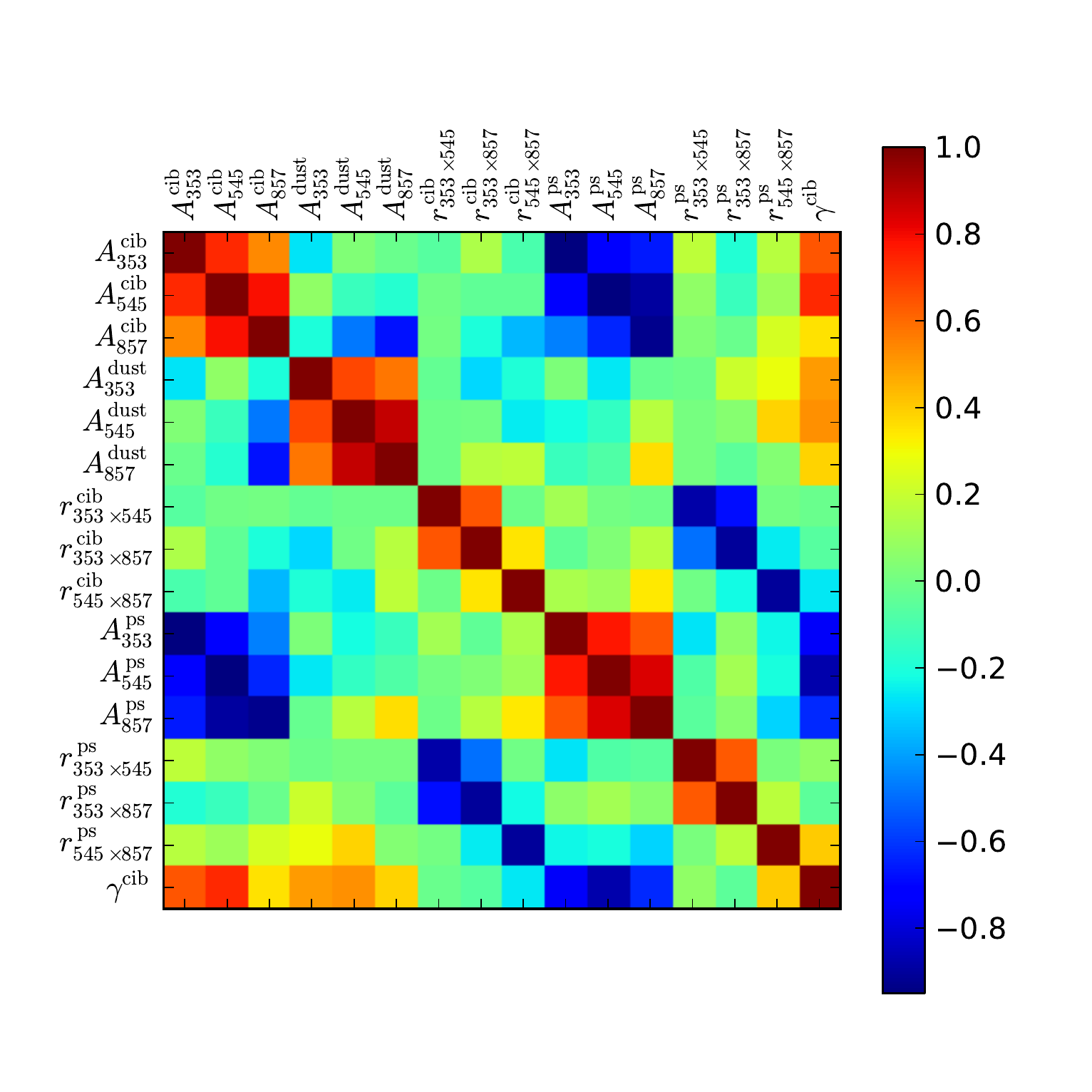}  
        \includegraphics[width=75mm, bb=0 0 441 391]{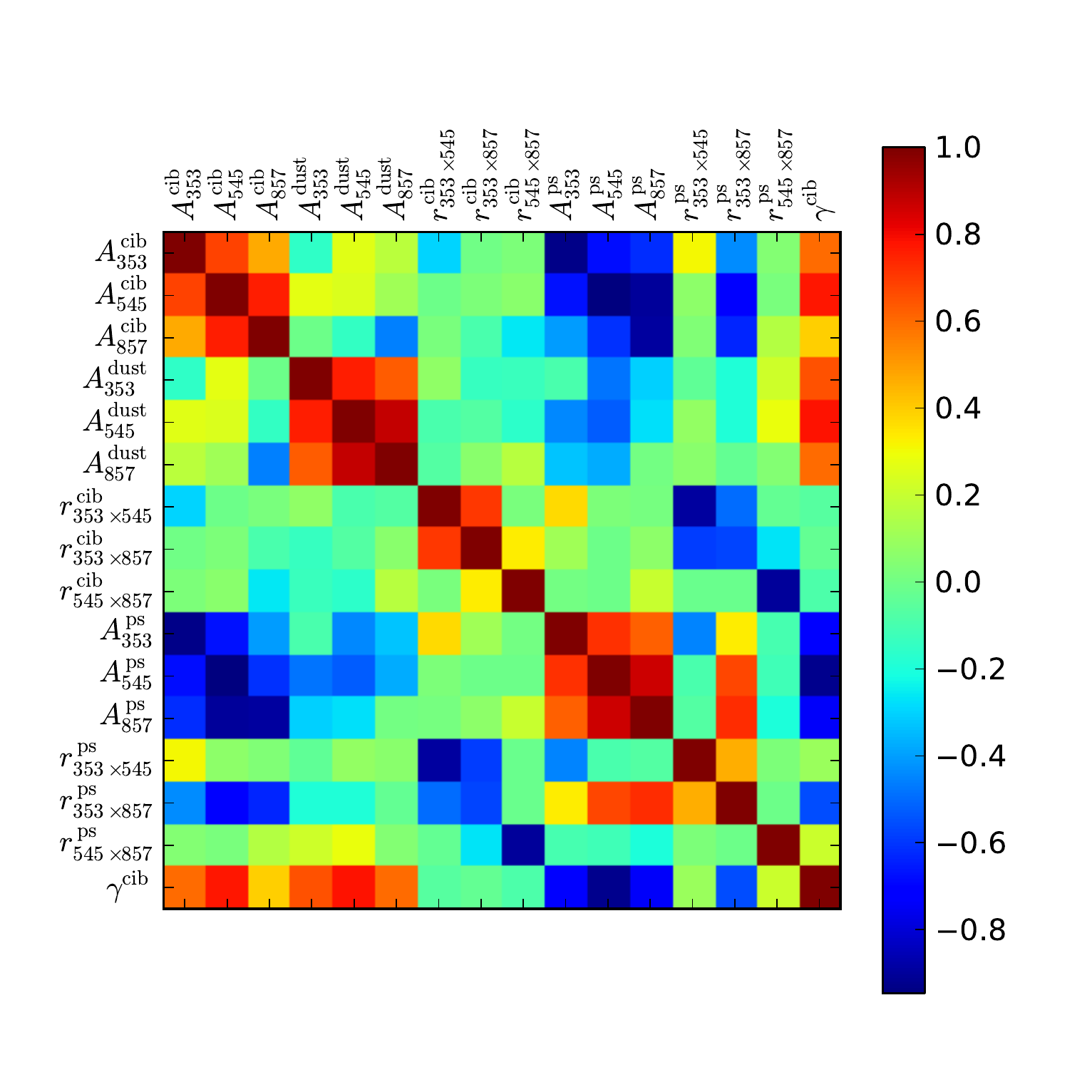} 
        \includegraphics[width=75mm,bb=0 0 441 391]{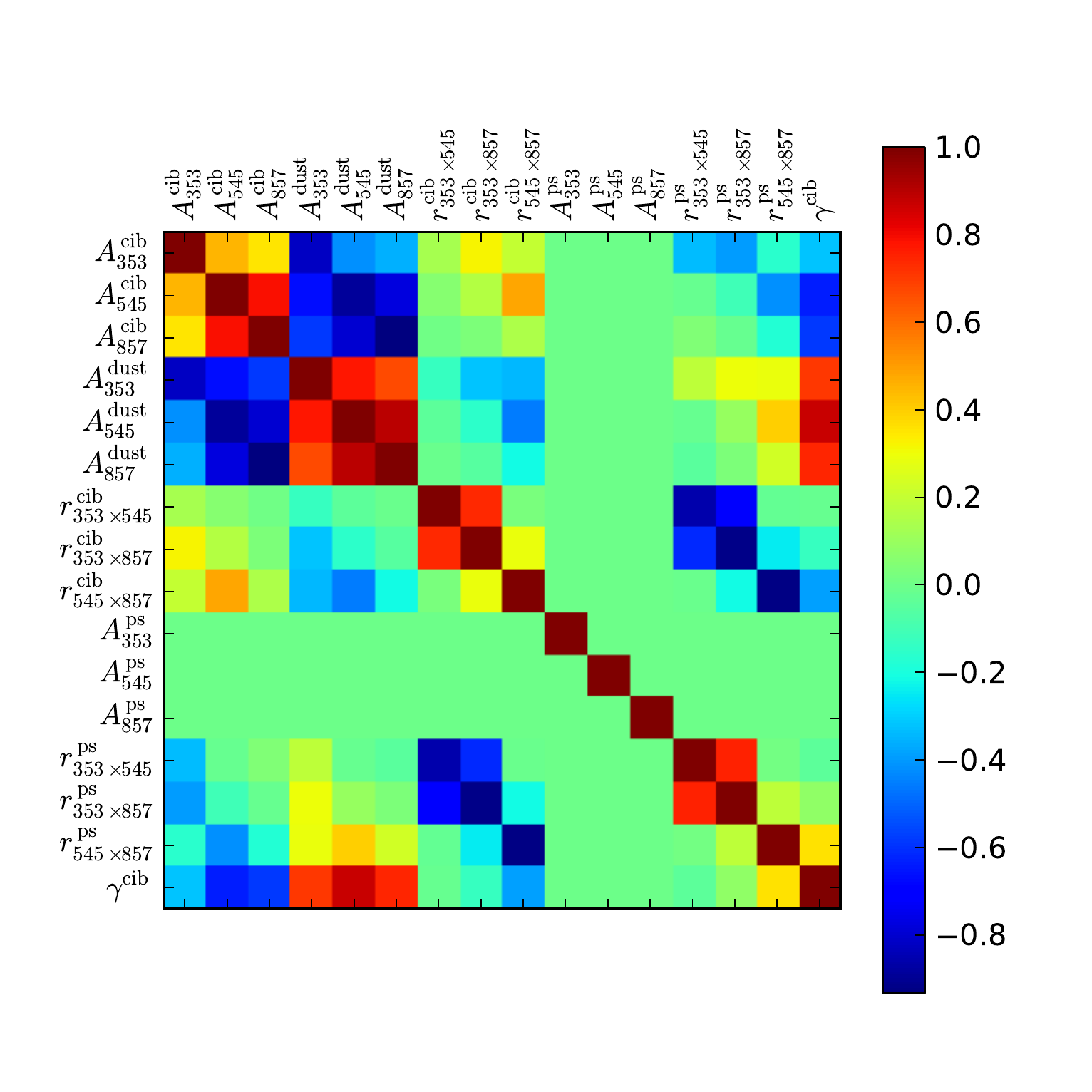}  
       \caption{Correlation matrix for the 16 parameters describing
         the CIB (power-law model), dust, and Poisson power levels for
         mask 40. The top panel shows the correlations estimated from
         MCMC samples of the likelihood for the \Planck\ data,
         marginalising over the calibration parameters. The middle
         plot shows the correlations calculated from the Fisher
         matrix, for fixed calibration parameters, and is very similar
         to what we find from the likelihood analysis. The bottom
         panel shows the effect of fixing the Poisson power levels in
         the Fisher matrix analysis. }
     \label{fig:fisher}
  \end{center}
\end{figure}

As an example of the results of the Fisher analysis, the forecasted
(marginal) parameter errors are shown for mask 40 and the power-law
CIB model  in~\reftab{fisher}. We find that the amplitude parameters
can be constrained to about $4\,\%$ precision or better, the
correlation coefficients for the CIB and Poisson powers to better than
$2\,\%$, and the power-law index of the CIB spectra to around $8\,\%$
precision.

The correlation matrix of the 16 model parameters is shown in the
middle panel of~\reffig{fisher}, and is very similar to what we
estimate from the MCMC samples in the full likelihood analysis (top
panel, which includes marginalisation over the calibration
parameters).  There are strong degeneracies amongst some
parameters. For example, the dust amplitudes are positively correlated
across frequencies, as are the CIB and Poisson amplitudes. This
behaviour arises because over the frequency range that we analyse, 
Galactic emission is nearly perfectly correlated across frequencies,
and the CIB emission is also strongly correlated. Sample variance fluctuations
in the CIB, for example, affect all frequency spectra similarly,
positively correlating the errors on the CIB amplitude parameters
across frequencies. A further significant degeneracy exists between the
CIB amplitudes and the Poisson amplitudes at each frequency. As expected, these are
strongly anti-correlated because their power spectra have  similar shapes.
 We note also the CIB power-law
index, $\gamma^{\rm cib}$, is positively correlated with the CIB
amplitudes. This is a consequence of our choice of the pivot scale at
$\ell=2000$. For \Planck, most of the statistical weight on the
amplitude and shape of the clustered CIB spectrum comes from
multipoles smaller than this, for which an increase in $\gamma^{\rm
  cib}$ can be compensated by an increase in the amplitude.
Similarly, $\gamma^{\rm cib}$ is positively correlated with the dust
amplitudes, which get most weight from low multipoles, but is
anti-correlated with the Poisson amplitudes, which get most weight
from $\ell > 2000$.

\subsubsection{Effect of priors on levels of Poisson power}
\label{sec:pardeg}

Separating the clustered CIB from the Poisson power is difficult.
With \Planck, we are able to measure the power spectra at multipoles
$\ell \simlt 3000$, where theoretical models predict clustered spectra that
are shallower than Poisson (see Fig. \ref{fig:cibtemp}).  Power
spectrum measurements at higher multipoles do not necessarily eliminate
the degeneracy because 
 in physical models the power spectrum of the clustered CIB
steepens at non-linear scales, exacerbating the
degeneracy with the Poisson power. Discriminating between a clustered
and Poisson contribution, purely based on shape information of the
power spectra, is therefore sensitive to both the multipole range and the
assumed shapes of the clustered CIB spectra.  

An alternative is to impose priors on the Poisson power estimated from
source counts (Sec.~\ref{sec:ps}), which can be obtained from
high-resolution, deep observations over limited areas of the sky. This
approach is also not straightforward, since it 
requires accurate number counts, with accurate error estimates (including sample variance) extending to low flux-density levels.  To assess the impact of constraining
the Poisson amplitudes, we 
augment our Fisher analysis with the following priors on the $A^{\rm
  ps}_{\nu}$: Gaussian priors with relative errors of 20\,\% of the
fiducial value\footnote{Note that in analysis of~\cppcib, 
priors on the Poisson power levels were imposed based on the \bcount\
models,  with estimated errors of 20\,\%.}; and very narrow priors that fix the Poisson levels
completely. In both cases we do not apply any priors on the Poisson
correlation coefficients, $r^{\rm ps}_{\nu\times \nu'}$. These examples are meant to be illustrative. The 
formal errors on the point source amplitudes predicted from number counts given in Eq.~(\ref{eqn:PSpred}) are 
significantly smaller than 20\%. However, as mentioned in Sect. \ref{sec:ps}
 these formal errors may well underestimate the true errors.
(We shall see later that there is a large discrepancy at 353\,GHz in the Poisson power levels measured
from \Planck\ and those predicted by the counts of \citealt{Weiss2009} and \citealt{Karim2013}.)

Our results with these priors are presented in~\reftab{fisher}
and~\reffig{fisher}. With the Poisson power levels constrained at the
20\,\% level, the parameter constraints and correlation properties are
hardly changed (We do not show the associated correlation matrix
in~\reffig{fisher} for this reason.). However, when we fix the Poisson amplitudes,
 the uncertainties on both the CIB and dust amplitudes
improve by at least a factor of two, and similarly for the power-law
index $\gamma^{\rm cib}$. There are also some changes in the
correlations between parameters. For example, the CIB and dust
amplitudes are markedly anti-correlated, whereas there is no
significant correlation without the tight priors on the $A^{\rm
  ps}_\nu$. Moreover, the amplitude and power-law index of the CIB
switch from being positively correlated to anti-correlated. This
is because the statistical weight in the determination of
the amplitude and shape of the clustered CIB shifts to smaller scales
when the Poisson power levels are fixed.

These results suggest that imposing priors on the Poisson power levels
derived from source counts in our analysis of the \Planck\ data has
little impact unless the source count predictions are accurate to much
better than 20\,\%. Even then, significant degeneracies between
parameters remain.

\section{Angular power spectra and parameter results from PLANCK}
\label{sec:cs}

\begin{figure*}
  \begin{center}
  \begin{tabular}{cc}
    \includegraphics[width=85mm]{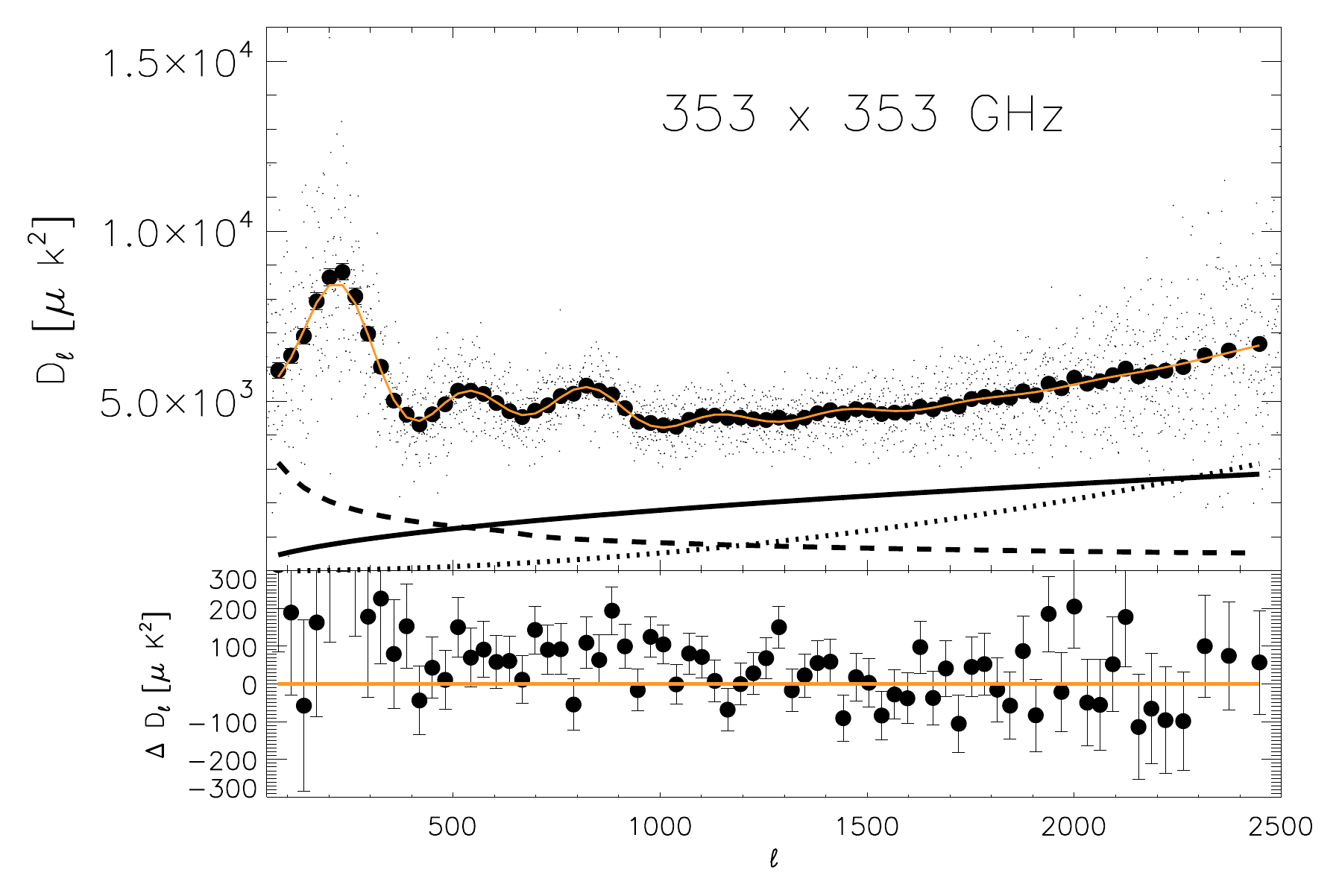} &                 \includegraphics[width=85mm]{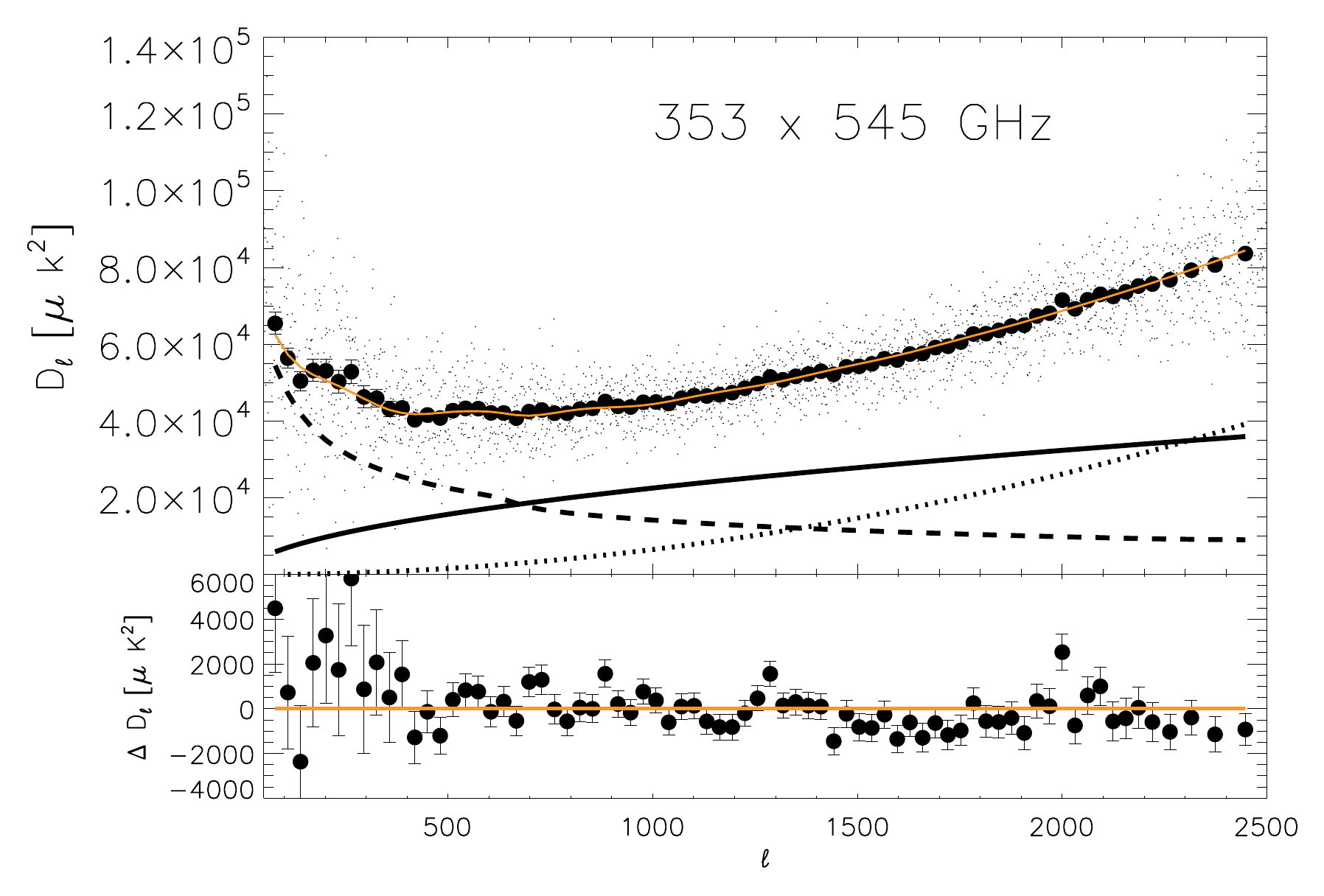} \\
    \includegraphics[width=85mm]{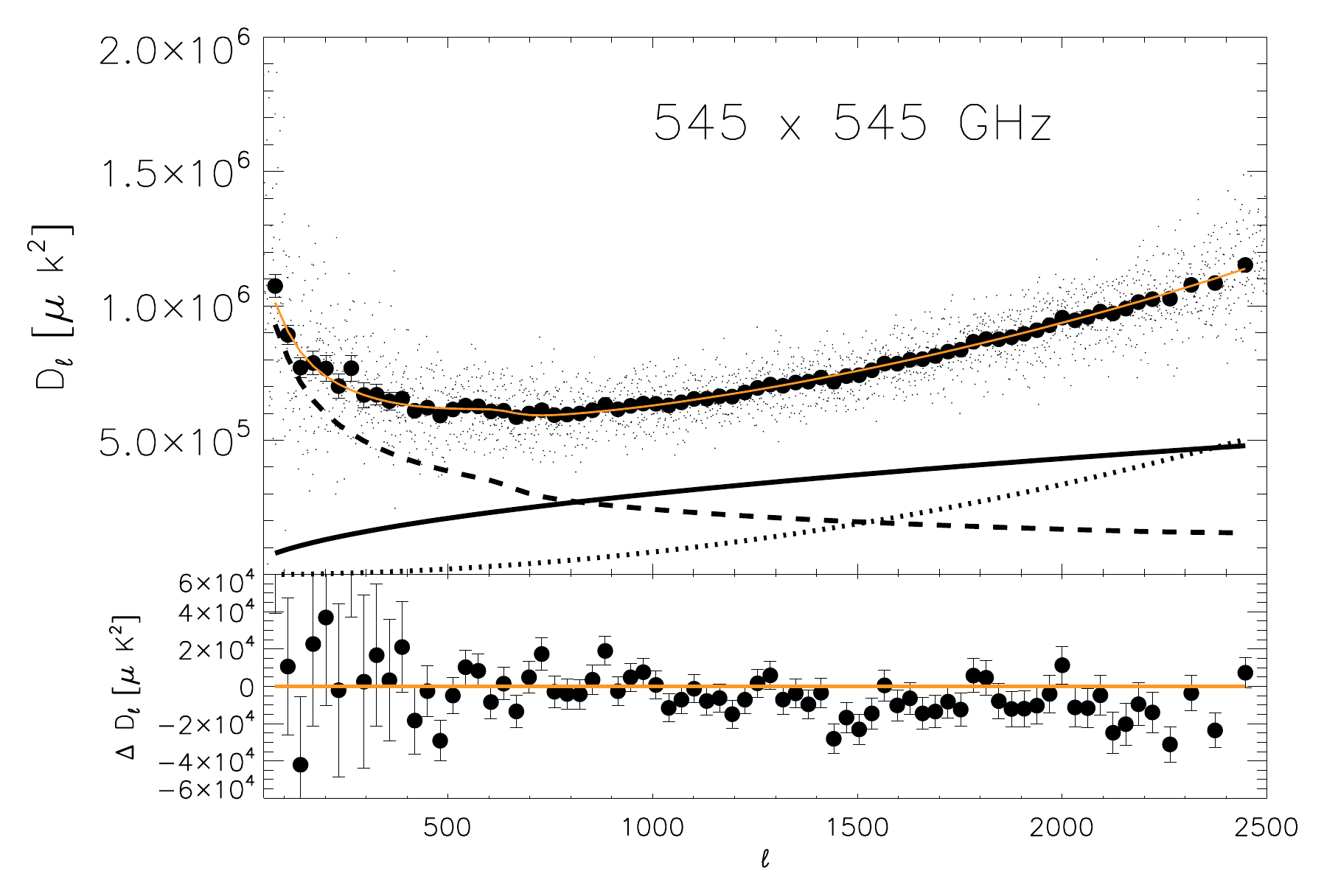} &                 \includegraphics[width=85mm]{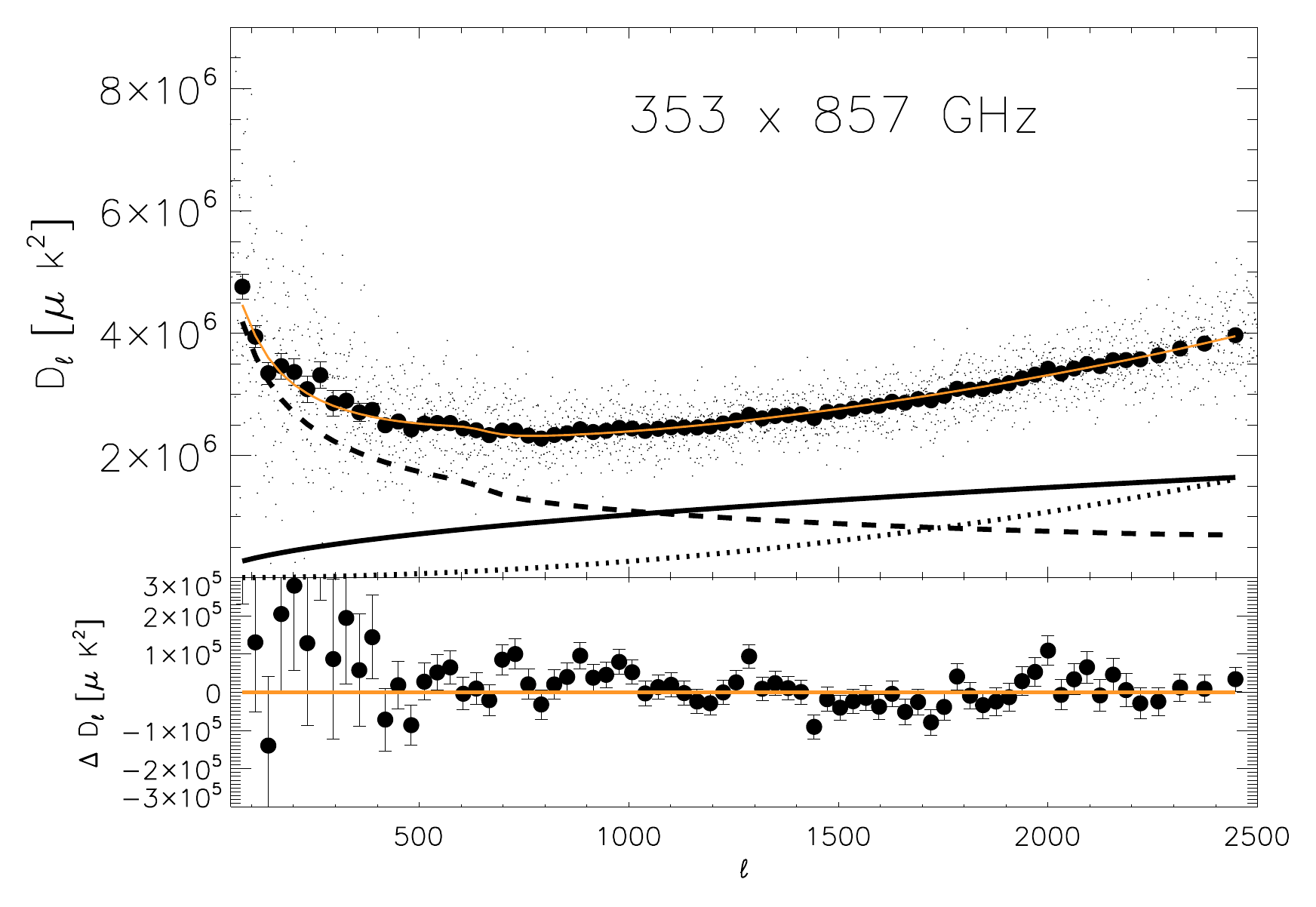} \\
    \includegraphics[width=85mm]{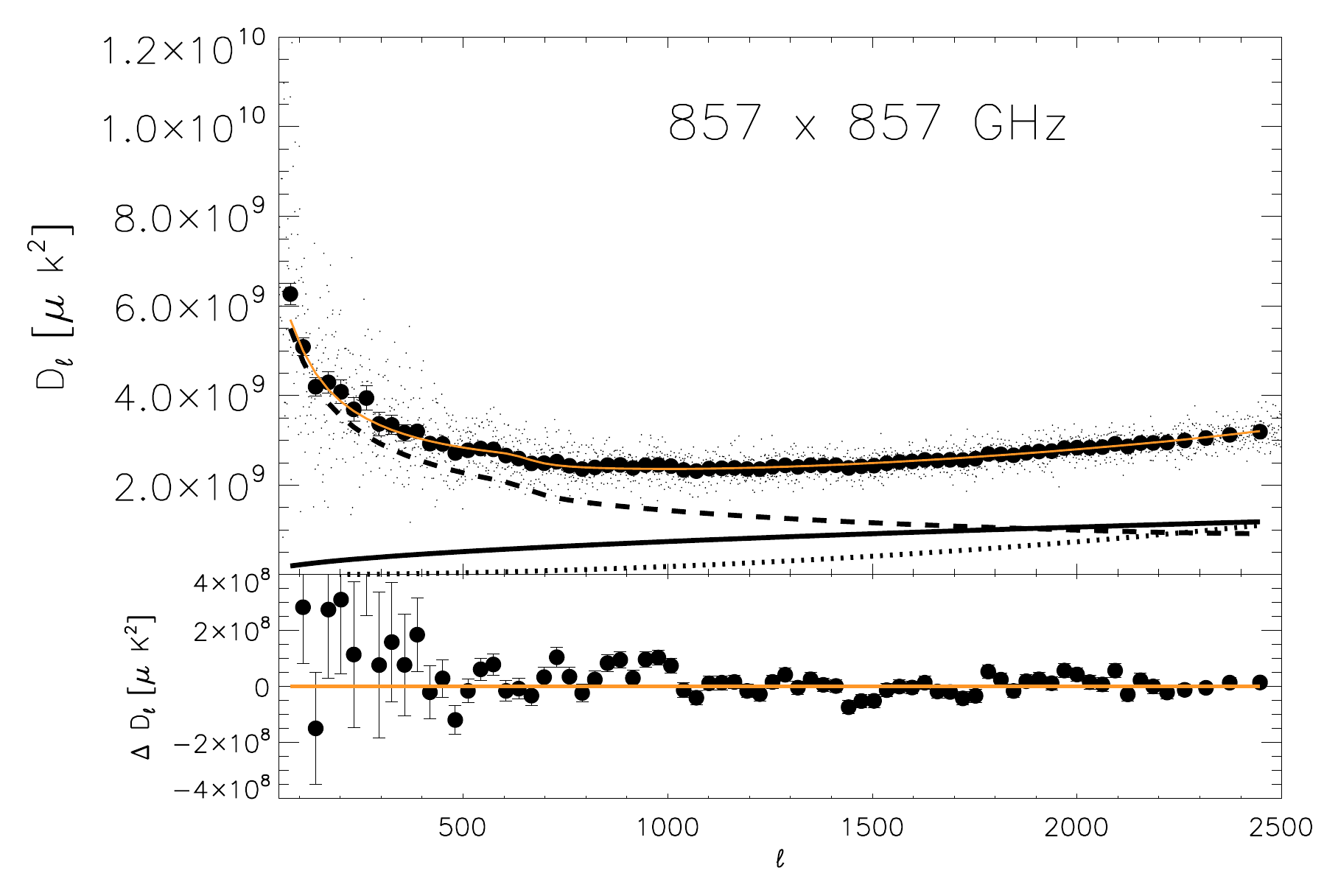} &                 \includegraphics[width=85mm]{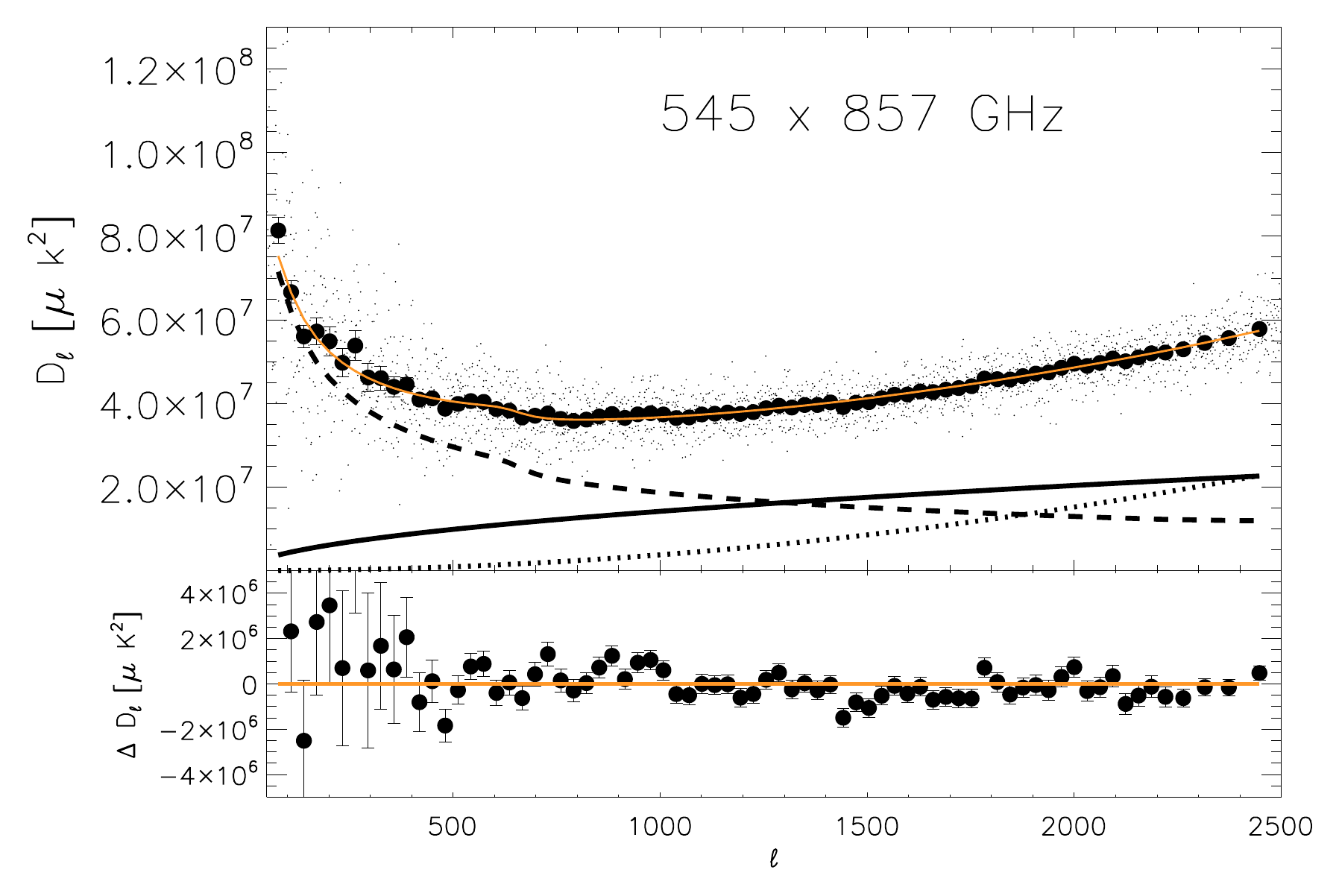} 
   \end{tabular}
       \caption{Measured angular power spectra for mask 40, constructed from cross-correlating Year-1 and Year-2 maps. The spectra are
 binned in flat bandpowers (for $\mathcal{D}_\ell$) with $\Delta \ell = 31$. The lines show the best-fit parametric models (based on the power-law CIB model): total (orange solid lines); clustered CIB (black solid lines); Galactic dust (black dashed lines); and Poisson power from the shot noise of extragalactic sources (black dotted lines). The error bars on the binned spectra are $\pm1\,\sigma$ errors determined from the diagonal components of the (binned) analytic covariance matrices. In each plot, the lower panels show the residuals of the binned spectra from the best-fit model. The small dots in the upper panels show
the power-spectra multipole-by-multipole.}
     \label{fig:cib}
  \end{center}
\end{figure*}

\begin{figure*}
  \begin{center}
  \begin{tabular}{cc}
    \includegraphics[width=85mm]{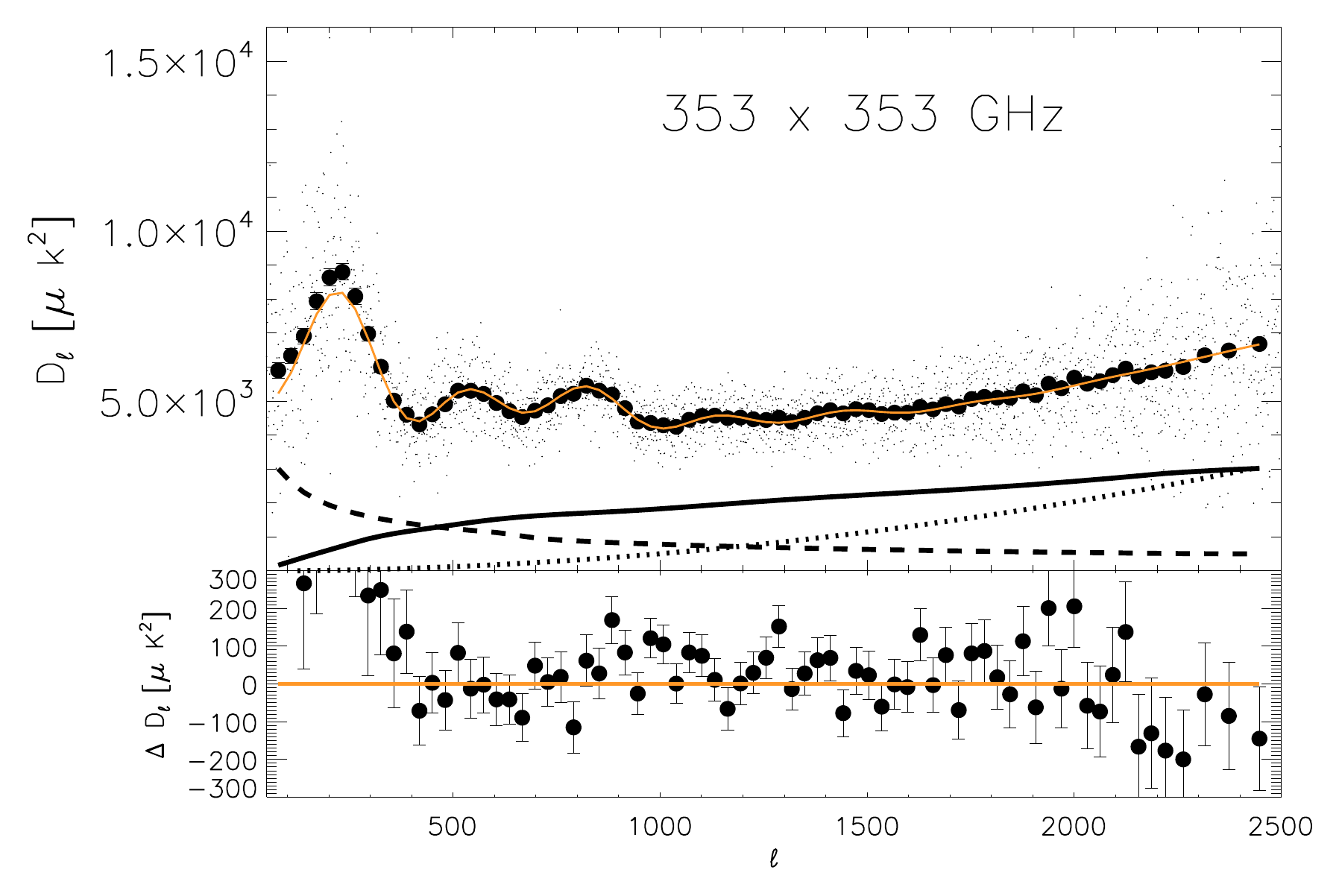} &                 \includegraphics[width=85mm]{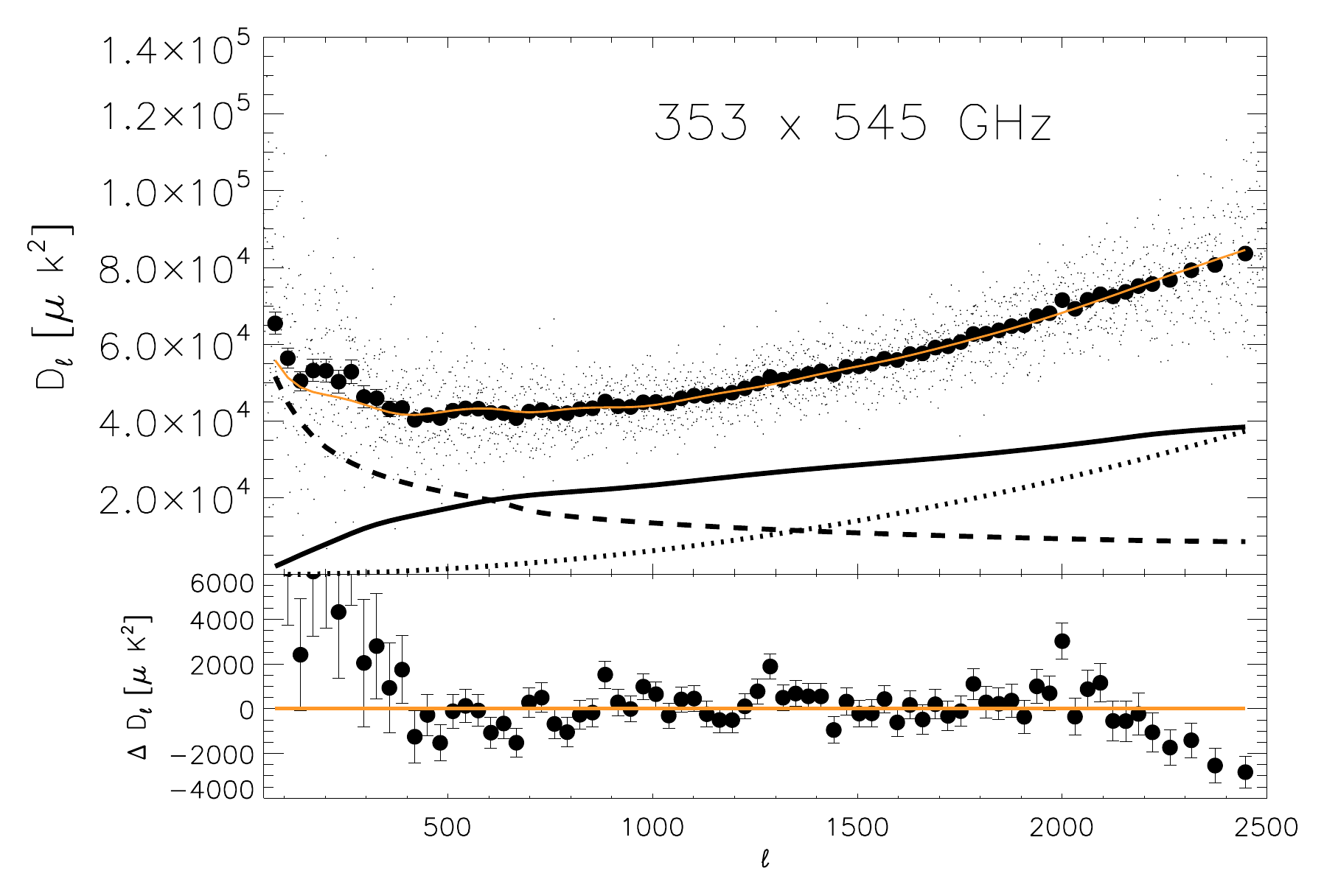} \\
    \includegraphics[width=85mm]{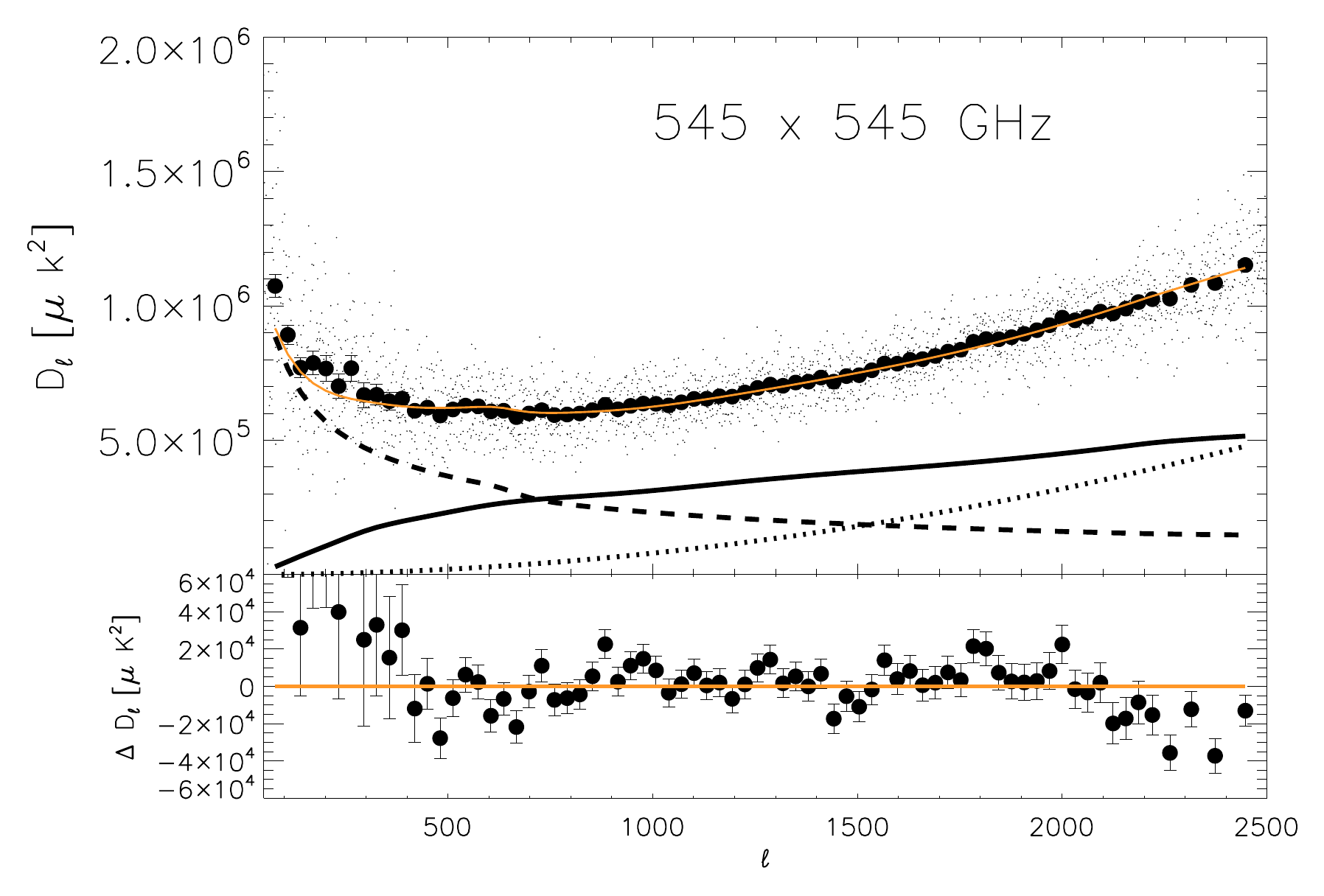} &                 \includegraphics[width=85mm]{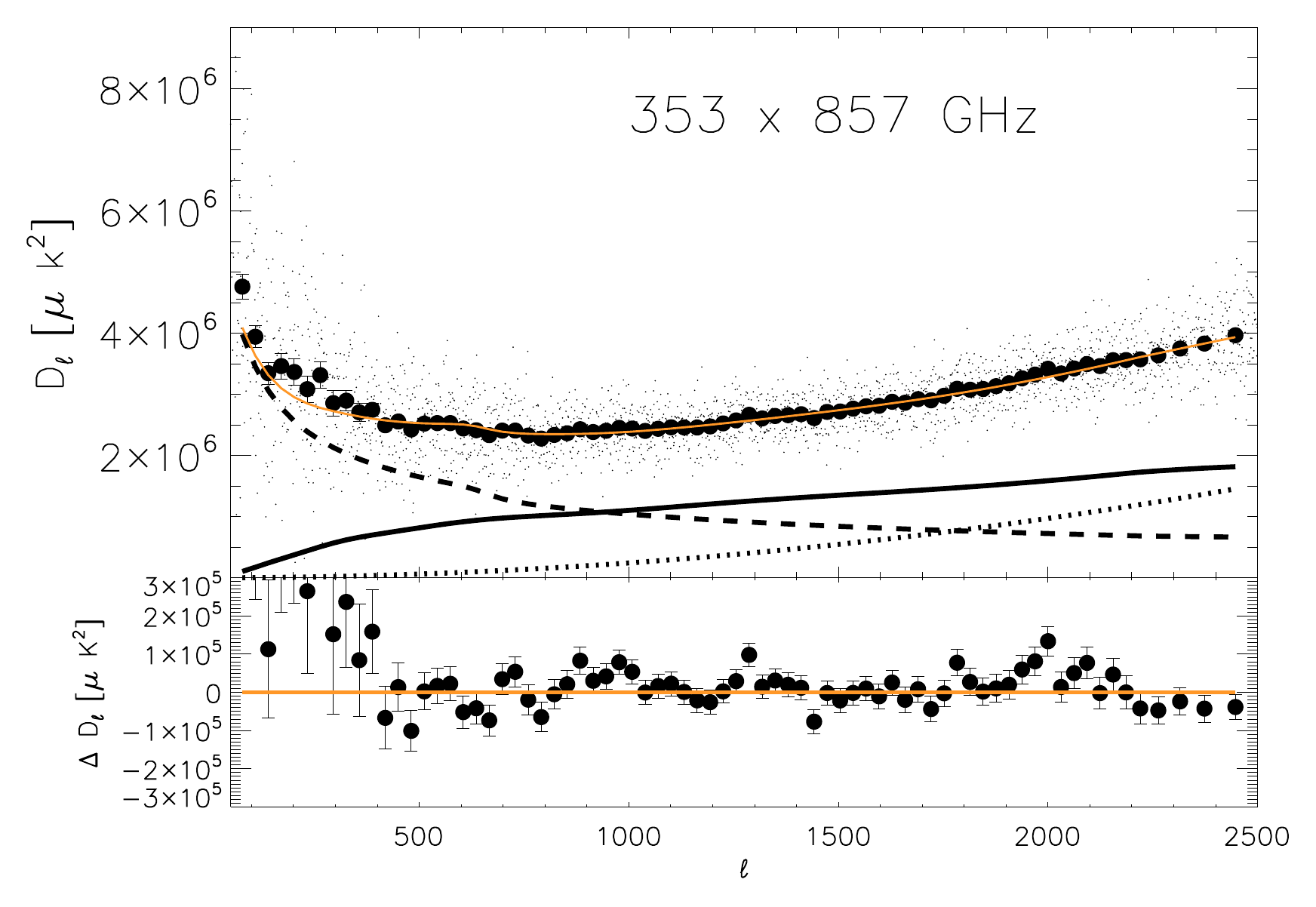} \\
    \includegraphics[width=85mm]{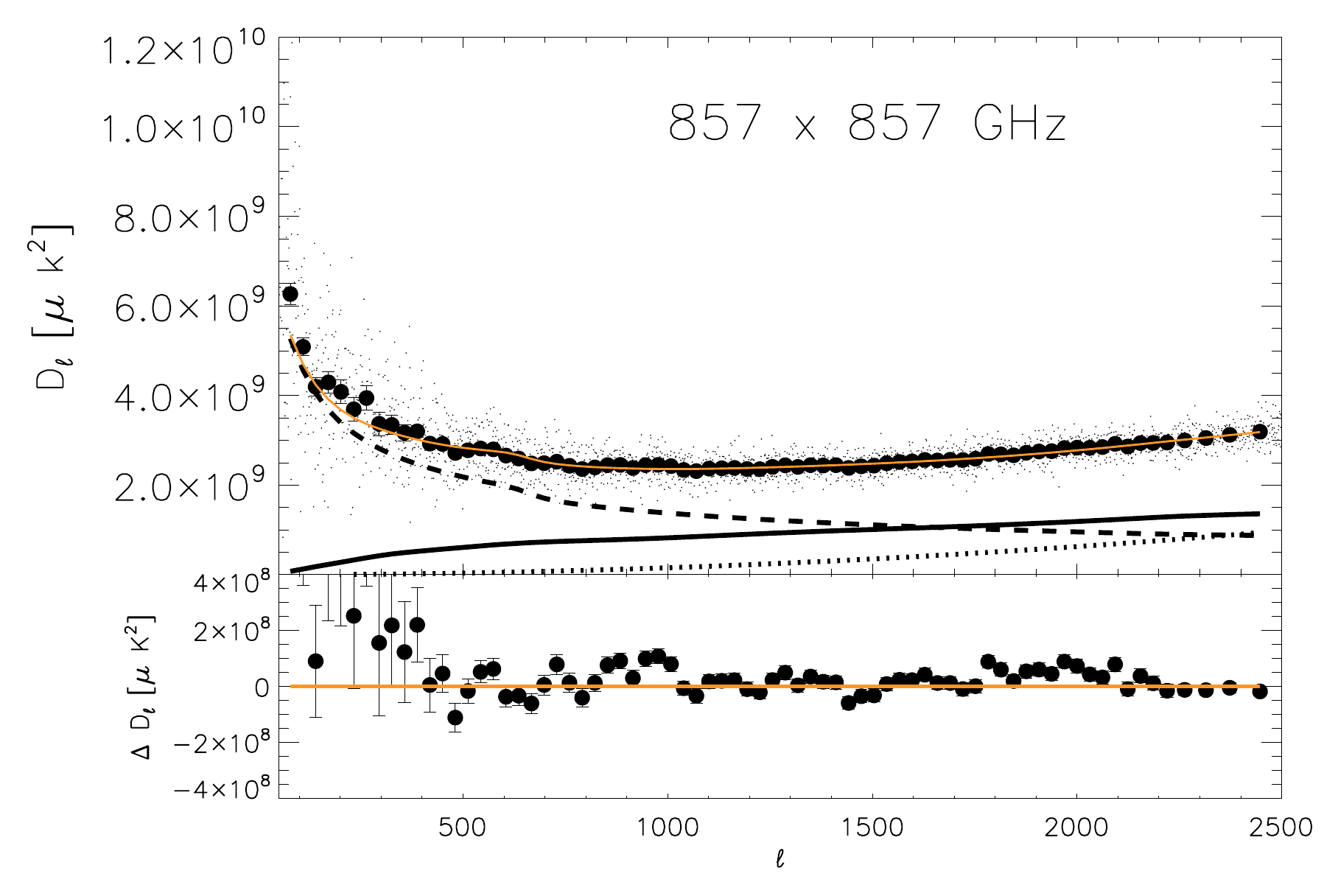} &                 \includegraphics[width=85mm]{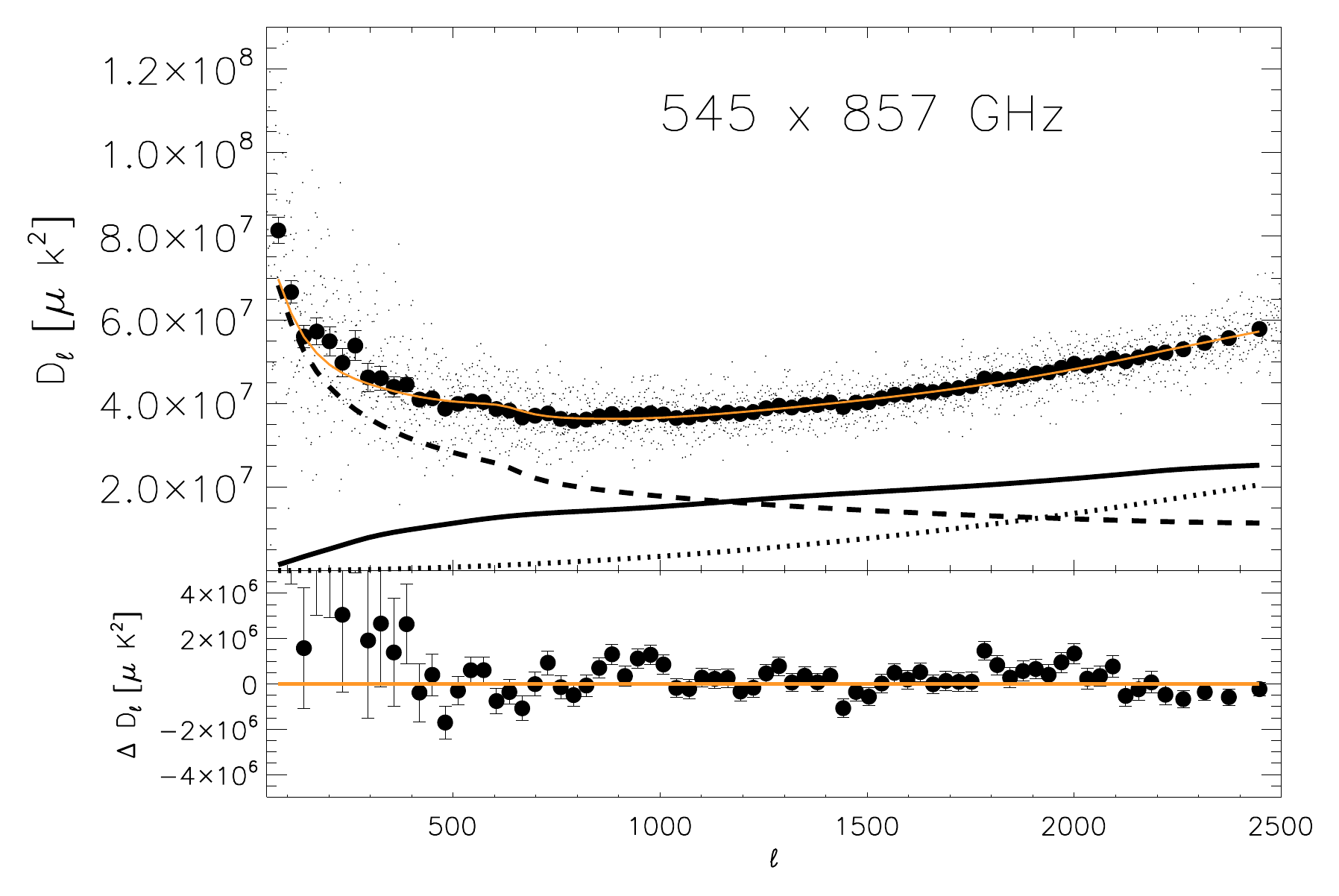} 
   \end{tabular}
       \caption{Same as~\reffig{cib}, but with the clustered CIB signal modelled with the \cppcib-based halo model template.}
     \label{fig:cib2}
  \end{center}
\end{figure*}

\begin{table*}
\begingroup
\newdimen\tblskip \tblskip=5pt
\caption{Best-fit parameter values and 68\,\% marginalised
  errors from the full likelihood analysis. Results are shown for the power-law CIB model for masks 30,
  40, and 50, and for the halo model template for mask 40 only
  (last column). We also include the results of a joint-mask analysis
  of the power-law model
  that uses spectra for mask 40 and 50. The number of degrees of
  freedom is 14\,687 for the power-law model, 14\,688 for the template
  model, and 29\,390 for the joint-mask analysis.
The frequency correlation coefficients $r^{\rm cib}$ and $r^{\rm ps}$
are omitted here; they are presented in~\reftab{correlation}. We also list the reduced
$\chi^2$ (denoted $\hat \chi^2$) for each fit. }
\label{t:result}                            
\nointerlineskip
\vskip -3mm
\footnotesize
\setbox\tablebox=\vbox{
   \newdimen\digitwidth
   \setbox0=\hbox{\rm 0}
   \digitwidth=\wd0
   \catcode`*=\active
   \def*{\kern\digitwidth}
   \newdimen\signwidth
   \setbox0=\hbox{+}
   \signwidth=\wd0
   \catcode`!=\active
   \def!{\kern\signwidth}
\halign{\tabskip 0pt#\hfil\tabskip 1.5em&
#\hfil&
\hfil#\hfil&
\hfil#\hfil&
\hfil#\hfil&
\hfil#\hfil&
\hfil#\hfil\tabskip 0pt\cr                            
\noalign{\doubleline}
Type& Parameter& \multispan5\hfil Best fit $\pm68\,\%$ error\hfil\cr
\omit&\omit& mask 30& mask 40& mask 50& Joint$^{\rm mask 40+50}$& mask
40 (halo model)\cr
 \noalign{\vskip 3pt\hrule\vskip 5pt}
\omit&$\hat \chi^2$ & 1.005 & 0.985& 0.977& 1.024& 0.994\cr
 \noalign{\vskip 3pt\hrule\vskip 5pt}
 CIB& $A_{353}^{\rm cib} (\times10^3)$& $2.50\pm0.07$& *$2.56\pm0.07$&*$2.42\pm0.07$& $2.57\pm0.07$& *$2.63\pm0.06$\cr
 \omit&$A_{545}^{\rm cib} (\times10^5)$& $4.50\pm0.20$& *$4.47\pm0.21$&*$4.25\pm0.21$& $4.69\pm0.18$& *$4.74\pm0.22$\cr 
\omit &$A_{857}^{\rm cib} (\times10^9)$& $1.09\pm0.07$& *$1.09\pm0.07$&*$1.00\pm0.07$& $1.11\pm0.07$& *$1.22\pm0.08$\cr 
\omit& $\gamma^{\rm cib}$& $0.51\pm0.02$& *$0.53\pm0.02$&*$0.54\pm0.03$&
$0.56\pm0.02$& --\cr
 \noalign{\vskip 3pt\hrule\vskip 5pt}
Dust&$A_{353}^{\rm dust} (\times10^3)$&  $4.31\pm0.19$& *$8.76\pm0.22$&$16.44\pm0.29$& *$8.99\pm0.21$, $16.15\pm0.22$& *$8.14\pm0.22$\cr
\omit&$A_{545}^{\rm dust} (\times10^6)$& $1.28\pm0.06$& *$2.61\pm0.11$&*$4.96\pm0.21$& *$2.77\pm0.10$, *$5.00\pm0.18$& *$2.50\pm0.11$\cr 
\omit&$A_{857}^{\rm dust} (\times10^9)$& $6.85\pm0.45$& $15.08\pm0.97$&$28.48\pm1.80$& $15.19\pm0.96$, $27.99\pm1.74$& $14.45\pm0.91$\cr 
 \noalign{\vskip 3pt\hrule\vskip 5pt}
PS&$A_{353}^{\rm ps} (\times10^3)$&  $2.15\pm0.07$& *$2.10\pm0.07$&*$2.25\pm0.07$&$2.10\pm0.06$&*$2.06\pm0.05$\cr
\omit&$A_{545}^{\rm ps} (\times10^5)$& $3.54\pm0.19$& *$3.42\pm0.20$&*$3.64\pm0.20$&$3.50\pm0.18$&*$3.33\pm0.15$\cr 
\omit&$A_{857}^{\rm ps} (\times10^8)$& $7.20\pm0.51$& *$7.34\pm0.53$&*$7.76\pm0.56$&$7.20\pm0.51$&*$6.37\pm0.45$\cr 
 \noalign{\vskip 3pt\hrule\vskip 5pt}
Cal& $cal_{353}$& $1.00\pm0.01$&*$1.00\pm0.01$& *$1.00\pm0.01$& $1.00\pm0.01$& *$1.00\pm0.01$\cr
\omit& $cal_{545}$& $1.05\pm0.04	$& *$1.03\pm0.04$&*$1.04\pm0.04$& $1.07\pm0.04$& *$1.05\pm0.04$\cr
\omit& $cal_{857}$& $1.01\pm0.06$& *$1.01\pm0.06$&*$1.01\pm0.06$& $1.02\pm0.06$& *$1.02\pm0.06$\cr
      \noalign{\vskip 3pt\hrule\vskip 5pt}
}}
\endPlancktablewide 
\endgroup
\end{table*}

\begin{table*}
\begingroup
\newdimen\tblskip \tblskip=5pt
\caption{Goodness-of-fit tests for the individual auto- and
  cross-frequency spectra, binned with $\Delta \ell=31$. Results are
  shown for mask 40, and in each case are compared to the
  best-fit power-law and halo models for the clustered CIB for the appropriate sky fraction from~\reftab{result}.
  For the joint-mask analysis, the goodness of fit
  of the spectra for mask 40 are shown, but the model is
  the best-fit using both masks. The quantity $\Delta\chi^2 =
  \chi^2-N_{\rm dof}$ is the difference from the expected $\chi^2$,
where the number of degrees of freedom $N_{\rm
    dof} = 74$ in all cases (the number of multipole bins for each spectrum).
The fourth and eighth columns list $\Delta\chi^2$ in units of the expected
dispersion, $\sqrt{2N_{\rm dof}}$, and the fifth and ninth columns list the
probability to exceed (PTE) the tabulated  value of $\chi^2$. 
The spectrum $C_\ell^{\rm clean}$ is a combination of the $857\times
857$, $545\times 857$, and $545\times 545$ spectra designed to remove most of
the Galactic dust emission (see~\refsec{dustclean}).
 }
\label{t:chi2_binned}                            
\nointerlineskip
\vskip -3mm
\footnotesize
\setbox\tablebox=\vbox{
   \newdimen\digitwidth
   \setbox0=\hbox{\rm 0}
   \digitwidth=\wd0
   \catcode`*=\active
   \def*{\kern\digitwidth}
   \newdimen\signwidth
   \setbox0=\hbox{$<$}
   \signwidth=\wd0
   \catcode`!=\active
   \def!{\kern\signwidth}
\halign{\tabskip 0pt#\hfil\tabskip 1.5em&
\hfil#\hfil&
\hfil#\hfil&
\hfil#\hfil&
\hfil#\hfil&
\hfil#\hfil&
\hfil#\hfil&
\hfil#\hfil&
\hfil#\hfil\tabskip 0pt\cr                            
\noalign{\doubleline}
Spectrum& $\chi^2$&$\hat \chi^2$&$\Delta\chi^2/\sqrt{2N_{\rm dof}}$& PTE[\%]& $\chi^2$&$\hat \chi^2$&$\Delta\chi^2/\sqrt{2N_{\rm dof}}$& PTE[\%]\hfil\cr
&\multispan4\hfil Power-law\hfil&\multispan4\hfil Power-law (Joint) \hfil\cr
  \noalign{\vskip 3pt\hrule\vskip 5pt}
 $353\times353$&101.12&$1.37$&2.23&!2.0*& *95.02& $1.28$&1.73&!5.0*\cr
 $545\times545$&126.95&$1.72$&4.35&!0.01&102.25&$1.38$&2.32&!1.7*\cr 
 $857\times857$&168.49&$2.28$&7.76&$<$0.01&168.04&$2.27$&7.73&$<$0.01\cr
 $353\times545$&100.23&$1.35$&2.16&!2.3*&*99.40&$1.34$&2.09&!2.6*\cr 
  $353\times857$&111.51&$1.51$&3.08&!0.3*&112.86&$1.53$&3.19& !0.2*\cr 
 $545\times857$&124.94&$1.69$&4.17&!0.02&139.32&$1.88$&5.37& $<$0.01\cr
 $C_\ell^{\rm clean}$&114.47&1.55&3.33&!0.2*&101.73&1.37&2.28&!1.8*\cr
 \noalign{\vskip 3pt\hrule\vskip 5pt}
&\multispan4\hfil halo model template \hfil&\multispan4\hfil halo model template (Joint) \hfil\cr
  \noalign{\vskip 3pt\hrule\vskip 5pt}
  $353\times353$&119.88&1.62&*3.77&!0.06&125.81&1.70&*4.26&!0.02\cr
 $545\times545$&159.35&2.15&*7.02&$<$0.01& 158.18&$2.14$&*6.92&$<$0.01\cr 
 $857\times857$&233.84&3.16&13.14&$<$0.01&202.66&2.74&10.58&$<$0.01\cr
 $353\times545$&157.99&2.14&*6.90&$<$0.01&159.88& $2.16$&*7.06&$<$0.01\cr 
  $353\times857$&135.73&1.83&*5.07&!0.3*&130.38&$1.76$&*4.63&$<$0.01\cr
 $545\times857$&164.05&2.22&*7.40&$<$0.01&157.40&$2.13$&*6.86&$<$0.01\cr
  $C_\ell^{\rm clean}$&389.60&5.26&25.94&$<$0.01&387.07&5.23&25.73&$<$0.01\cr
        \noalign{\vskip 3pt\hrule\vskip 5pt}
}}
\endPlancktablewide 
\endgroup
\end{table*}

\subsection{Power spectra}
\refFig{cib} shows the six auto- and cross-frequency angular power spectra that we measure from cross-correlating the 
yearly maps for mask 40 (shown in~\reffig{mask}).
(The beam and mask are deconvolved from the spectra, which are binned
assuming flat bandpowers in $\mathcal{D}_\ell$).  The figure also
shows the best-fit model for these spectra obtained from the full
likelihood analysis assuming the power-law model for the clustered
CIB, the decomposition of this best-fit model into individual
components, and the residuals with respect to the best-fit model.

The power spectra at 545 and 857\GHz\ are signal dominated for
multipoles $\ell < 2000$, while the 353\GHz\ spectrum is signal
dominated only at $\ell < 1000$. On large and intermediate scales, the
353\GHz\ spectrum is dominated by dust and CMB, while at 857\GHz\ dust
is the dominant component for $\ell < 1500$. At 545\GHz, the clustered
CIB is the dominant signal over the multipole  range $1000 < \ell < 2000$.

On the full likelihood, the power-law CIB model appears to fit the data well. For the fiducial sky
fraction of $40\,\%$, the best-fit model has a reduced $\hat \chi^2=0.985$ for $14\,687$ degrees of freedom (corresponding to all
unbinned auto- and cross-frequency spectra in the multipole range $\ell =
50$--$2500$) which is acceptable to within $1.25\,\sigma$ (although see below for a discussion of the
goodness of fit of the binned spectra). We recover calibration constraints, $cal_\nu$, that are
 close to unity and are consistent with our priors. As mentioned in~\refsec{data}, we have used conservative priors and have not imposed any priors on the relative calibration. The central values that we find for $cal_{545}$ from our likelihood analyses are somewhat larger than the more precise determinations reported in~\citet{plancklowl2016}, which are based on Solar dipole calibrations, but our posterior distributions for $cal_{545}$ are still consistent with the dipole-based calibrations given the size of our errors. Our main parameter results are given in~\reftab{result}, and the best-fit
models are compared with the measured spectra in~\reffig{cib}
and~\reffig{cib2} for the power-law and halo model CIB templates,
respectively. Goodness-of-fit tests of individual auto- and
cross-frequency spectra with respect to the best-fit models of the
full likelihood analysis are reported in Table~\ref{t:chi2_binned}.

In the rest of this section, we discuss various aspects of these
results. Section~\ref{sec:resultcib} discusses the results for the
clustered CIB for both the power-law model and the \cppcib-based halo-model template.
 Section~\ref{sec:jointmask} describes the results of a
``joint-mask'' analysis that further leverages the statistical
anisotropy of the Galactic dust emission. Our results are compared
with previous measurements from \Planck\ and \textit{Herschel}
in~\refsec{compare}, and in~\refsec{psprior} we compare the Poisson
amplitudes infered from our spectra with expectations based on
source counts. \refapp{con} discusses additional tests of the stability of our results to various
data cuts and methodological changes.

\subsection{CIB models}
\label{sec:resultcib}

Using the power-law model for the clustered CIB power we find,
consistently across the three frequencies and sky fractions, that the
clustering component contributes about 54\,\% of the total CIB power
at $\ell=2000$. We obtain a tight constraint on the CIB slope of
$\gamma^{\rm cib}=0.53\pm0.02$. This value is consistent with that
obtained in the 2013 analysis of cosmological parameters
from \Planck\ data alone~\citep{planck2013-p08}, $\gamma^{\rm
cib}=0.40\pm0.15$, using the 100, 143, and 217\,GHz channels. Both
analyses probe the clustered CIB over a similar range of
multipoles. However, at  frequencies $\nu \le 217$ \GHz\
the high amplitude of the CMB and other foreground components,  
in addition to dust emission, 
makes it  difficult to extract information 
on the CIB at multipoles $\ell \simlt 1000$.

Although the best-fit models gives acceptable $\chi^2$ using the full
likelihood multipole-by-multipole, one can see visually from Figs. \ref{fig:cib}
and \ref{fig:cib2} that the binned spectra show systematic residuals, particularly 
for the $545\times 857$ and $857\times 857$ spectra (which are the
spectra most contaminated by Galactic dust). \reftab{chi2_binned}
lists $\chi^2$ for each of the binned spectra and best-fit models
shown in Figs.~\ref{fig:cib}
and~\ref{fig:cib2}. For the power-law CIB template, the
$545\times 857$ and $857\times 857$ spectra have $\chi^2$ values that are $4\,\sigma$ and
$8\,\sigma$ high, respectively.  The fits are even worse on mask 50, 
where the dust power is roughly double that for mask 40. We have a partial 
understanding of the source of these anomalous $\chi^2$ values (see Table \ref{t:dresult} below). A significant
contribution to the high $\chi^2$ values comes from our
modelling of the  Galactic dust contributions to the covariance matrices. As discussed
in~\refsec{likelihood}, by assuming that the dust is a
statistically-isotropic Gaussian random field, we are 
 over-counting the effective number of modes and therefore underestimating the dust contribution
to the covariance matrices. 
We defer further discussion of this issue
to~\refsec{ngnoise} where we develop a heuristic model for the sampling variance 
of Galactic dust that improves  the $\chi^2$ of the
high-frequency spectra significantly, while having very little 
impact on the model parameters reported in~\reftab{result}.

\begin{figure}
  \begin{center} \includegraphics[width=88mm]{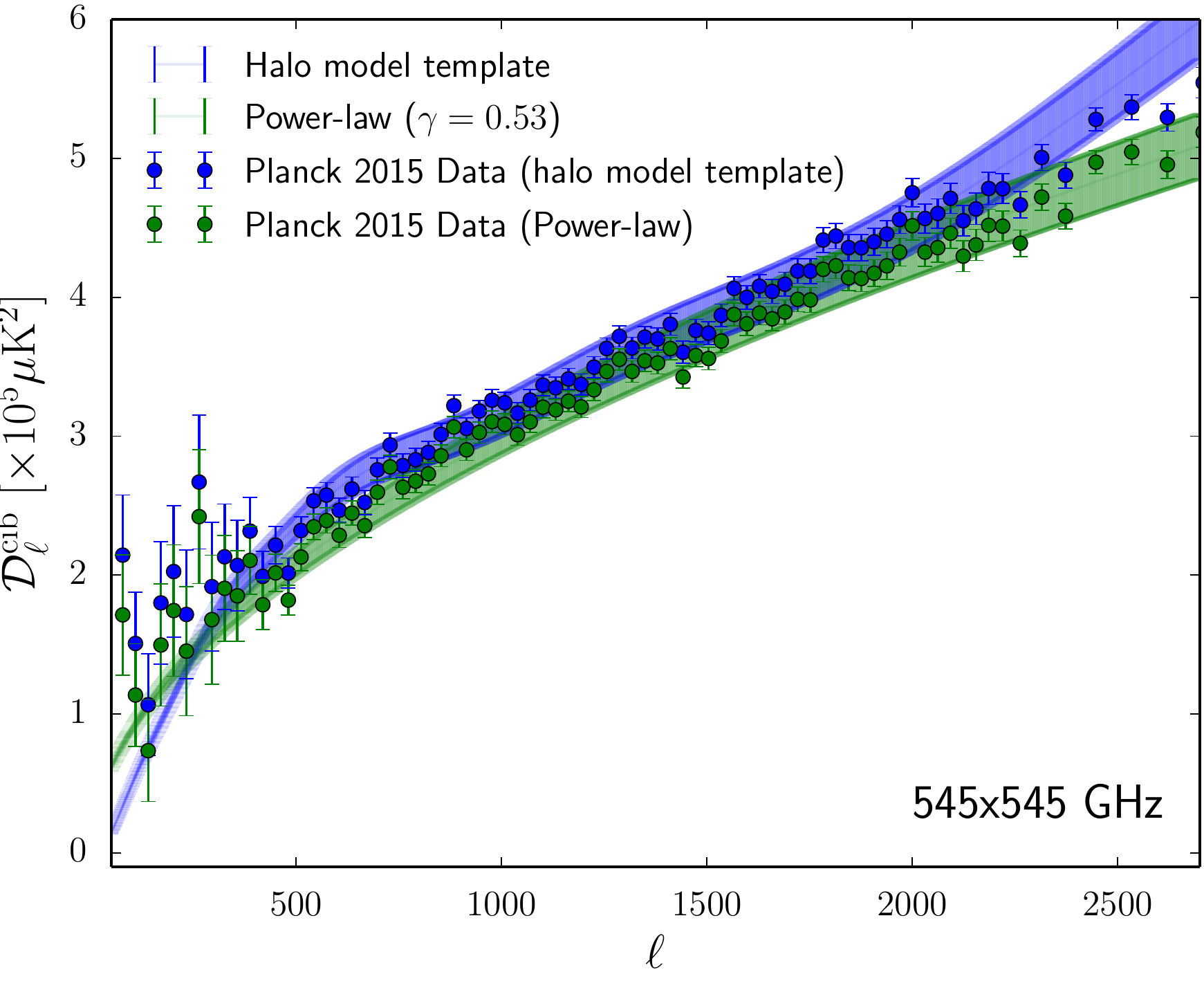}
  \includegraphics[width=88mm]{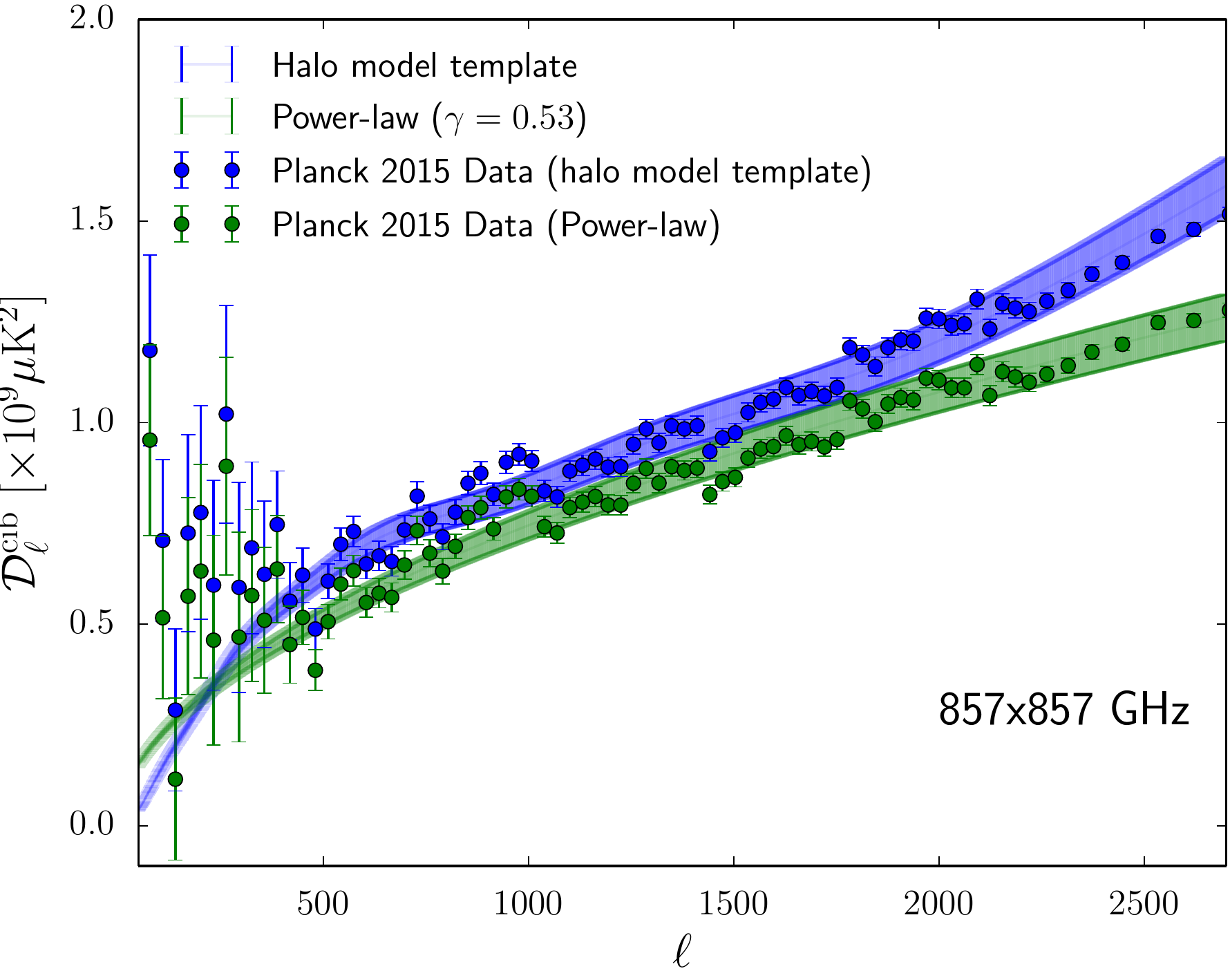}
   \caption{Clustered
        CIB power spectrum at 545\GHz\ (top) and 857\GHz\ (bottom). The data points are
        the measured spectra for mask 40 after subtracting the
        best-fit Poisson and Galactic dust power in the appropriate CIB
        model. The power-law model, with best-fit slope $\gamma^{\rm
        cib}=0.53$, is shown in green and the halo model template in
        blue.  The thickness of the lines
        correspond to the marginalized $1\,\sigma$ error on
        $A^{\rm cib}$.}  \label{fig:compare_cib} \end{center}
\end{figure}

We now consider the effect of replacing the power-law template for the
clustered CIB power with the halo model template described
in \refsec{halomodel}. Results are presented in Fig. \ref{fig:cib2} and in Tables \ref{t:result} 
and \ref{t:chi2_binned}  for mask 40. The \cppcib\ halo model gives a consistently poorer fit 
to the data than the power-law model  on both the full likelihood ($\Delta \chi^2 = 135$) and on 
the binned spectra.  The two CIB models are compared
in~\reffig{compare_cib} for the  $545\times 545$ and $857 \times 857$ spectra;
this shows clearly the steepening of the halo model template both on large and small
scales, which is disfavoured by the data. 

The model-fitting approach adopted in this paper relies on the accuracy of the dust
and CIB templates. Most of the statistical power of the \Planck\ data lies in the multipole
range $\ell \sim 1000$--$2000$, and so the likelihood will find solutions that minimise the 
residuals in this multipole range.  Using a template of the wrong shape will affect the
residuals at both low and high multipoles. At face value, this is what is happening in 
Fig. \ref{fig:compare_cib}. Both CIB templates produce reasonably good fits in the multipole
range $\ell=1000$--$2000$, but the halo model template leaves systematic residuals at both high and 
low multipoles. This suggests that the  \cppcib\  halo model template is a poor fit to the clustered
CIB component at high frequencies. We will discuss this important point in further detail in
Sec.~\ref{sec:compare} after comparing our dust-corrected spectra with 
those of \cppcib\  and with the {\it Herschel}/SPIRE measurements of
~\citet[][hereafter \viero]{Viero2013}, the latter extending
up to multipoles $\ell\approx 30\, 000$. Note also that the power-law template, despite giving better fits to the data, is only a phenomenological model and there is no reason to expect constant power-law behaviour  on all scales. The power-law template deviates from the halo model template on large scales ($\ell \simlt 300$), where the physical modelling that underlies the halo model template should be rather robust. One should therefore be cautious not to over-interpret the apparently better fit of the power-law template on large scales, particularly given the difficulties in dust removal there and the large sample variance. We present a fuller discussion of the behaviour of these models on large scales in~\refsec{dustclean}.


\subsection{Joint-mask analysis}
\label{sec:jointmask}

The Fisher matrix analysis of Sec.~\ref{sec:fisher} reveals
degeneracies between the clustered and Poisson power of the CIB and the
dust power.  To reduce the degeneracy with the dust power, we have
experimented with combining information from different regions of the
sky in an attempt to provide a more accurate separation between the
statistically-isotropic CIB components and the anisotropic Galactic dust emission.
We have therefore performed a ``joint-mask'' likelihood analysis using
spectra computed on two different masks. We adopt one set of CIB and
point source amplitudes for both masks (since these describe statistically-isotropic
components) but different dust component parameters for each mask
(since the dust emission is anisotropic).  To do this, we simply
include in the data vector the 12 spectra measured on the two masks
and compute the appropriate covariance matrix to form a likelihood.

The result of using mask 40 and mask 50 in the joint-mask analysis are
given in the sixth column in \reftab{result}. We can see that the
parameters from the joint mask analysis are very similar, with slightly reduced errors, 
to those determined individually on mask 40 and mask 50, showing that our
results are stable. The main effect is to increase the dust amplitudes
determined from mask 40 at 353 and 545 GHz by about $1\sigma$ (see \reffig{us_v_pip} below). We
found similarly stable parameters using mask 30 and mask 40 in a joint
masks analysis.

\subsection{Comparison with previous measurements}
 \label{sec:compare}

\begin{figure*}
  \begin{center}
  \includegraphics[width=170mm]{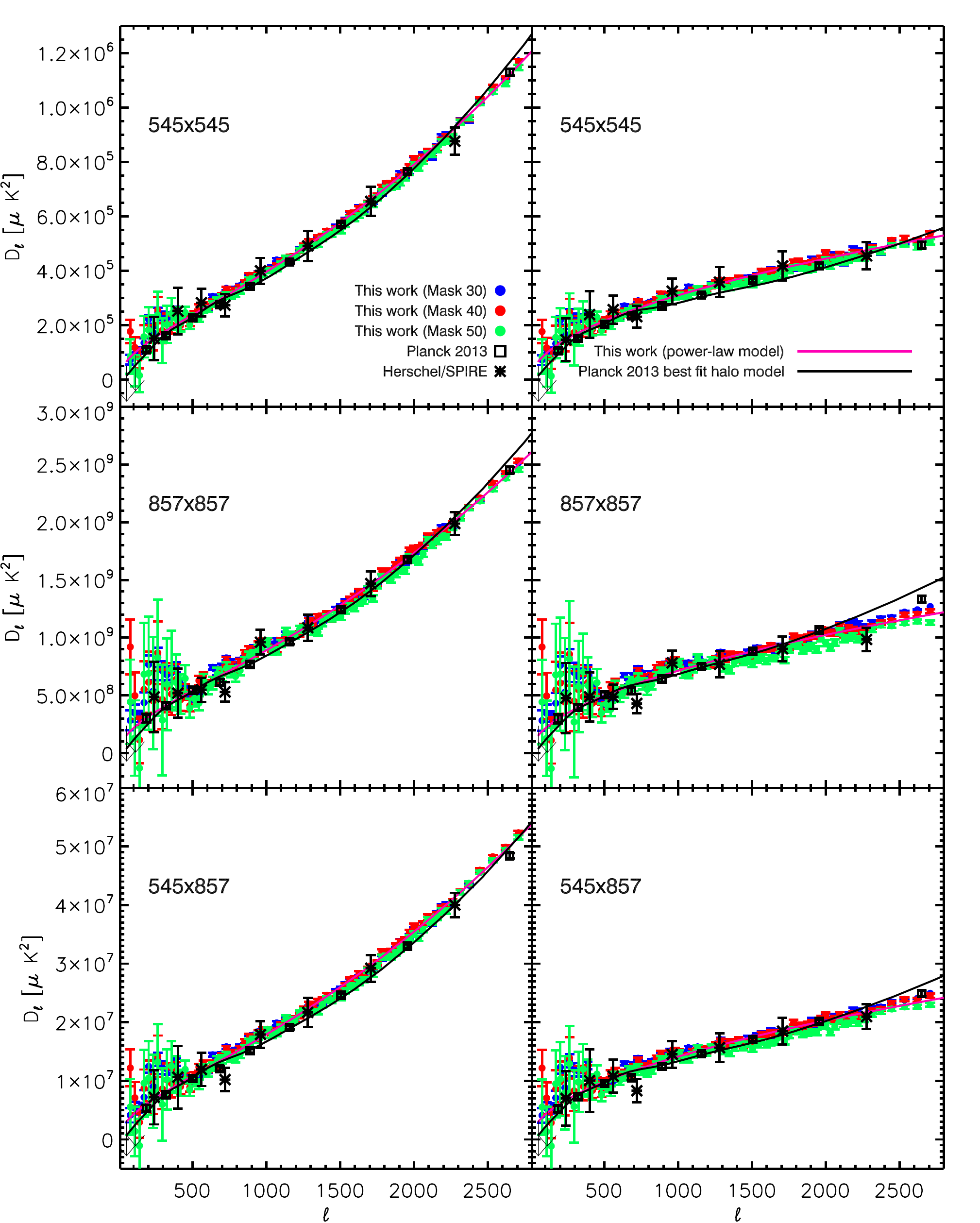}
       \caption{Comparison of our CIB power spectra for $545\times
         545$, $857\times 857$, and $545\times 857$ (top to bottom,
         respectively) for masks 30 (blue), 40 (red), and 50 (green) with those from 
\Herschel/SPIRE data (black asterisks;~\viero), and from the 2013
\Planck\ data (black squares;~\cppcib). The \Herschel/SPIRE
measurements have been colour-corrected and recalibrated to match the
\Planck\ 545 and 857\GHz\ bands, as described in the text. The
measurements on the left are the total CIB (clustered plus Poisson)
power. Also plotted are our best-fit CIB models for mask 40 assuming the power-law clustering template (red lines) and the best-fit halo models from~\cppcib\ (black lines). The measurements on the right are estimates of the clustered CIB power, obtained by subtracting the reported best-fit Poisson power levels.
}
     \label{fig:dx11-p2013-herschel}
  \end{center}
\end{figure*}

\begin{figure*}
  \begin{center}
    \includegraphics[width=170mm]{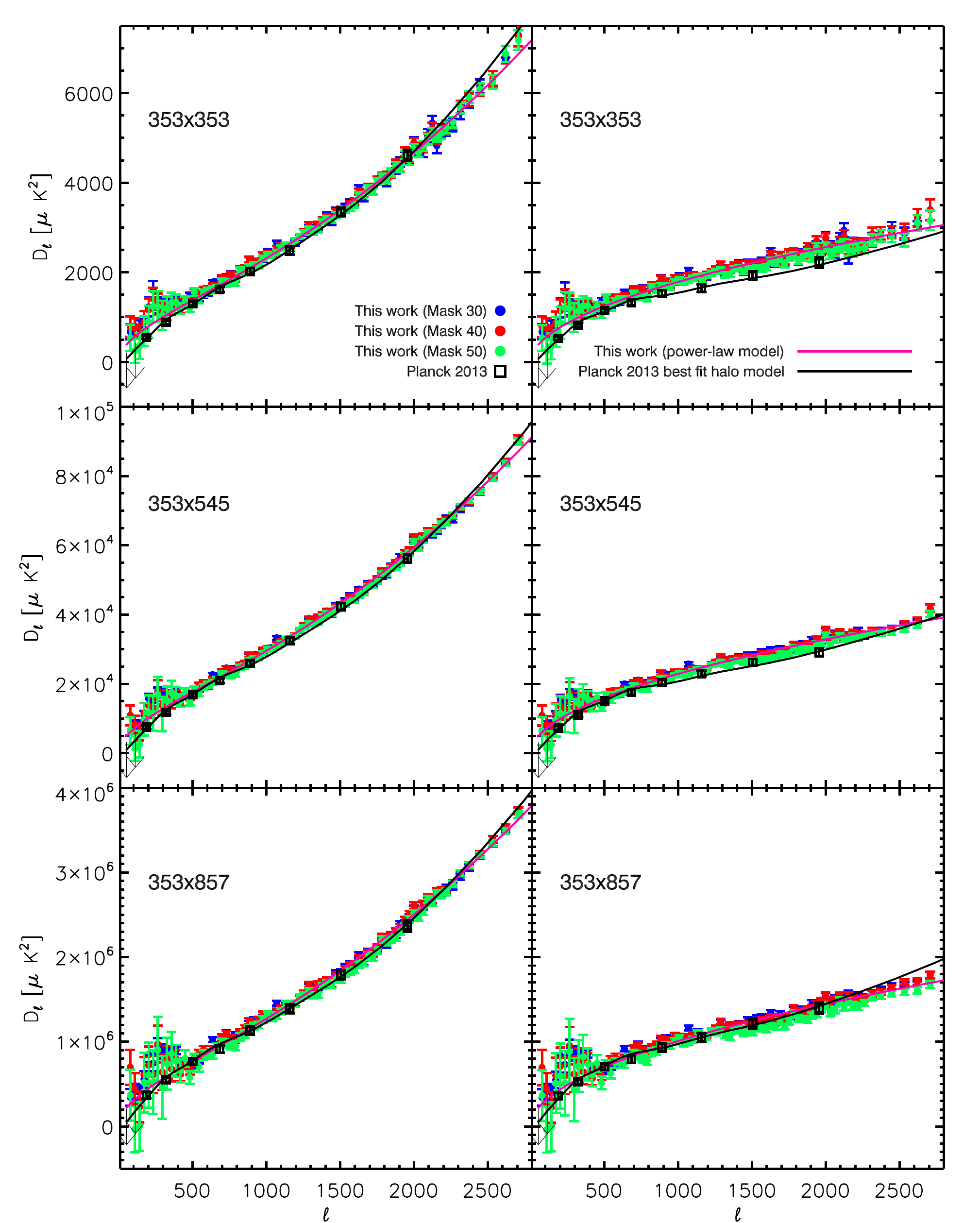}
       \caption{Same as~\reffig{dx11-p2013-herschel}, but for the
         $353\times 353$, $353\times 545$, and $353\times 857$ spectra
         (top to bottom, respectively). Comparisons are done only between this work and~\cppcib.  
}
     \label{fig:dx11-p2013}
  \end{center}
\end{figure*}

\begin{figure*}
  \begin{center}

           \includegraphics[width=75mm]{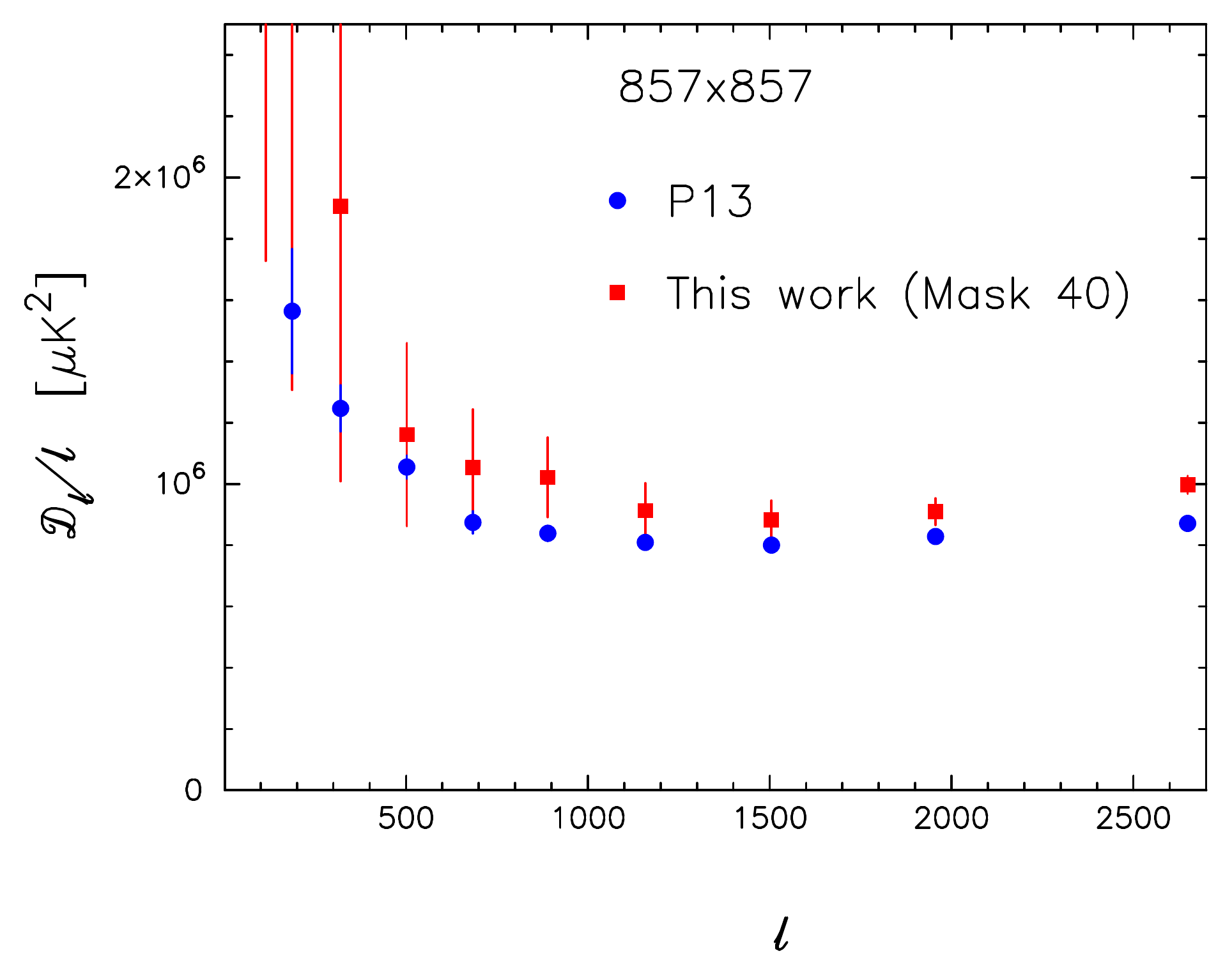}             \includegraphics[width=75mm]{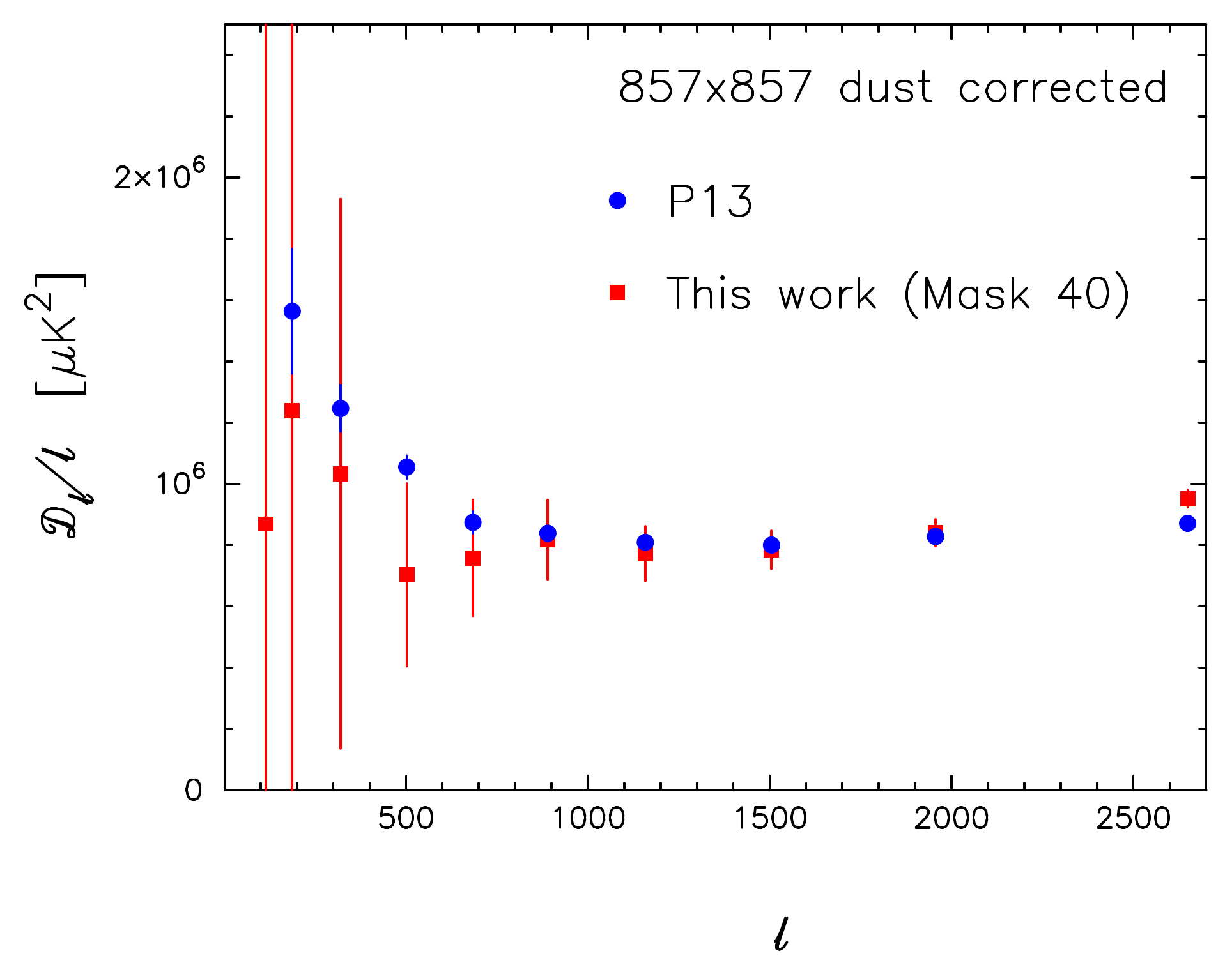}   \\
           \includegraphics[width=75mm]{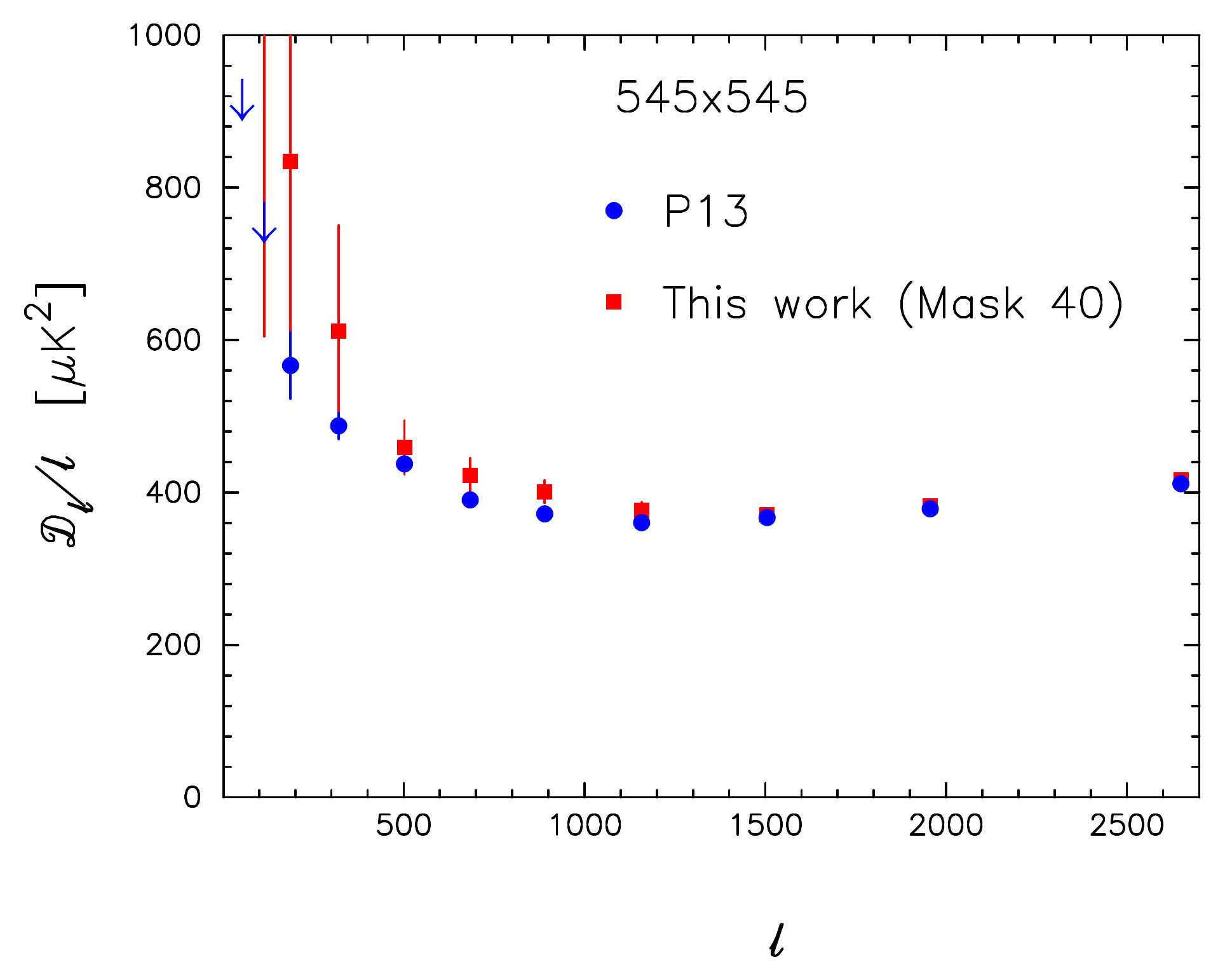}             \includegraphics[width=75mm]{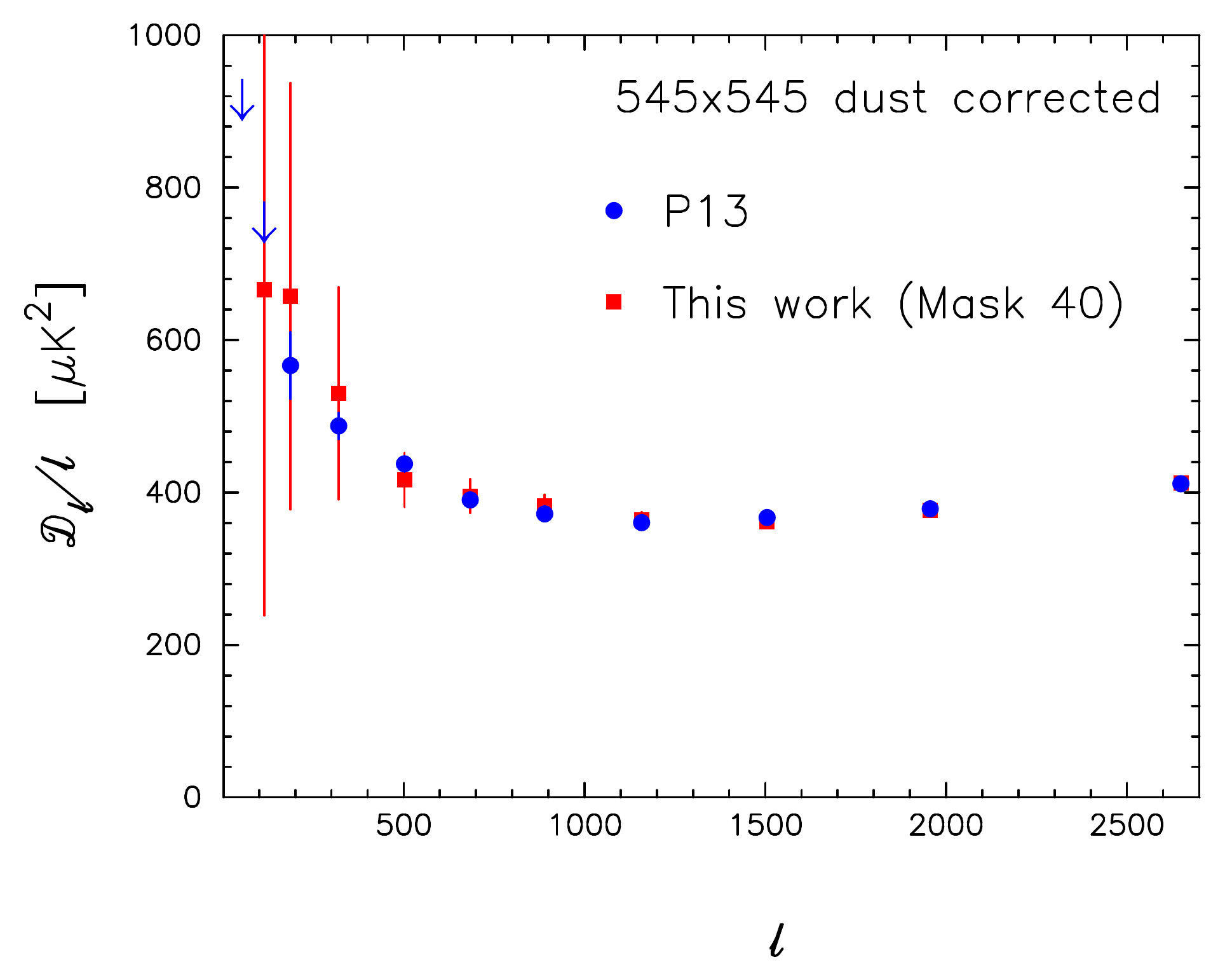}   \\
           \includegraphics[width=75mm]{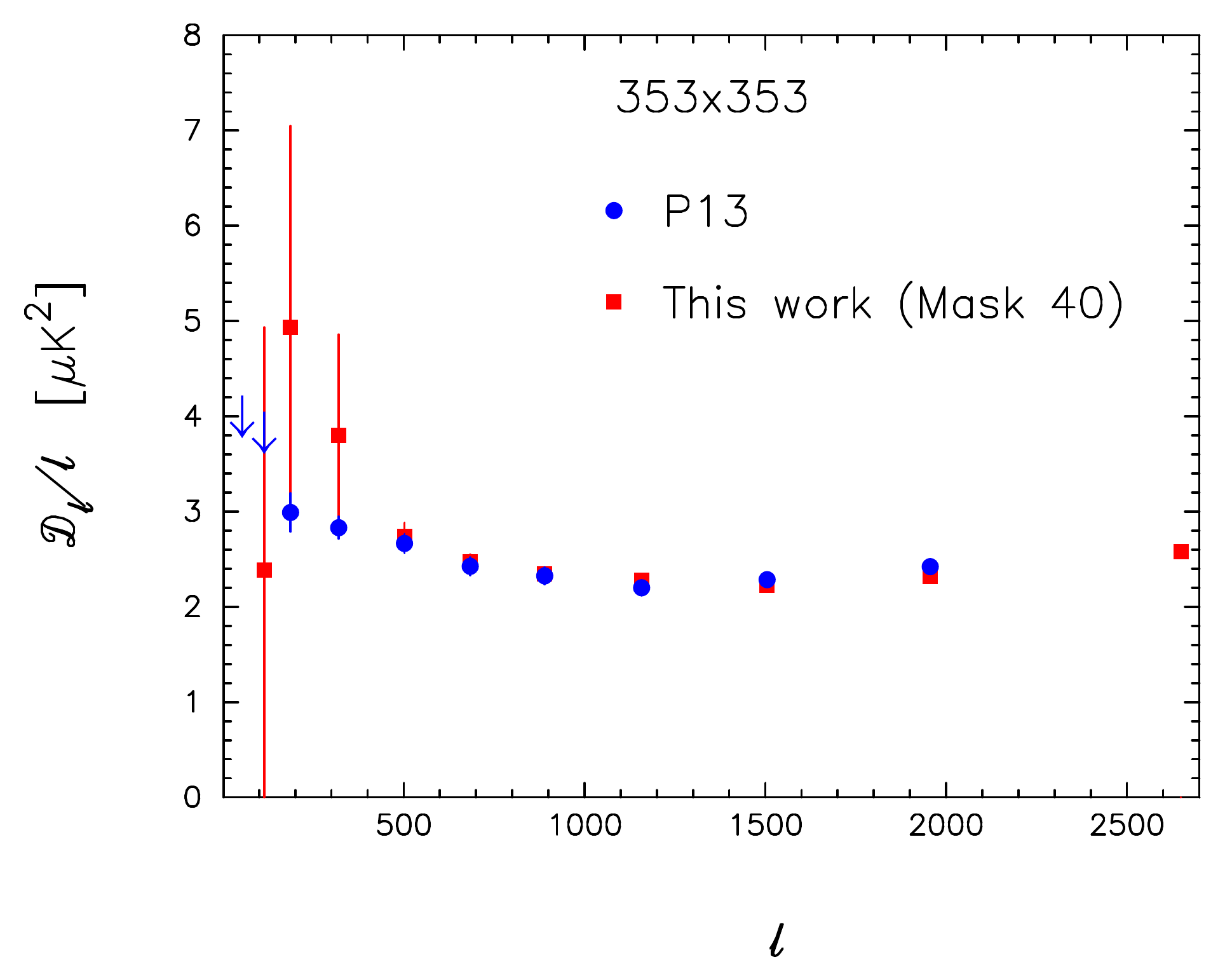}             \includegraphics[width=75mm]{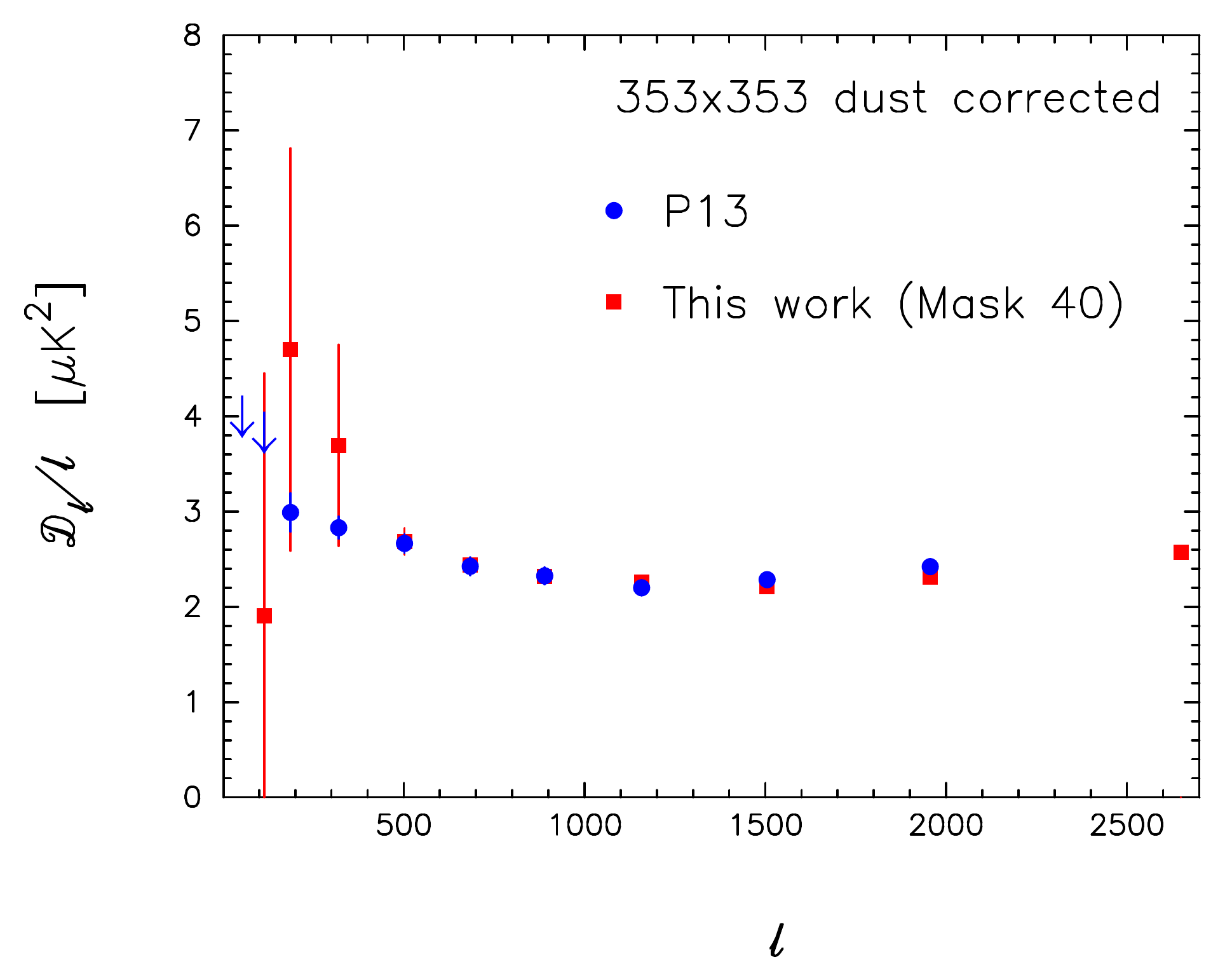}   

       \caption{\textit{Left}: Our dust-cleaned spectra (red points) for mask 40,
         from Fig.~\ref{fig:dx11-p2013-herschel},  but binned in exactly the same way as the~\cppcib\ spectra shown by the blue 
points. \textit{Right}: Our dust-cleaned spectra after making small adjustments to the amplitudes of the dust templates chosen to
minimise differences with the \cppcib\ spectra.}
     \label{fig:us_v_pip}
  \end{center}
\end{figure*}

In Figs.~\ref{fig:dx11-p2013-herschel} and~\ref{fig:dx11-p2013} we
compare CIB auto- and cross-frequency power spectrum measurements at
353, 545, and 857\GHz\ with the most recent measurements from {\it
  Herschel}/SPIRE (\viero) and the results from the \Planck\
Collaboration based on the 2013 data~(\cppcib). For the data points of
\cppcib, we multiply them by $1.021^2$,  $0.98^2$, and $0.96^2$ for
353, 545 and 857\GHz, respectively, to account for the recalibration
between the 2013 and 2015 \Planck\ data. We use the {\it
  Herschel}/SPIRE power spectra with only extended sources masked so
that the Poisson contribution is comparable in all three analyses. In
addition, we also plot our best-fitting CIB model (assuming the
power-law template for the clustered component) and the best-fitting
halo model from~\cppcib. We plot the total CIB power spectrum
measurements, i.e., clustered plus Poisson, and also an estimate of
the clustered power. For the latter, we subtract the reported
best-fitting Poisson power levels for each measurement. Following
\cppcib, we multiply the power spectra of \Herschel/SPIRE (data and
best-fit Poisson power levels) at 350 and $500\,\mu\text{m}$ by 1.016 and 0.805, respectively, to colour correct for comparison with measurements with \Planck\ at 857 and 545\GHz. We further multiply the \Herschel/SPIRE spectra by calibration factors that account for the SPIRE/\Planck\ relative photometric calibration: $1/1.047^2$ and $1/1.003^2$ at 545 and 857\GHz, respectively~\citep{Bertincourt2015}.

\begin{figure*}
  \begin{center}

           \includegraphics[width=55mm]{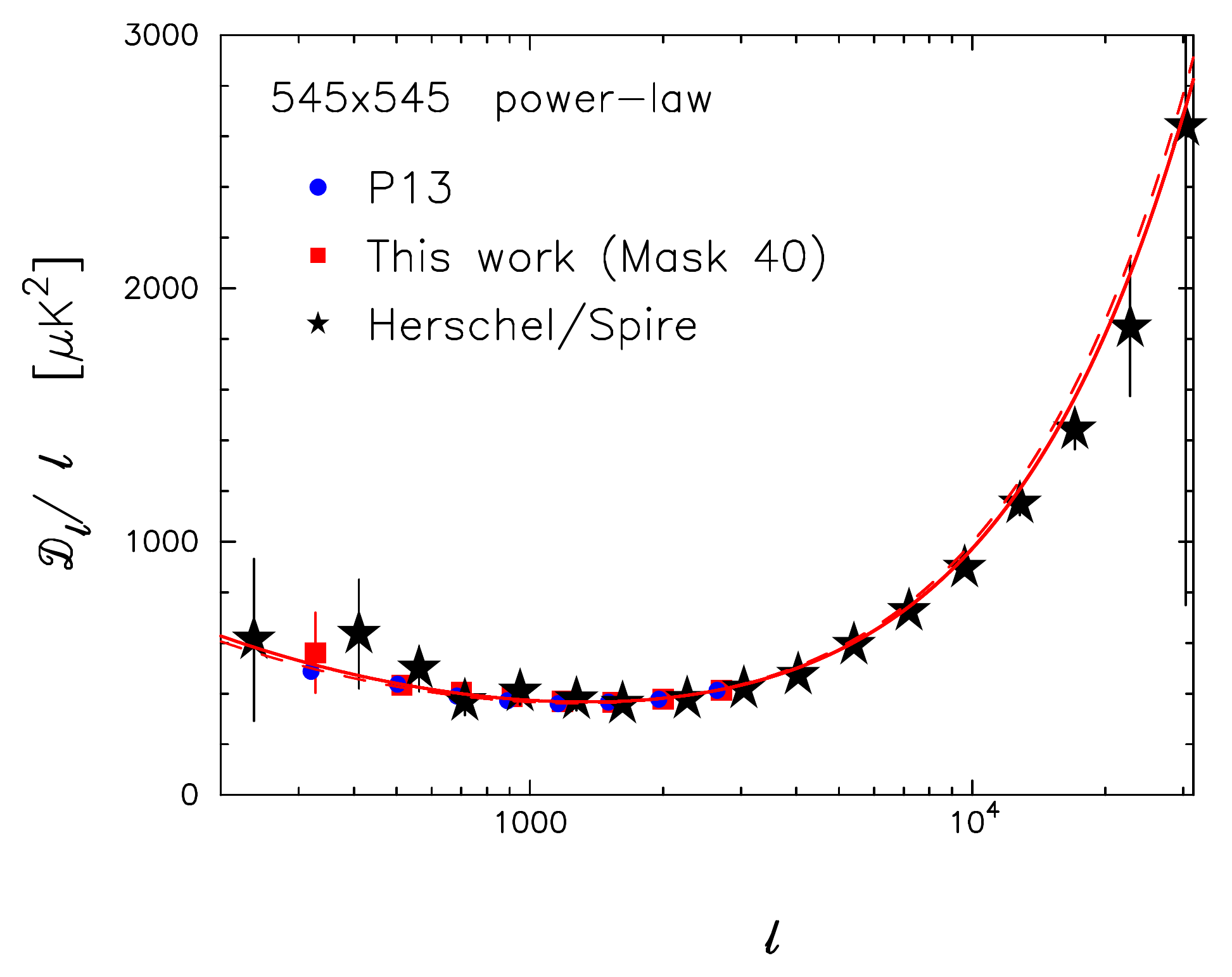}  \includegraphics[width=55mm]{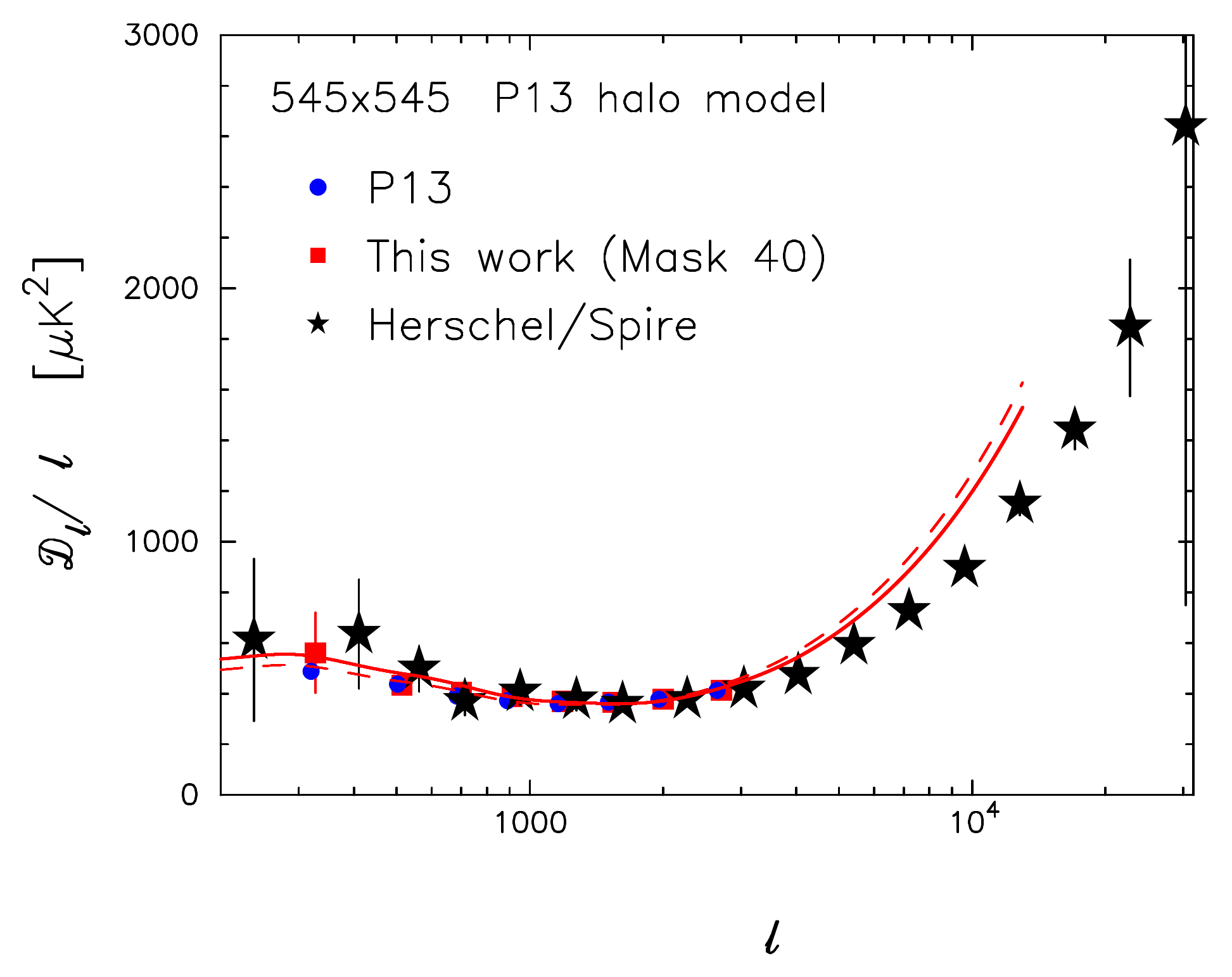}   \includegraphics[width=55mm]{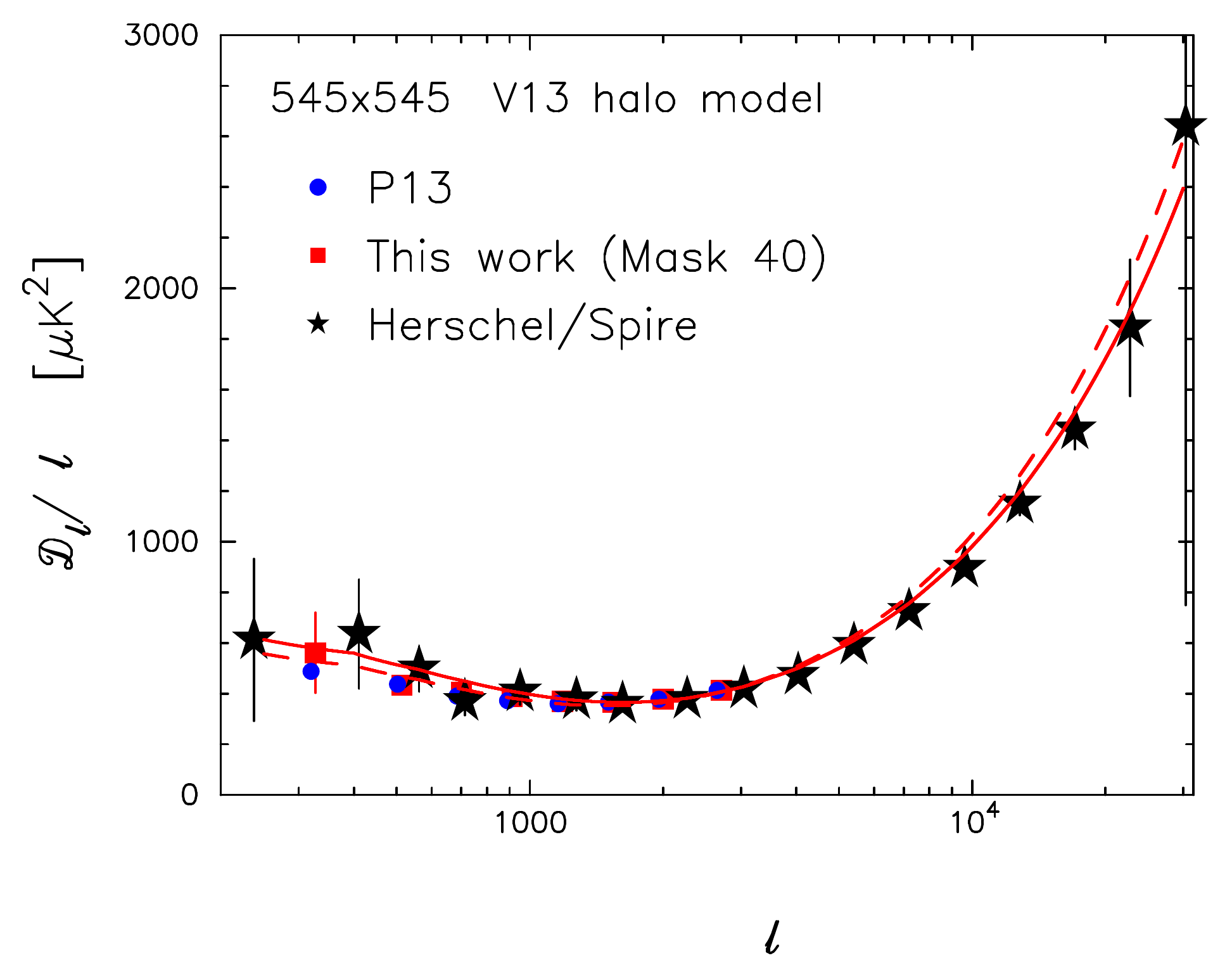}   \\
           \includegraphics[width=55mm]{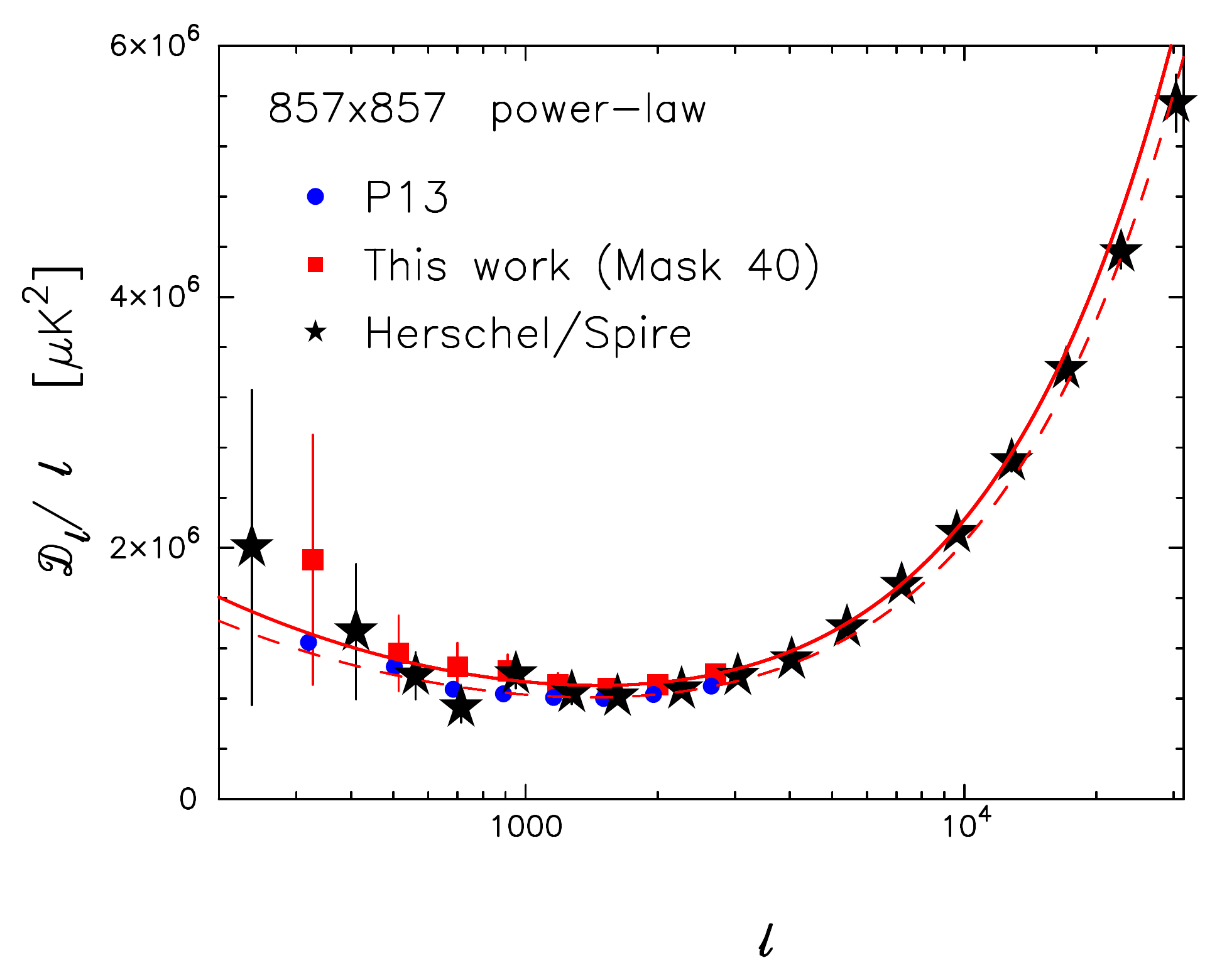}  \includegraphics[width=55mm]{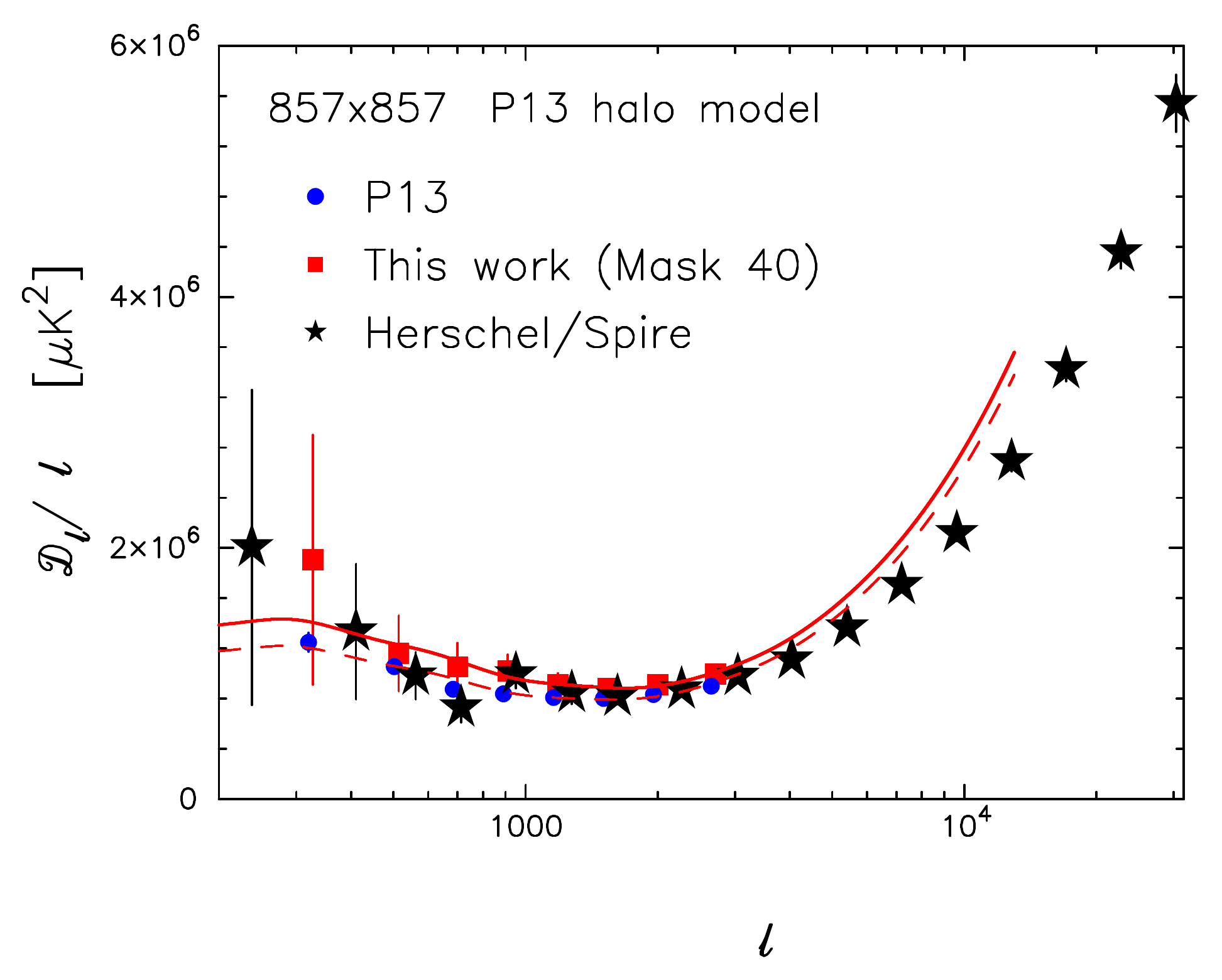}   
\includegraphics[width=55mm]{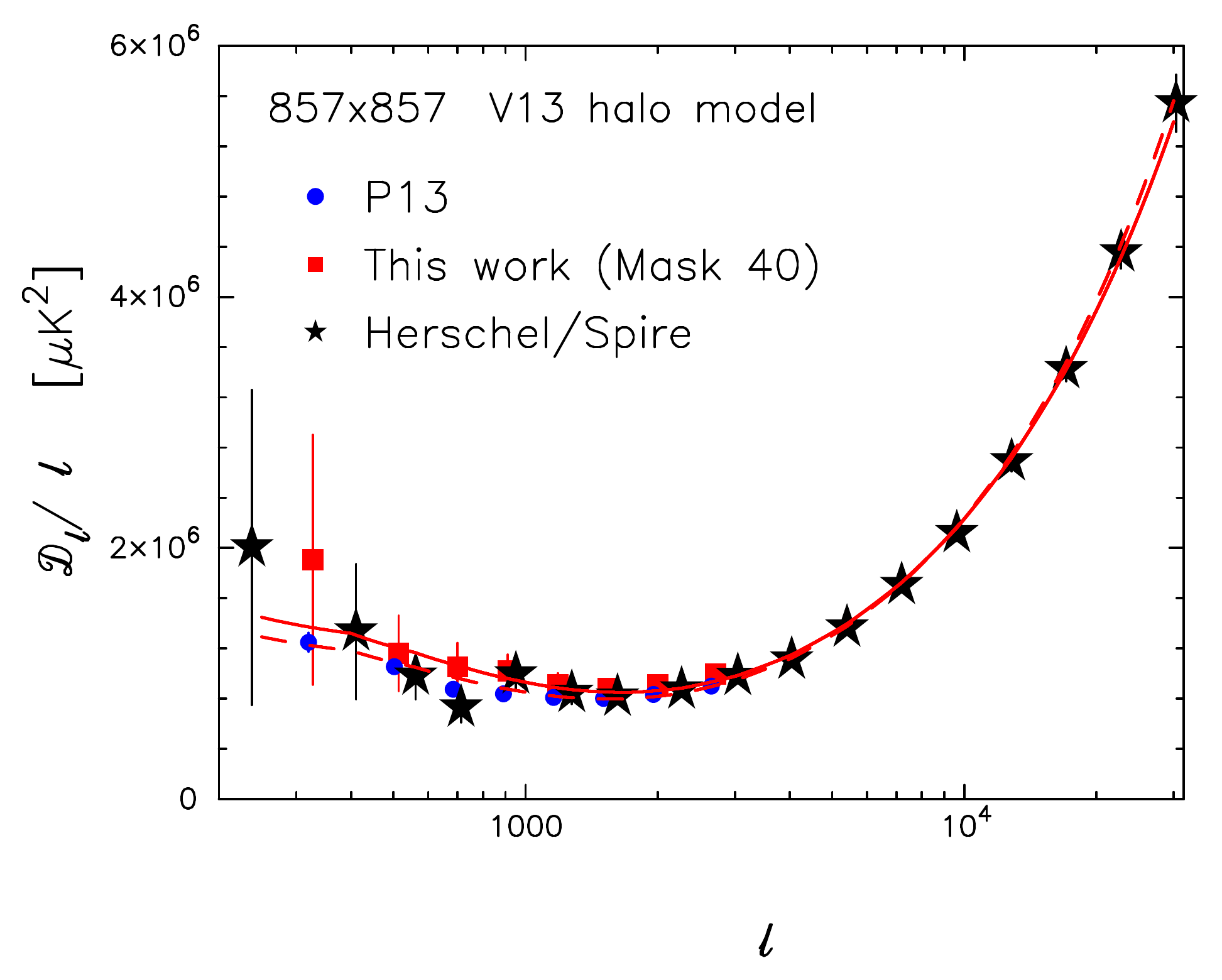}   

       \caption{Our spectra (red squares) as plotted in the left-hand
         panels of Fig.~\ref{fig:us_v_pip} (i.e., with no additional
         correction to the dust amplitudes) compared to the \cppcib\
         spectra (blue points) and the \Herschel/SPIRE spectra from
         \viero\ (black asterisks). The lines show best fits of a
         Poisson amplitude and clustered CIB template amplitude to
         each spectrum. The panels to the left show fits assuming a 
power-law clustered CIB template. The central panels show fits to the \cppcib\ halo model template. Panels to the right show
fits to a halo model template from \viero\ (as described in the text). The solid lines in the left and centre panels show fits to
our spectra while the dashed lines show fits to the \cppcib\ spectra. These yield nearly identical solutions. The solid lines in the right-hand panels show fits of the \viero\ template to our spectra combined with the \viero\ spectra, while the dashed lines show fits
to the \cppcib\ spectra combined with the \viero\ spectra. }
     \label{fig:fits_to_templates}
  \end{center}
\end{figure*}

An important result shown in Figs.~\ref{fig:dx11-p2013-herschel}
and~\ref{fig:dx11-p2013} is the remarkable stability of our results
computed  on masks 30, 40, and 50. Over this range of masks, the dust
power varies by a factor of around four, yet the recovered CIB spectra are 
almost identical. This shows that our likelihood technique can recover the CIB over larger areas of sky, and provides incontrovertible evidence for the 
existence of an isotropic CIB component.  

\begin{table}
\caption{CIB and Poisson power correlation coefficients across
  frequencies: (1) as measured in this work ($50\le \ell\le2500$); (2)
  \cppcib\  ($150\le \ell \le1000$), by averaging $C^{\nu\times
    \nu'}_\ell/(C^{\nu\times \nu}_\ell \times
  C^{\nu'\times\nu'}_\ell)^{1/2}$; and (3) \viero\ ($100\le \ell
  \le10000$). The measurements are also compared with model
  predictions (4) at $\ell = 1000$ from~\citet{Bethermin2013}.}
\begin{center}
\begin{tabular}{ccccc}
\hline\hline
$r^{\rm cib}_{\nu\times\nu'}$ & (1) & (2) & (3) & (4)  \\
\hline
$353\times545$&$0.975\pm0.005$&$0.983\pm0.007$&--&0.99 \\
$353\times857$ & $0.892\pm0.009$& $0.911\pm0.003$& --& 0.93\\
$545\times857$ & $0.949\pm0.003$& $0.949\pm0.005$& $0.95\pm0.03$ & 0.97\\
\hline
$r^{\rm ps}_{\nu\times\nu'}$ & (1) & (2) & (3) & (4)  \\
\hline
$353\times545$ & $0.980\pm0.008$ &$0.941\pm0.034$&--& 0.96 \\
$353\times857$ & $0.860\pm0.011$ & $0.803\pm0.054$&--& 0.83\\
$545\times857$ & $0.970\pm0.004$ & $0.897\pm0.052$& $0.98\pm0.03$& 0.92 \\
\hline
\end{tabular}
\end{center}
\label{t:correlation}
\end{table}

\begin{table*}
\begingroup
\newdimen\tblskip \tblskip=5pt
\caption{Mean values of the Poisson power from the full likelihood
  analysis and the 68\,\% confidence limits. The prediction from the
  \bcount\ model, the \cppcib\  measurements (see Table 6 and Table 9
  of \cppcib, respectively),  and the best-fit Poisson power from this
  work (\reftab{result}) are compared here. The final column gives
  upper limits on the Poisson power levels computed from the
  \Herschel/SPIRE spectra of \viero, as described in the text.
The numbers are in units of
 Jy$^2$\,sr$^{-1}$ and, where appropriate, have been colour-corrected to a
 CIB SED.}
\label{t:psprior}                            
\nointerlineskip
\vskip -3mm
\footnotesize
\setbox\tablebox=\vbox{
   \newdimen\digitwidth
   \setbox0=\hbox{\rm 0}
   \digitwidth=\wd0
   \catcode`*=\active
   \def*{\kern\digitwidth}
   \newdimen\signwidth
   \setbox0=\hbox{+}
   \signwidth=\wd0
   \catcode`!=\active
   \def!{\kern\signwidth}
\halign{\tabskip 0pt#\hfil\tabskip 1.5em&
\hfil#\hfil&
\hfil#\hfil&
\hfil#\hfil&
\hfil#\hfil&
\hfil#\hfil\tabskip 0pt\cr                            
\noalign{\doubleline}
Frequency [GHz]&Number count prediction&\bcount\ prediction&\cppcib\  measurement&Mask 40 result& \viero\ limits\hfill\cr
  \noalign{\vskip 3pt\hrule\vskip 5pt}
353&*$181\pm15$*&*$225\pm45$**&*$262\pm8$**&*$226\pm8$**&--\cr
545&$1729\pm82$*&$1454\pm291$*&$1690\pm45$*&$1539\pm90$*&$< 1775 \pm 63$\cr
857&$7355\pm221$&$5628\pm1126$&$5364\pm343$&$5929\pm428$&$< 6240  \pm 80$\cr
      \noalign{\vskip 3pt\hrule\vskip 5pt}
}}
\endPlancktablewide 
\endgroup
\end{table*}

We find generally good agreement among the the various analyses for the combined
clustering+Poisson signal (shown in the left-hand panels in
Figs.~\ref{fig:dx11-p2013-herschel} and~\ref{fig:dx11-p2013}) and also
for the inferred clustered CIB component (shown in the right-hand
panels).  The latter comparison is more model dependent, requiring
fidelity of the respective likelihoods and clustered CIB templates.
It is also  affected by priors imposed on the Poisson
amplitudes. Here the agreement between \cppcib\ and our results is less
good, particularly for the $353\times353$ and $353\times545$
spectra. The red lines in Figs. \ref{fig:dx11-p2013-herschel} and
\ref{fig:dx11-p2013} show our best fits for the power-law CIB
template. These provide good fits to our spectra. The black lines show
the best-fit halo model as presented in \cppcib. For the combined
clustering+Poisson signal, the power-law and halo model fits are
nearly identical, but they look quite different for the clustered CIB
component. This poses an interesting (and physically important)
problem: \cppcib\ claim that their halo model provides a good fit to their
spectra, whereas we find that the power-law CIB template provides a
consistently better fit to our spectra (as shown in Figs.
\ref{fig:cib}, \ref{fig:cib2}, and \ref{fig:compare_cib}).  Before
tackling this apparent inconsistency, we first present a more detailed
comparison of our spectra with those of \cppcib. To reduce the numbers of
figures, we will present results for the $353\times353$,
$545\times545$ and $857\times857$ spectra only.\footnote{The behaviour of
  the cross-frequency spectra can be inferred straightforwardly from
  the comparisons of the auto-frequency spectra.}

The panels to the left in Fig.~\ref{fig:us_v_pip} show our dust-cleaned spectra (mask 40) compared to the spectra from
\cppcib\  (from their Table D1, with corrections for the change in
calibrations from the 2013 to 2015 maps) binned using exactly the same
binning as in \cppcib. For our spectra, the dominant source of error
in these broad bandpowers is the subtraction of the dust template,
which is highly correlated across multipoles.
The panels to the right show what happens if the dust amplitudes are allowed to vary, minimising differences between our spectra and
those of \cppcib. For all of the spectra, small shifts upwards in the dust
amplitudes ($0.6\,\sigma$ for $353\times 353$, $1.2\,\sigma$ for
$545\times 545$, and $1.5\,\sigma$ for $857\times857$) bring the spectra
into good agreement. It is important to point out that there is some
sensitivity of the dust solutions to the shape assumed for the
clustered CIB template. Figure~\ref{fig:us_v_pip} show our results for
the power-law CIB template and so it is possible that the small
systematic underestimation of the dust amplitudes compared to
\cppcib\  may be a consequence of a mismatch between the power-law
template and the true clustered CIB spectrum. However, any shifts in the dust template
amplitudes are small and well within our error budget. We conclude, therefore,
that our dust-cleaned spectra, computed over large areas of the sky,
are in good agreement with those of \cppcib.

In fact, the small differences in the spectra shown in
Fig. \ref{fig:us_v_pip} have little impact on the physical
interpretation of the clustered CIB component. This is illustrated in
Fig.~\ref{fig:fits_to_templates}, which compares the $545\times 545$
and $857\times 857$ spectra shown in the left-hand panels of
Fig. \ref{fig:us_v_pip} (i.e., with no further dust correction to our
spectra) with the \Herschel/SPIRE spectra from \viero\
plotted up to multipoles $\ell \sim 30\,000$. The panels to the left
show simple $\chi^2$ fits of the power-law clustered CIB template plus
a point source
amplitude to each of our spectra (solid lines) and to the \cppcib\  spectra
(dashed lines). (We note that for our spectra, these $\chi^2$ fits to the binned
dust-cleaned spectra give CIB and Poisson amplitudes in very good agreement with
those obtained from the full likelihood analysis.) The fits to both sets of spectra
are nearly identical and are in good agreement with the \Herschel/SPIRE spectra at
high multipoles. The central panels show what happens if we switch from the power-law CIB template
to the \cppcib\  halo model template.\footnote{The \cppcib\ halo model templates are tabulated up to a maximum multipole
of $\ell=13\,000$.}
Again, the fits to both sets of spectra are almost identical
and overshoot the \Herschel/SPIRE spectra by a wide margin. In fact, the $\chi^2$ values for the \cppcib\ 
spectra strongly favour the power-law CIB template over the \cppcib\
halo model. For nine data points, 
the power-law CIB model fits to the \cppcib\  spectra give $\chi^2 =
3.9$ and $6.4$, respectively, for the 
$545\times545$ and $857\times857$ spectra, whereas the \cppcib\  halo model fits give $\chi^2=33$ and
$\chi^2 = 20$ for these spectra. Although the best-fit halo model in \cppcib\  gives an acceptable fit
to the full set of spectra over the frequency range $3000$--$217$ GHz ($\chi^2 = 100.7$ for 98 data points)
it does not give good fits to the $545\times 545$ and $857\times857$ spectra.

The panels to the right in Fig.~\ref{fig:fits_to_templates} show
$\chi^2$ fits to the \Planck\ and \Herschel/SPIRE spectra
using a $350\,\mu\text{m}\times 350\,\mu\text{m}$ halo model template from  \viero.\footnote{Specifically, we use the halo model template 
  {lss14$\_$halo$\_$model$\_$350x350$\_$flux$\_$cut$\_$300mJy.txt},
  available from \url{http://www.astro.caltech.edu/~viero/viero_homepage/toolbox.html}.}
This
template provides good fits to the \Planck\ and \Herschel/SPIRE spectra
over the full range of multipoles $\ell \approx 200$--$30000$. Unsurprisingly,
over most of the multipole range probed by \Planck, the \viero\ template
is closely approximated by a power-law with index $\gamma^{\rm cib} =
0.53$. The main difference between the \cppcib\ and \viero\ CIB templates is in
the relative amplitudes of the one and two-halo contributions. The
two-halo contribution in the \viero\ template has a lower amplitude relative to the one-halo
term,  leading to better fits to the data.

\reftab{correlation} compares the Poisson and clustered CIB correlation coefficients
determined from our analysis with those measured by \cppcib\ and \viero. Both components are
highly correlated between $545$ and $857$ GHz, as predicted by the \citet{Bethermin2013}
models; see column (4) of \reftab{correlation}. Both our analysis and \cppcib\ show a decorrelation
of the $353$\,GHz CIB  and Poisson amplitudes from $857$\,GHz, again in qualitative
agreement with the \citet{Bethermin2013} models.

\subsection{Consistency of Poisson amplitudes with source counts}
\label{sec:psprior}

Table~\ref{t:psprior} compares our Poisson point source amplitudes
(for the power-law CIB model fits) with the values computed in
Sec.~\ref{sec:ps} from measurements of source counts. At $857$ and
$545$\,GHz, our likelihood analysis favours Poisson amplitudes
somewhat lower than the source count
predictions. We note that at these frequencies, the predictions are
dominated by the counts from~\citet{Bethermin2012b}.
At $353$\,GHz, our best-fit Poisson power level is about $25\,\%$ larger than source count 
prediction based on fits to the counts from~\citet{Geach2016}, and more than a factor of two higher
than expected from the APEX/LABOCA and ALMA counts plotted in
Fig.~\ref{fig:sourcecount}. However, the source counts of~\citet{Geach2016} show no obvious turnover
at faint flux densities and so our predicted Poisson power levels are strongly dependent on the 
extrapolation of the source counts to faint levels.

Table~\ref{t:psprior} also lists the Poisson point source amplitudes
determined by \cppcib\ from their halo model fits. The difference in the
clustered CIB template will introduce systematic differences in the
recovered Poisson amplitudes.  In addition,~\cppcib\ introduced flat priors
on the point source amplitudes of $\pm 20\%$ around the \bcount\
predictions (listed in the third column of 
~\reftab{psprior}). In fact, the \cppcib\ MCMC fits hit the upper end of their
priors at $353$\,GHz and $545$\,GHz. Despite these differences,
qualitatively the \cppcib\ spectra give similar results, namely a larger Poisson amplitude
 at $353$\,GHz and somewhat lower Poisson amplitudes at $545$\,GHz
and $857$\,GHz compared to the number count predictions. 

In fact, since the \Herschel/\Planck\ spectra extend up to $\ell \approx
30\,000$ one can derive strict upper limits to the Poisson point source
amplitudes. We fitted the three highest multipole bins ($\ell = 17\,064$--$30\,369$) of the $350\times350\micron$ and $500\times 500\micron$ `extended sources masked' spectra given in Table 10 of \viero\ to a
constant. This gives the limits listed in the final column of Table~\ref{t:psprior}. The actual Poisson amplitudes could be 
substantially lower  depending on the relative
contribution of the one-halo CIB term to the measured spectra.
\footnote{For example, the fits 
of the \viero\  template shown in \reffig{fits_to_templates} give Poisson amplitudes of
$1130 \pm 35$ and $4410 \pm 185$ ${\rm Jy}^2 {\rm sr}^{-1}$ for the $545\times545$
and $857\times857$ spectra respectively.These are substantially lower than the mask 40 
numbers listed in Table \ref{t:psprior} exacerbating the discrepancy with the
number count predictions. This may indicate that the one halo term in 
\viero\ template is too steep at high multipoles.}
 The upper 
limits on the Poisson amplitudes derived from the \Herschel/SPIRE
spectra are consistent with those determined from the
\Planck\ spectra, further suggesting that the number count predictions
at $545$\,GHz and (particularly) $857$\,GHz overestimate the Poisson amplitudes.

\section{Addressing deficiencies in the likelihood analysis}
\label{sec:gal}

The previous section demonstrated that our likelihood analysis
produces stable solutions for the CIB power spectra over large areas
of the sky, which are in good agreement with the spectra measured in~\cppcib. However, there are two noteable problems with our approach.  The
first concerns the high $\chi^2$ values reported in Table
\ref{t:chi2_binned}. The $\chi^2$ values increase with frequency,
strongly suggesting that our modelling of the contribution of Galactic
dust to the covariance matrices is deficient. Section
\ref{sec:ngnoise} develops a simple model for the sample variance of this
statistically-anisotropic component. The second problem concerns the
large sample variance at low multipoles inherent in subtracting dust
at the power-spectrum, rather than the map, level.  This can be seen
clearly in Fig. \ref{fig:us_v_pip}. If one wants to recover the CIB
spectrum accurately at multipoles $\ell \simlt 500$, there really is no
alternative other than to subtract dust at the map level using a
tracer of Galactic emission. As an alternative to \ion{H}{i} cleaning,
Sec.~\ref{sec:dustclean} explores whether we can exploit the small
differences between the Galactic dust and CIB spectral energy 
distributions in the \Planck\ maps  to subtract dust whilst retaining 
information on
the CIB.

\subsection{Sample variance of statistically-anisotropic Galactic dust}
\label{sec:ngnoise}

As discussed in~\refsec{likelihood}, we have modelled  the sample
variance from Galactic dust as if it were a statistically-isotropic
Gaussian field, with power spectrum appropriate to the given mask. In reality, 
the statistical
anisotropy in the dust emission will increase the sample variance of
the dust power spectrum, and also the covariance between different
multipoles. We noted in~\refsec{resultcib} that this may explain the
poor $\chi^2$ values that we find for the $545\times 857$ and
$857\times 857$ spectra and the oscillatory residuals at $\ell \simlt 1000$
seen in Figs.~\ref{fig:cib} and~\ref{fig:cib2}. In this section, we present
some tests of this hypothesis, and develop a simple heuristic model that
allows us to account for statistically-anisotropic dust in the power
spectrum covariance matrix.

We showed in~\refsec{dust} that in regions of low dust emission, the
dust power spectrum measured from mask-difference spectra is very
similar in shape. However, the amplitude of the power spectrum varies
strongly across the sky. This motivates a simple model for the
dust emission as a statistically-isotropic, Gaussian process modulated
by a large-scale field, i.e., we assume that the dust emission 
$d(\hat{\mathbfit{n}})$ varies as 
\begin{equation}
d( \hat{\mathbfit{n}}) =
m(\hat{\mathbfit{n}})\left[1+g(\hat{\mathbfit{n}})\right],
\label{eqn:mod1}
\end{equation}
 where
$g(\hat{\mathbfit{n}})$ is a statistically-isotropic, zero-mean,
Gaussian random field and $m(\hat{\mathbfit{n}})$ is a more
slowly-varying modulation field. Note that this model has the local
variance of dust emission proportional to the square of the local mean
emission, consistent with the findings in ~\cite{Miville2007}.

We estimate the modulation field by computing the local variance from smoothed, yearly maps at 857\GHz\ within circular regions of radius $6\degr$ centred on \healpix $N_{\rm side} = 256$ pixels. The covariance matrix for this modulated-Gaussian model can be calculated following the standard calculation for a masked, statistically-isotropic Gaussian field (e.g.,~\citealt{Efstathiou2004}) but with the mask $w(\hat{\mathbfit{n}})$ replaced by $w(\hat{\mathbfit{n}}) m(\hat{\mathbfit{n}})$. In the limit of wide bandpowers, the effective number of modes that determines the power spectrum variance is 
\begin{equation}
\nu_\ell = (2\ell+1) \frac{\langle w^2 m^2 \rangle^2_\Omega}{\langle w^4 m^4 \rangle_\Omega} \, ,
\label{eqn:mod}
\end{equation}
and is lower than the variance computed assuming no modulation, which is given by Eq.~\ref{eqn:mod}  with $m(\hat{\mathbfit{n}})=1$ (\citealt{Hivon2002}; see also~\citealt{Challinor2005}).
Here, the angle brackets, $\langle \cdot \rangle_\Omega$, denote averages over the sky. The reduction in the number of modes increases the power spectrum variance, and the covariance between different multipoles.
We defer a fuller discussion of the modulated-Gaussian model, and the details behind the derivation of Eq.~\ref{eqn:mod} and its extension for unbinned spectra, to~\refapp{dust}.

\begin{figure*}
  \begin{center}
        \includegraphics[width=120mm]{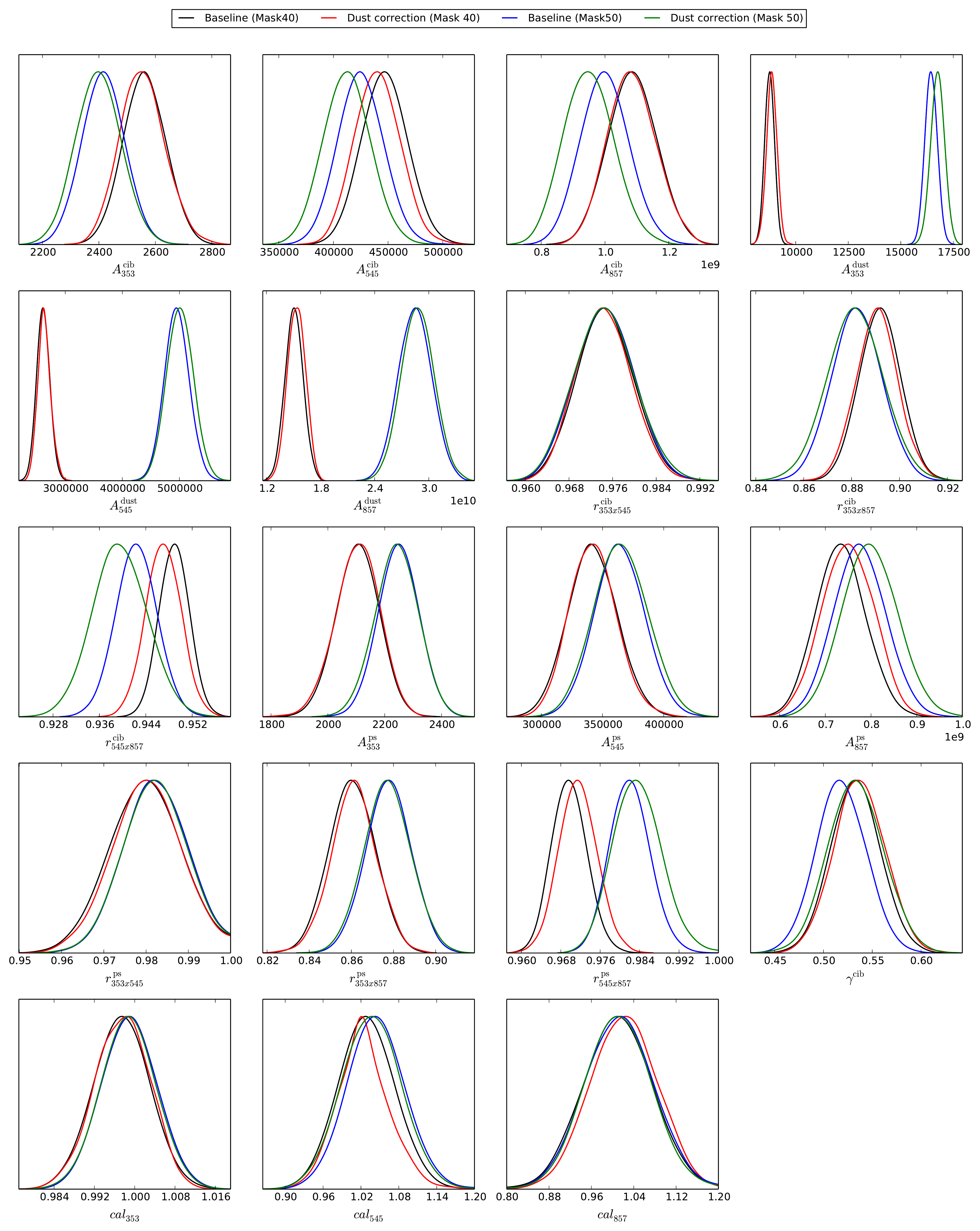}
       \caption{Posterior distributions of the model parameters (for the power-law clustered CIB model) for masks 40 (black) and 50 (blue), and the same after accounting for the statistical anisotropy of Galactic dust in the power spectrum covariance matrix (red for mask 40 and green for mask 50). Note that the results from mask 40 are very similar before and after the correction to the covariance matrix.}
            \label{fig:result}
  \end{center}
\end{figure*}

\begin{figure*}
  \begin{center}
        \includegraphics[width=85mm]{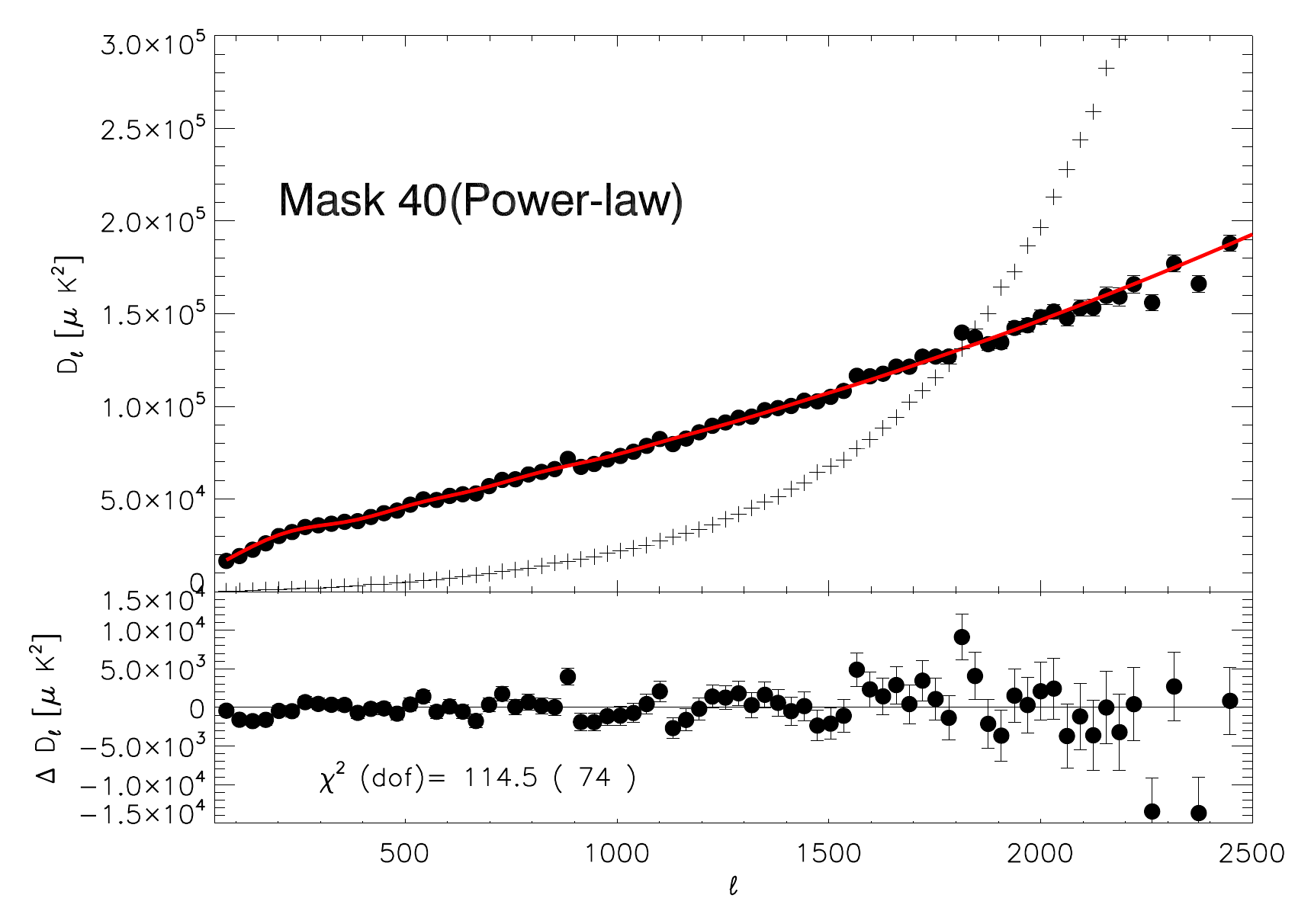}
        \includegraphics[width=85mm]{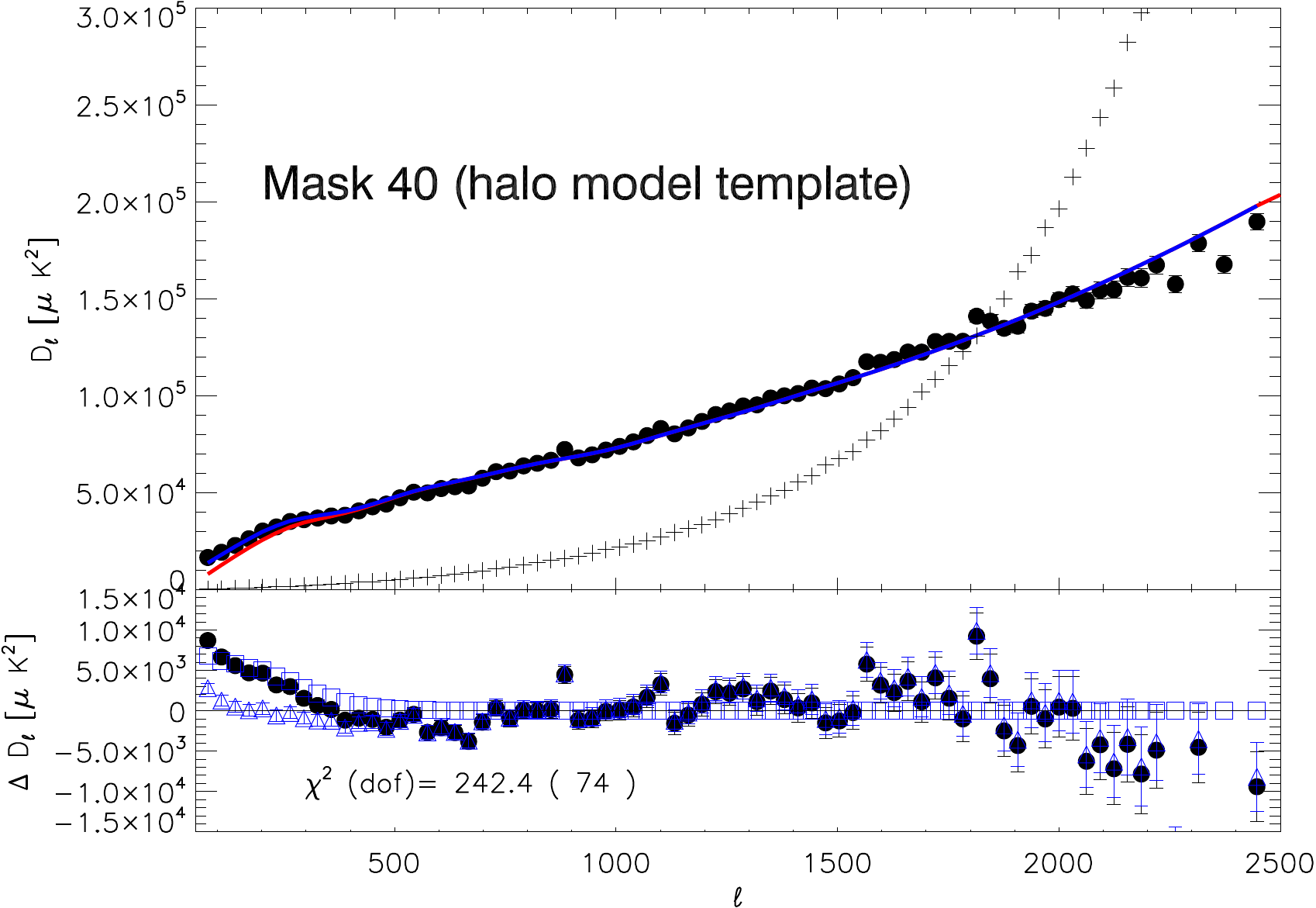}  
  \caption{Dust-nulled spectrum for mask 40 computed from
    Eq.~(\ref{eqn:dustclean}) with values of $\alpha$ taken
    from~\reftab{dustclean}. We show results for the analysis with a
    power-law template for the clustered CIB (left) and the halo model
    template (right). The red lines show the model predictions using
    the best-fit parameters of the respective analysis. The data is
    binned using the same scheme as in Figs.~\ref{fig:cib} and~\ref{fig:cib2}. 
    The small black crosses show the noise level of
    the dust-nulled map of Eq.~(\ref{eqn:dustclean1}) computed from the half-ring difference noise
    maps. The bottom panel of each plot shows the residuals of the
    data with respect to the model. Error bars are computed from the
    diagonals of the binned covariance matrices. There are clear
    excesses at $\ell < 300$ in the dust-nulled spectrum with respect
    to the model prediction for the halo model analysis.  These
    excesses are due to small variations in the SED of the Galactic
    dust emission over the sky.  The blue squares in the bottom panel
    of the right-hand plot show the power spectrum of the same
    combination of the 857\GHz\ and 545\GHz\ maps measured on the
    \Planck\ models of dust emission discussed in~\refsec{dust}. This
    spectrum has been multiplied by a factor of 0.7. With this small
    rescaling, the power spectrum from the SED variations in the model
    reproduces the excess observed in the data on large scales. Adding
    the modelled power from dust-cleaning errors to the CIB model gives the
    blue line plotted in the top panel. The residuals of the data
    with respect to this model are shown as the blue triangles in the
    bottom panel. }
     \label{fig:dustcleaned}
  \end{center}
\end{figure*}

We constructed a new likelihood using the modulated-Gaussian model of
Eq. \ref{eqn:mod1} and used this to derive new parameter constraints.
The marginalised posterior distributions for the power-law CIB model
are shown in~\reffig{result} for masks 40 and 50. For most parameters,
the changes are small with relative shifts in the peaks of the
posteriors much less than $1\,\sigma$. As expected, the changes are
larger for mask 50 than mask 40. In fact, at 857\GHz\ where accurate
dust modelling is most critical, the dust correction to the covariance
moves the parameters for mask 50 closer to those found for mask
40. There are also substantial improvements in the $\chi^2$ of the
individual auto- and cross-frequency spectra with the new likelihood,
as shown in~\reftab{dresult}.  These results provide further evidence
that it is our modelling of dust sampling variance that is mainly responsible
for the high $\chi^2$ values. Our heuristic model goes a long way in
accounting for the high $\chi^2$ values, but does not completely
eliminate the problem for the $857\times857$ spectra.

\begin{table*}
\begingroup
\newdimen\tblskip \tblskip=5pt
\caption{\small Goodness-of-fit tests for the individual auto- and cross-frequency spectra, binned with $\Delta \ell = 31$, with respect to the best-fit model (power-law clustered CIB) from the full likelihood analysis for masks 40 and 50. The quantities tabulated are the same as in~\reftab{chi2_binned}. Results are shown for the standard covariance matrix (``original''), which reproduce values from~\reftab{chi2_binned} for convenience, and with the correction for statistically-anisotropic Galactic dust (``dust corrected''). The number of degree of freedom is 74 in all cases. }
\label{t:dresult}                            
\nointerlineskip
\vskip -3mm
\footnotesize
\setbox\tablebox=\vbox{
   \newdimen\digitwidth
   \setbox0=\hbox{\rm 0}
   \digitwidth=\wd0
   \catcode`*=\active
   \def*{\kern\digitwidth}
   \newdimen\signwidth
   \setbox0=\hbox{+}
   \signwidth=\wd0
   \catcode`!=\active
   \def!{\kern\signwidth}
   \newdimen\signwidth
   \setbox0=\hbox{<\,}
   \signwidth=\wd0
   \catcode`?=\active
   \def?{\kern\signwidth}
\halign{\tabskip 0pt#\hfil\tabskip 1.5em&
\hfil#\hfil&
\hfil#\hfil&
\hfil#\hfil&
\hfil#\hfil&
\hfil#\hfil&
\hfil#\hfil&
\hfil#\hfil&
\hfil#\hfil\tabskip 0pt\cr                            
\noalign{\doubleline}
Spectrum& $\chi^2$&$\hat \chi^2$&$\Delta\chi^2/\sqrt{2N_{\rm dof}}$& PTE[\%]& $\chi^2$&$\hat \chi^2$&$\Delta\chi^2/\sqrt{2N_{\rm dof}}$& PTE[\%]\hfil\cr
&\multispan4\hfil Mask 40 (original) \hfil&\multispan4\hfil Mask 40 (dust corrected) \hfil\cr
  \noalign{\vskip 3pt\hrule\vskip 5pt}
$353\times353$&101.12&$1.37$&2.23&?2.0*&*90.59&$1.22$&1.36&9.2\cr
$545\times545$&126.95&$1.72$&4.35&?0.01&*90.13&$1.22$&1.33&9.8\cr 
$857\times857$&168.49&$2.28$&7.76&<\,0.1*&131.32&$1.77$&4.71&0.1\cr
$353\times545$&100.23&$1.35$&2.16&?2.3*&*91.63&$1.24$&1.45&8.1\cr 
$353\times857$&111.51&$1.51$&3.08&?0.3*&*96.70&$1.31$&1.87&3.9\cr 
$545\times857$&124.94&$1.69$&4.17&?0.2*&107.52&$1.45$&2.76&0.7\cr
 \noalign{\vskip 3pt\hrule\vskip 5pt}
&\multispan4\hfil Mask 50 (original) \hfil&\multispan4\hfil Mask 50 (dust corrected) \hfil\cr
  \noalign{\vskip 3pt\hrule\vskip 5pt}
$353\times353$&*96.44&$1.30$&1.84&?*4.1&*83.06&$1.22$&!0.74&22.0\cr
$545\times545$&*92.16&$1.25$&1.49&?*7.5&*70.16&$0.95$&*-0.32&60.5\cr
$857\times857$&162.86&$2.20$&7.39&*<\,0.1&107.07&$1.45$&!2.72&*0.7\cr
$353\times545$&*89.77&$1.21$&1.30&?10.2&*70.81&$0.96$&*-0.26&58.4\cr 
$353\times857$&105.48&$1.43$&2.59&?*1.0&*80.40&$1.09$&!0.53&28.6\cr
$545\times857$&134.27&$1.81$&4.95&*<\,0.1&*93.40&$1.26$&!1.59&*6.3\cr
        \noalign{\vskip 3pt\hrule\vskip 5pt}
}}
\endPlancktablewide 
\endgroup
\end{table*}

\subsection{Dust cleaning}
\label{sec:dustclean}

Ideally, one would like to remove the dust emission at the map level
to eliminate the dust sample variance entirely. \cppcib\ used
\ion{H}{i} as a
tracer of dust emission and this appears to work well in clean regions
of the sky. At present, the \cppcib\ analysis provides the most accurate
estimates of the dust cleaning over the multipole range
$\ell=145$--$590$.\footnote{We note
  that because of the limited angular resolution of the \ion{H}{i} maps
  ($16\arcmin$) for the GASS field (\cppcib\ also used other \ion{H}{i} fields having higher angular resolution of $10\arcmin$, but the GASS field is dominating the analysis),
   the dust subtraction in \cppcib\ is performed at the map
  level up to $\ell=590$. At larger multipoles, dust is corrected at
  the power spectrum level using a power-law model for the dust spectrum with
  an amplitude fitted to the dust spectrum inferred from the
  \ion{H}{i} maps
  over the multipole range $\ell=120$--$590$. Mis-subtraction of dust at the
  map level will always lead to excess power, which is why \cppcib\ quote
  upper limits to the CIB power spectrum at $\ell<
  145$. Mis-subtraction of dust at the power spectrum level at higher
  multipoles can introduce systematic errors of either sign.}

\begin{table*}
\begingroup
\newdimen\tblskip \tblskip=5pt
\caption{Dust-cleaning coefficients $\alpha$ and the $\chi^2$ of the
  cleaned spectrum.  These are shown for various masks and for the
  power-law and halo model templates for the clustered CIB power. The
  $\chi^2$ values, and associated probabilities to exceed (PTE), are
  with respect to the best-fitting model of the respective full
  likelihood analysis.  The last three columns refer to the fits when
  a model for the residual dust power from SED variations across the
  sky (i.e., the blue squares shown in~\reffig{dustcleaned}) are
  accounted for. The number of degrees of freedom is 74 in all cases.
}
\label{t:dustclean}                            
\nointerlineskip
\vskip -3mm
\footnotesize
\setbox\tablebox=\vbox{
   \newdimen\digitwidth
   \setbox0=\hbox{\rm 0}
   \digitwidth=\wd0
   \catcode`*=\active
   \def*{\kern\digitwidth}
   \newdimen\signwidth
   \setbox0=\hbox{+}
   \signwidth=\wd0
   \catcode`!=\active
   \def!{\kern\signwidth}
\halign{\tabskip 0pt#\hfil\tabskip 1.5em&
\hfil#\hfil&
\hfil#\hfil&
\hfil#\hfil&
\hfil#\hfil&
\hfil#\hfil&
\hfil#\hfil&
\hfil#\hfil&
\hfil#\hfil&
\hfil#\hfil\tabskip 0pt\cr                            
\noalign{\doubleline}
Mask&$\alpha$&  $\chi^2$ &PTE[\%] & $\alpha$&  $\chi^2$ &PTE[\%]& $\alpha$&  $\chi^2$ &PTE[\%] \hfill\cr
  &\multispan3\hfil Power-law model\hfil&\multispan3\hfil Halo model\hfil&\multispan3\hfil  Halo model+dust residuals\hfil\cr
  \noalign{\vskip 3pt\hrule\vskip 5pt}
30& 0.0134&113.1&0.7&0.0133&348.3&$<0.1$&0.0133&222.4&$<0.1$\cr
40& 0.0130&114.5&0.2&0.0130&389.6&$<0.1$&0.0130&236.4&$<0.1$\cr
50& 0.0129&112.3&0.3&0.0129&492.8&$<0.1$&0.0129&305.1&$<0.1$\cr
      \noalign{\vskip 3pt\hrule\vskip 5pt}
}}
\endPlancktablewide 
\endgroup
\end{table*}

As an alternative, we have experimented with 
using a linear combination of the 545 and 857\,GHz maps to null
 dust emission. Consider the `dust-nulled' map
\begin{equation}
M^{\rm cleaned}= M_{545} - \alpha M_{857}, \label{eqn:dustclean1}
\end{equation}
where $\alpha=\sqrt{A^{\rm dust}_{545}/A^{\rm dust}_{857}}$ is the
ratio of the dust power spectrum amplitudes at these frequencies
determined as the best-fit of the full likelihood analysis. Here,
$M_{545}$ is a 545\GHz\ map, and similarly for $M_{857}$.  If the
dust were fully coherent between 545 and 857\GHz, 
$M^{\rm cleaned}$ should contain no dust signal, only CIB, and the power spectrum of the cleaned map
should contain no sample variance from dust. Note that this
removes from the spectrum both the sample variance from the dust
itself, and from chance correlations between dust and the CIB. We
generate two such maps corresponding to the Year-1 and Year-2
data. Since the maps have different beams, the subtraction is actually
done in the power spectrum domain by forming
\begin{multline}
\hat{C}_\ell^{\rm cleaned} = \hat{C}_\ell^{545,{\rm Y1}\times545,{\rm Y2}} - \alpha \hat{C}_\ell^{857,{\rm Y1}\times545,{\rm Y2}} \\
-\alpha \hat{C}_\ell^{545,{\rm Y1}\times857,{\rm Y2}} + \alpha^2\hat{C}_\ell^{857,{\rm Y1}\times857,\rm{Y2}} \, .
\label{eqn:dustclean}
\end{multline}
The covariance of the $\hat{C}_\ell^{\rm cleaned}$ can be constructed
from linear combinations of the elements of the 545 and
857\GHz\ blocks of the full covariance matrix. By construction, if
dust is included in the full covariance matrix with the same scaling factor
$\alpha$, the covariance of the $\hat{C}_\ell^{\rm cleaned}$ will
contain no dust.

We perform this dust cleaning on three different masks with the
cleaning coefficients $\alpha$ summarised in~\reftab{dustclean}. Note
that there are small differences in the values of $\alpha$ when
fitting the power-law and halo model templates for the clustered CIB
in the likelihood; we report both sets of values
in~\reftab{dustclean}. Figure~\ref{fig:dustcleaned} shows the
dust-nulled spectra computed from Eq.~(\ref{eqn:dustclean}) on mask
40 for the power-law and halo model templates. We also show the
best-fit models determined from the respective likelihood analyses,
which contain only clustered CIB and Poisson power by construction,
and the residuals of the measured dust-nulled spectra from these. The
$\chi^2$ of the dust-nulled spectra with respect to the models are
reported in~\reftab{dustclean}. Comparing to the raw $545\times 545$
spectrum in~\reffig{cib}, we see that the cleaning reduces the total
power by a factor of around 50 on the largest scales (this is a
reduction in CIB power as well as dust power), and the variance
of the spectra is significantly reduced. We note that the differences
$\delta \alpha$ in cleaning coefficients for the different clustered
CIB templates imply only very small differences in the level of dust
power that remains in the dut-nulled spectrum: a fraction $(\delta \alpha
/ \alpha)^2$ of the dust power in the $545\times 545$ spectrum, i.e.,
around $50 \mu{\rm K}^2$ on the largest scales compared
to around $ 10^4\,\mu{\rm K}^2$ from the
CIB. Figure~\ref{fig:dustcleaned} shows that the power-law model
provides a better fit to the dust-nulled spectra than the \cppcib\ halo
model. At high multipoles, the best-fit \cppcib\ halo model sits high
compared to the data, in agreement with the results shown in Fig.~\ref{fig:compare_cib}.

\begin{figure}
  \begin{center}
        \includegraphics[width=85mm]{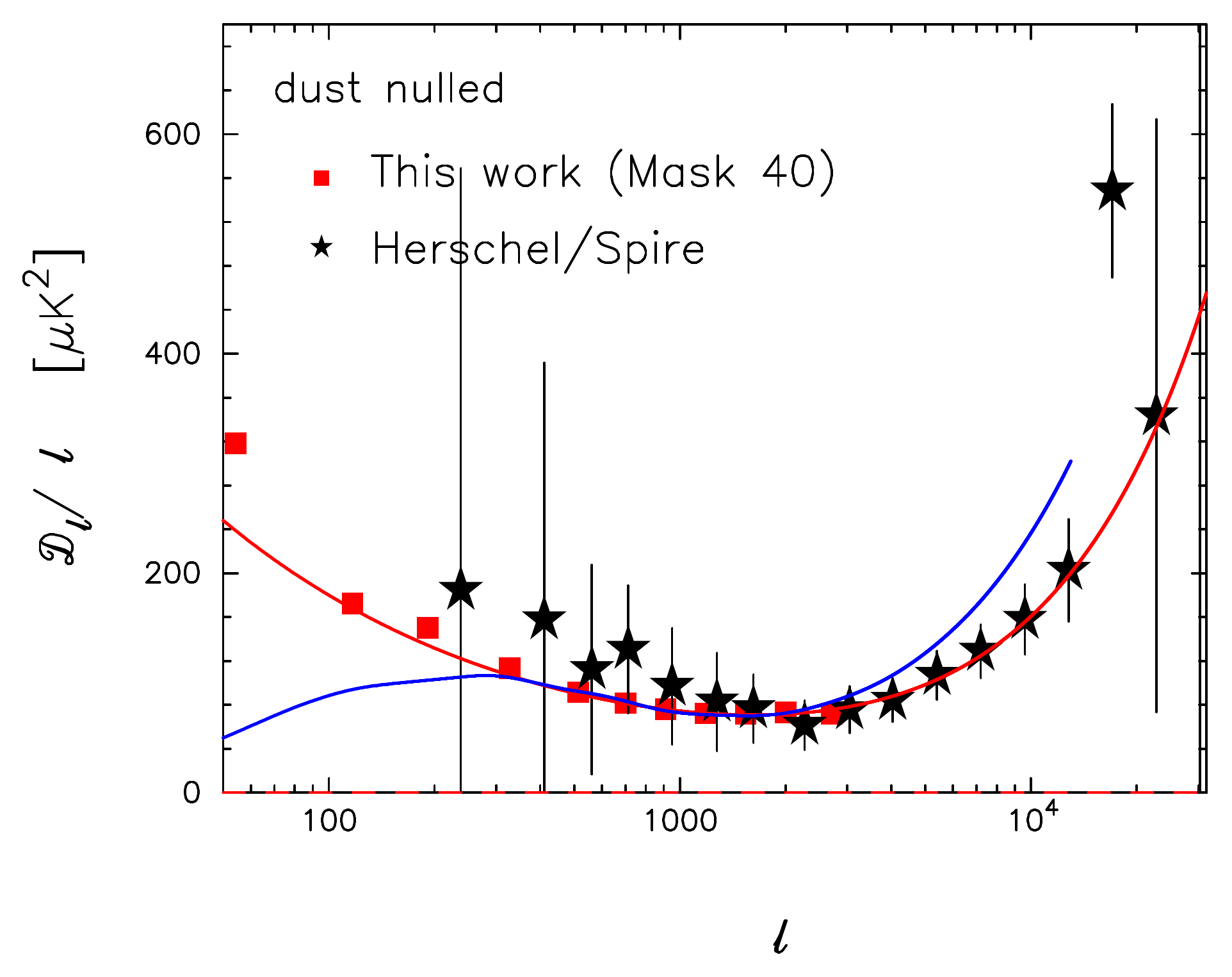}
  \caption{Dust-nulled \Planck\ spectrum for mask 40 (red squares).
    The black stars show the equivalent dust-nulled spectral combination
    (Eq.~\ref{eqn:dustclean}) for the \Herschel/SPIRE data from~\viero\ (Note that this refers to the CIB-only spectra in \viero.). The red line shows
a fit to our data assuming a power-law clustered CIB component. The blue line shows
a fit assuming the \cppcib\ halo model.}
     \label{fig:857cleaned}
  \end{center}
\end{figure}

\begin{figure}
  \begin{center}
        \includegraphics[width=80mm, angle=0]{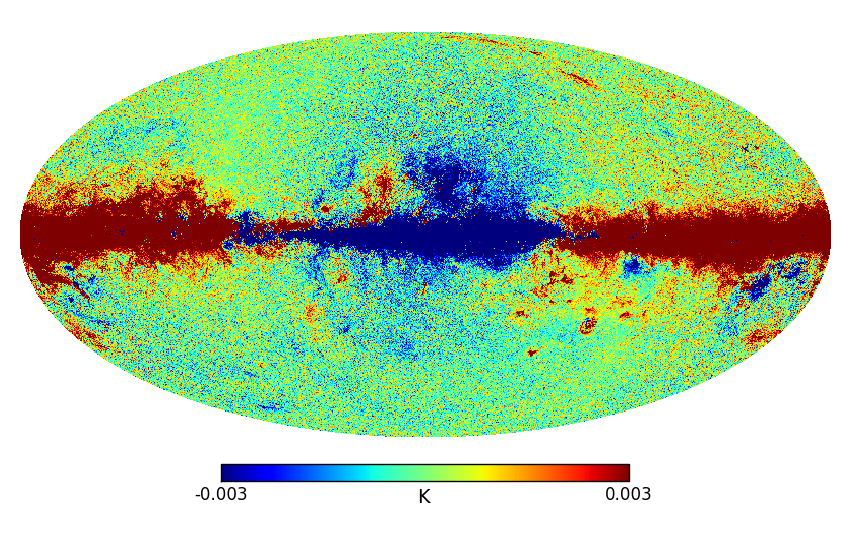} 
        \includegraphics[width=80mm,angle=0]{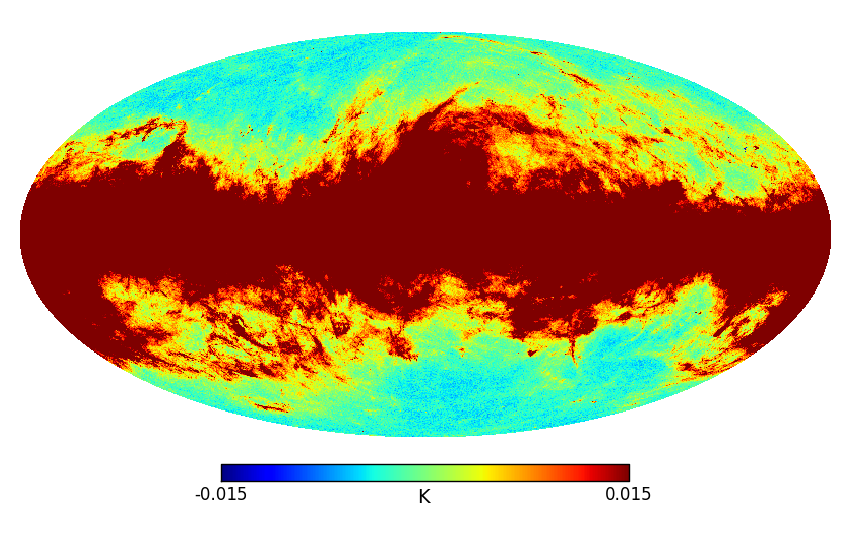} 
       \caption{\textit{Top}: Dust-cleaned map, $M_{545}-\alpha
         M_{857}$.
The dust-cleaning coefficient $\alpha =0.0130$ is taken from~\reftab{dustclean} for mask 40 and the power-law CIB template.
The colour scale is limited to $[-3\,{\rm mK},+3\,{\rm mK}]$ to better reveal the small dust residuals at high Galactic latitude (the residuals in the Galactic plane are saturated). These arise from local variations in the SED of the dust emission. \textit{Bottom}: the original 545\,GHz map, with Galactic mask 40 overlaid, to show how effective the frequency differencing is at cleaning Galactic dust emission at high latitude. Note the different ranges of the colour scale between the two panels. In both cases, the mean measured outside of Galactic mask 40 has been removed.}
     \label{fig:dustcleanmap}
  \end{center}
\end{figure}

Figure~\ref{fig:857cleaned} compares our dust-nulled spectrum with the equivalent
spectral combination (Eq. \ref{eqn:dustclean}) computed from the \Herschel/SPIRE measurements
of \viero. (The errors on the \Herschel/SPIRE points are computed by adding the errors of the individual spectra in 
quadrature as we do not have estimates of the covariances between the component spectra.  The errors plotted in
Fig.~\ref{fig:857cleaned} are therefore likely to be overestimated.) The \Planck\ and \Herschel/SPIRE spectra match well
in the overlap region $1000 \simlt \ell \simlt 3000$. The solid red line shows the best fit power-law model to our
spectrum. This fit provides a good match to the \Herschel/SPIRE spectrum at high multipoles. The blue line shows
the best-fit \cppcib\ halo model. This overshoots the \Herschel/SPIRE spectrum at high multipoles. The trends illustrated in Fig. 
\ref{fig:857cleaned} are qualitatively similar to those shown in Fig.~\ref{fig:fits_to_templates}.

The \cppcib\ halo model and power-law model give almost identical fits over the multipole range $\ell=300$--$1000$. At lower multipoles,
the halo model falls off more steeply. Subtracting dust at the power spectrum level, as in our standard analysis, 
cannot discriminate between these models at low multipoles because of
the high sample variance of Galactic dust
(see Fig.~\ref{fig:fits_to_templates}). The \ion{H}{i} cleaning applied by \cppcib\ recovers the CIB power spectrum accurately
down to multipoles of $\ell \approx 200$, but even at these multipoles the differences between the two model fits shown in
Fig. \ref{fig:857cleaned} are quite small. The dust-nulled spectrum plotted in Fig.~\ref{fig:857cleaned}
shows  clear excess power at low multipoles that is highly discrepant with the halo model predictions. This raises
the question of whether the excess is a property of the clustered CIB, or whether it is symptomatic of problems with
dust subtraction at large scales. (Recall that errors in dust subtraction at the map level will always produce excess
power.)

In fact, it is very unlikely that the excess is related to the modelling of the CIB.
At large scales, probing the two-halo term, the  CIB should be proportional to the 
matter power spectrum with an amplitude that depends on the effective galaxy bias 
and dust emissivity per unit volume.  We experimented fitting the
parametric halo model
of \cppcib\ directly to the cleaned spectrum of Fig.~\ref{fig:857cleaned}, but
were unable to find good fits at low multipoles for physically reasonable parameters.

A more plausible explanation of the excess power at low multipoles is
the presence of SED variations of the Galactic dust emission over the
sky. The cleaned (Year-1) map, $M^{\rm cleaned}$, is shown
in~\reffig{dustcleanmap} along with the 545\,GHz map. Comparing these
maps demonstrates that the frequency combination in $M^{\rm cleaned}$
does remove dust very effectively away from the Galactic
plane. However, even at high Galactic latitude there are anisotropic
features that are clearly not CIB fluctuations. See, for example, the
filamentary structures near the North Galactic Pole, which coincide
with features in the dust temperature and spectral index maps
reconstructed from \Planck\ and IRAS $100\,\mu{\rm m}$ data
in~\citet{planck2013-p06b}. 

To demonstrate that small SED variations
produce residual power in the cleaned map with the same shape and
amplitude that we see in~\reffig{dustcleaned}, we form the same
frequency difference (i.e., Eq.~\ref{eqn:dustclean1} with $\alpha =
0.0130$) of the \Planck\ dust-model maps introduced
in~\refsec{dust}~\citep{planck2014-a14}. The SED variations in these
maps are derived from the estimates of the dust temperature and
spectral index of~\citet{planck2013-p06b}. Since we are interested in
large angular scales, we smooth the simulated dust maps as described in 
\refapp{dust}. The power spectrum of the masked,
frequency-differenced dust maps are shown in~\reffig{dustcleaned} (in
the lower panel of the right-hand plot). This spectrum has a very similar
shape to the excess power over the halo model that is measured in the data, with a
comparable amplitude. In the figure we have scaled the power spectrum
of the modelled dust maps by a factor of $0.7$ to bring it closer to
the measured excess. If we include this model for residual dust power,
the halo model fits well at low multipoles, although the fit at high
multipoles is still poor. The purpose of this comparison, however, is
not to develop a detailed model of dust-cleaning residuals at low
multipoles, but rather to illustrate that dust SED variations of the
order determined by \citet{planck2013-p06b} (whether one believes them
or not) can explain the excess power at low multipoles seen in our
dust-nulled  spectrum.

Figures~\ref{fig:dustcleaned} and~\ref{fig:857cleaned} strongly suggest
that the dust-nulled  maps are dominated by CIB emission at
multipoles $\ell \simgt 300$. A linear combination of the $545$ and $857$ GHz 
 maps therefore provides a simple way of generating a CIB-dominated
map at $\ell \simgt 300$ that can be used to cross-correlate with other data
sets. \cite{Larsen:2016} have used this approach to delens
the \Planck\ temperature maps using the CIB as a tracer of the matter
distribution. \citet{Larsen:2016} actually further reduce the dust contamination 
in the cleaned maps by masking the filamentary regions near the North
Galactic Pole visible in~\reffig{dustcleanmap}. 

Finally, we note that we do not expect any significant effect on the results presented in Sec.~\ref{sec:cs} from these 
small variations in the dust SED, since they are dwarfed by the large sampling variance in the dust power spectra
at low multipoles (see Fig. \ref{fig:dustspec}).

\section{Conclusions}
\label{sec:conclusion}

The main purpose of this paper has been to adapt the likelihood
approach used in the \Planck\ cosmological parameter analysis to
recover the power spectrum of CIB anisotropies at frequencies in the range
353--857\,GHz over large areas of sky. To achieve this, we have
adapted the \camspec power spectrum analysis software to high
frequencies. We use mask-difference power spectra, exploiting the
anisotropy of Galactic dust emission, to determine a template dust
power spectrum which we then fit to measured power spectra together
with amplitudes for the clustered and Poisson contributions to the CIB
power spectrum (and a number of other `nuisance' parameters). We have
validated our likelihood methodolgy using simulations and various
consistency tests.

We now summarise our main results.
\begin{enumerate}
\item The power spectrum of Galactic dust emission has a nearly
  universal shape over the frequency range 353--857\,GHz over large
  areas of sky (at least 60\,\%).
\item Our likelihood fits recover clustered+Poisson CIB spectra over 30, 40, and
 50\,\% of sky that are almost identical, even though the dust power
 varies by a factor of around four over this range of sky fractions. This
 demonstrates the stability of dust subtraction performed at the power
 spectrum level and also provides incontrovertible evidence for the
 statistical isotropy of the CIB fluctuations over large areas of sky.
\item We fit two models for the clustered CIB: a power-law model with index
$\gamma^{\rm cib}$, and a halo model template based on the best-fit
models of \cppcib.  For the power-law model we find $\gamma^{\rm
  cib}=0.53\pm0.02$, consistent with the values derived as
foreground solutions from the \Planck\ CMB likelihood at frequencies $\nu\le
217$\GHz~\citep{planck2013-p08}. Over the frequency range 353--857\,GHz, the power-law model is found to be a better fit to our spectra
than the \cppcib\ halo model.
\item Our CIB spectra are consistent with those determined by \cppcib\ over around
$2\,000\,\text{deg}^2$ in regions of low \ion{H}{i} column density and therefore low dust
emission. Our results are also consistent with the spectra measured by~\viero\ using \Herschel/SPIRE data.
\item Fitting either our spectra or the \cppcib\ spectra at 545 and 857\,GHz over
the multipoles probed by \Planck, we find that the \cppcib\ halo model
overpredicts the \Herschel/SPIRE spectra at high multipoles $\ell
\simgt 3\,000$.  A halo model from~\viero\ gives acceptable fits to the
\Planck\ and \Herschel/SPIRE data over the entire multipole range $200
\simlt \ell \simlt 30\,000$. The main difference between the~\cppcib\ and~\viero\
halo models lies in the relative amplitudes of the one- and two-halo
terms; the two-halo term in the~\viero\ model has a lower amplitude
so that the sum of the two terms is closely approximated by a single
power-law.
\item Our best fits give lower Poisson point source amplitudes at 545 and
857\,GHz than those inferred from source counts. Our Poisson amplitudes
at these frequencies are consistent with upper limits (but not with the best fits derived by using~\viero\ halo model) derived from the
\Herschel/SPIRE spectra at multipoles $\ell \sim 30\,000$.  At 353\,GHz, our
best-fit Poisson point source amplitude is about 25\,\% higher than inferred from
recent source counts. However, the source count prediction at this frequency is
sensitive to extrapolation of the counts to faint flux-densities.
\item Our spectra give high $\chi^2$ values relative to the best-fit
models. The excess $\chi^2$ values correlate strongly with frequency,
suggesting that they are caused by our modelling of the dust
contribution to the covariance matrices used to construct the
likelihood. We have developed a heuristic model to account for the
statistically-anisotropic nature of Galactic dust that substantially reduces, but
does not eliminate, the excess $\chi^2$ values. The best-fit dust and
CIB parameters hardly change if we use the heuristic model in the
likelihood.
\item The main drawback of cleaning dust at the power spectrum level is the
high sample variance at low multipoles ($\ell \simlt 500$). The only
way to reduce this sample variance is to subtract dust at the map
level.  We have experimented with linear combinations of the 545 and
857\,GHz maps designed to cancel Galactic dust, exploiting the small
differences between the dust and CIB SEDs. The power spectra of such
cleaned maps are insensitive to sky fraction and have a shape at
multipoles $\ell \simgt 500$ consistent with the CIB spectra
determined from the 545 and 857\,GHz \Planck\ maps with our standard analysis
(i.e., subtracting dust at the power spectrum level). This strongly
suggests that the dust-nulled linear combination is dominated by CIB
emission at multipoles $\ell\simgt 300$.  The dust-nulled power
spectrum from \Planck\ is in excellent agreement with the equivalent
combination of spectra measured by \Herschel/SPIRE. At $\ell \simlt
300$, the dust-nulled power spectrum shows excess power over the best-fitting halo
model that is consistent with small dust SED variations over the sky
inferred by \citet{planck2013-p06b}. Recovering the CIB spectrum
accurately at multipoles $\ell\simlt 300$ presents a formidable problem that has yet to be solved.\footnote{We note that \ion{H}{i} cleaning,
as applied by \cppcib\ in areas of low \ion{H}{i} column density, leads to upper limits to the 
CIB power spectrum at multipoles $\ell \simlt 200$.}

\end{enumerate}

\section*{Acknowledgements}
We thank the following for helpful discussions: Marina Migliaccio, Steven Gratton, and Diana Harrison on the likelihood analysis; Paolo Serra and Marco Tucci on the CIB modelling and halo-model codes; and Alexander Karim on the sources counts from ALMA and SCUBA. 
Some of the results in this paper have been derived using the {\tt HEALPix}
package. GL acknowledge financial support from  "Programme National de Cosmologie and Galaxies "(PNCG) of CNRS/INSU,
France, the OCEVU Labex (ANR-11-LABX-0060) and the *AMIDEX project (ANR-11-IDEX-0001-02) funded by the "Investissements d'Avenir" French government program managed by the ANR. S.Y.D. Mak acknowledges hospitality from the Laboratoire d'Astrophysique de Marseille, where part of this work was completed.

\bibliographystyle{mnras}
\bibliography{planckcib,planck_bib}

\appendix
\onecolumn

\section{Covariance matrices and likelihood validation}
\label{app:cov}
\subsection{Mask-coupling and covariance matrices}
The construction of the coupling matrices, $M_{\ell\ell'}$, which relate the theory angular power spectra to the mean of the pseudo-spectra, and the covariance of the pseudo-spectra follows, for example,~\citet{Efstathiou2004}. 

Consider the cross-spectrum between two maps, labelled by $i$ and $j$, with independent noise realisations. We allow for different masks, $w^i(\hat{\mathbfit{n}})$ and $w^j(\hat{\mathbfit{n}})$, to be applied to these maps before taking the cross-spectrum. The mean of the pseudo-spectrum between these maps is
\begin{equation}
\langle \tilde{C}_\ell^{ij} \rangle = \sum_{\ell'} M_{\ell\ell'}[w^i\times w^j]C_{\ell'}^{ij} \, ,
\end{equation}
where $C_\ell^{ij}$ is the theory cross-spectrum (smoothed by the appropriate beam transfer functions) and
\begin{align}
M_{\ell_1\ell_2}[w^i\times w^j] &= \frac{2\ell_2+1}{4\pi} \sum_{\ell_3} (2\ell_3+1) W^{ij}_{\ell_3}
\left(
\begin{array}{ccc}
\ell_1 & \ell_2 & \ell_3 \\
0 & 0 & 0 
\end{array}
\right)^2 \, \\
&= (2\ell_2+1) \Xi(\ell_1,\ell_2,W^{ij}) \, .
\label{eq:maskcouple}
\end{align}
Here, the cross-spectrum of the masks is
\begin{equation}
W_\ell^{ij} = \frac{1}{2\ell+1} \sum_m w^i_{\ell m} \left(w^j_{\ell m}\right)^\ast \, ,  
\end{equation}
where $w^i_{\ell m}$ are the spherical multipoles of mask $i$, and similarly for $w^j_{\ell m}$.

For the covariance matrices of the pseudo-spectra, we assume that the theory power spectra are smooth on the scale over which the cross-spectra of the masks have significant support. This allows us to approximate the covariance matrices as
\begin{multline}
\text{cov}\left(\tilde{C}_\ell^{ij}, \tilde{C}_{\ell'}^{pq}\right) \approx
\sqrt{C_\ell^{ip} C_{\ell'}^{ip}} \sqrt{C_{\ell}^{jq} C_{\ell'}^{jq}} \Xi(\ell,\ell',W^{(ip)(jq)})
+ \sqrt{C_\ell^{ip} C_{\ell'}^{ip}}  \Xi(\ell,\ell',W_\sigma^{(ip)(jq)})
+ \sqrt{C_\ell^{jq} C_{\ell'}^{jq}}  \Xi(\ell,\ell',W_\sigma^{(jq)(ip)}) \\
+ \Xi(\ell,\ell',W_{\sigma\sigma}^{(ip)(jq)}) + p \leftrightarrow q \, ,
\label{eq:cov}
\end{multline}
where
\begin{align}
W_\ell^{(ip)(jq)} &= \frac{1}{2\ell+1}\sum_m w^{(ip)}_{\ell m} \left(w^{(jq)}_{\ell m}\right)^\ast \, , \\
W_{\sigma,\ell}^{(ip)(jq)} &= \frac{1}{2\ell+1}\sum_m w^{(ip)}_{\ell m} \left(w^{(jq)}_{\sigma,\ell m}\right)^\ast \, , \\
W_{\sigma\sigma,\ell}^{(ip)(jq)} &= \frac{1}{2\ell+1}\sum_m w^{(ip)}_{\sigma,\ell m} \left(w^{(jq)}_{\sigma,\ell m}\right)^\ast \, ,
\end{align}
and $w^{(ij)}_{\ell m}$ are the spherical multipoles of the product of masks $i$ and $j$ and
\begin{equation}
w^{(ij)}_{\sigma,\ell m} = \delta_{ij} \int w^i(\hat{\mathbfit{n}}) w^j(\hat{\mathbfit{n}}) \sigma^2_i(\hat{\mathbfit{n}}) \Omega_{\text{pix}} \, d \hat{\mathbfit{n}} \, ,
\end{equation}
with $\sigma_i^2$ the noise variance in map $i$ in a pixel of area $\Omega_{\rm pix}$. The noise terms in Eq.~\ref{eq:cov} assume that the noise is Gaussian and uncorrelated between pixels. In practice, we generalise these terms to account approximately for correlated noise following the procedure described in Appendix A.8 of~\citet{planck2013-p08}.

\begin{figure}
 \begin{center}
       \includegraphics[width=180mm]{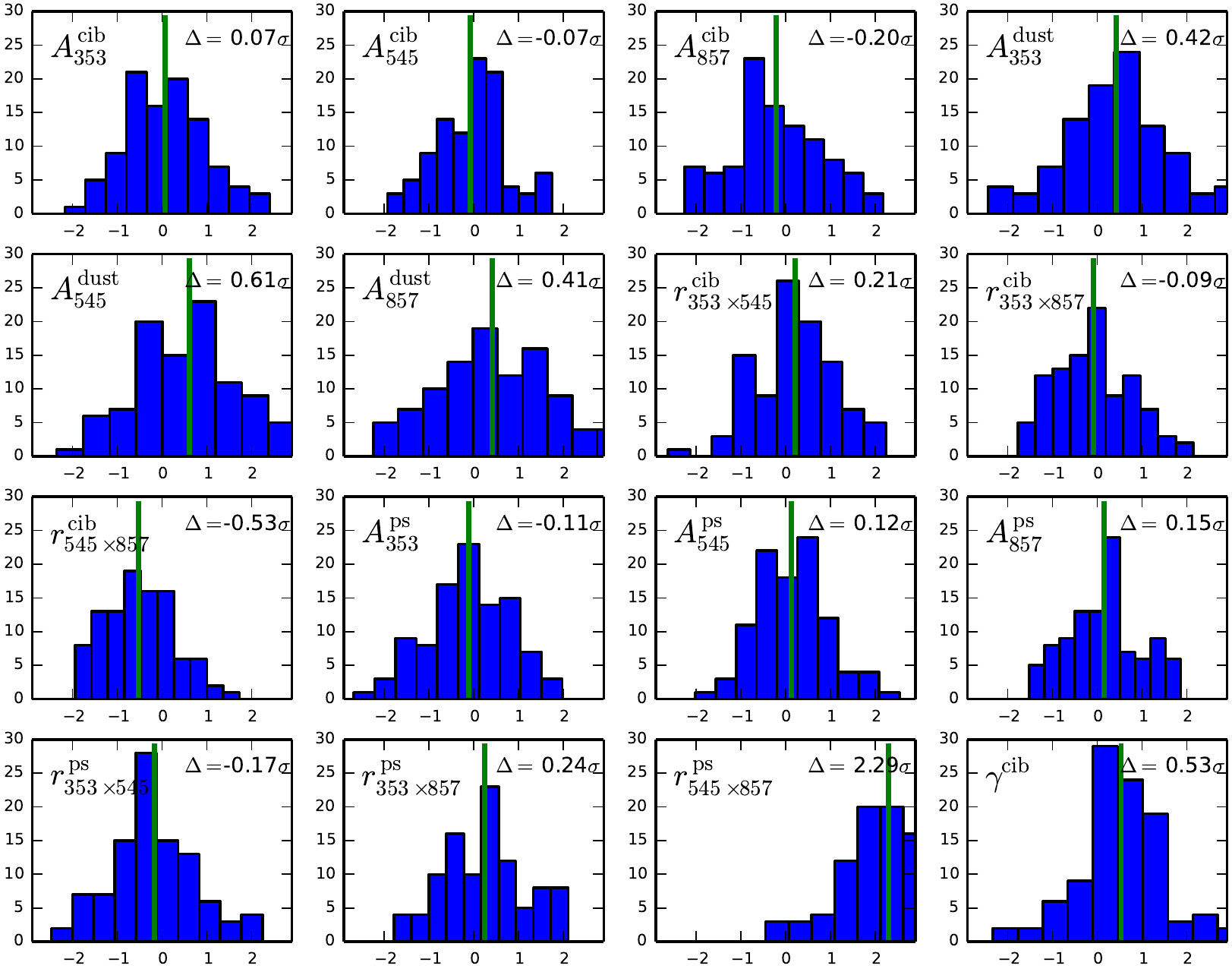}  
      \caption{Distributions of the differences between the parameter means and the input values across 100 simulations. The differences are expressed relative to the standard deviations determined from the data (for the same processing as the simulations). The averages of the differences across the simulations are shown by the green lines and correspond to the fractional biases quoted in each panel. }
    \label{fig:validation}
 \end{center}
\end{figure}

\subsection{Likelihood validation}
\label{app:val}
We have validated our likelihood by performing tests on simulated data. 
\subsubsection{Simulations}
\label{app:sim}
We generate simulations of the 353, 545, and 857\,GHz maps for the appropriate data splits (e.g., yearly or detector-sets) using Gaussian models of the signal and realistic \Planck\ noise. For each of these simulations, we repeat the same spectral and likelihood analysis as for the data. The template for the dust power spectrum and the covariance matrix are fixed to those that we use for the data.

We compute the six theory auto- and cross-frequency power spectra using the parameters of the best-fit power-law CIB model for mask 40 (Sec.~\ref{sec:cs}). We make simulated maps at 353, 545, and 857\,GHz by drawing Gaussian realisations from these spectra that are correctly correlated across frequencies (see, e.g.,~\citealt{Giannantonio2008} and~\citealt{Serra2014}). The maps are then smoothed with the appropriate beams, approximating these as circularly-symmetric, given by the auto-frequency beam transfer functions. Finally, noise realisations for the required data split from the (8th) \Planck\ full focal plane simulation set~\citep{planck2014-a14} are added to the beam-smoothed maps.
Maps are simulated at the same resolution as the data, i.e., \healpix $N_{\rm side}=2048$.

While these simulations can be used to test a number of steps in the construction of the likelihood, including the covariance matrices and the assumption that the measured power spectra are Gaussian distributed, they are not designed to test our assumptions of the dust modelling. In particular, by simulating dust as a statistically-isotropic, Gaussian field that is fully correlated across frequencies, the simulations conform with the assumptions made in the (baseline) likelihood by construction. It would be interesting in future work to use more realistic dust simulations to test these assumptions in detail.

\subsubsection{Parameter recovery}

For each simulation, we compute all auto- and cross-frequency spectra and construct a likelihood from which we sample the parameters of the power-law model using COSMOMC~\citep{Lewis:2002}. The posterior mean, $p_{\text{sim}}$, is estimated for each simulation, and compared to the input value, $p_{\text{input}}$. 
We quantify any bias through the quantities $\Delta \equiv (p_{\text{sim}}-p_{\text{input}})/\sigma_{\text{data}}$, where $\sigma_{\text{data}}$ is 
the standard deviation of the posterior distribution for the parameter determined from the data (processed in the same way as the simulations). Ideally, we expect $\Delta$ should scatter across simulations with a mean close to zero and standard deviation around unity. The former tests for bias in the recovered parameters while the latter tests that the posterior means scatter in accordance with the widths of the posterior distributions (i.e., whether the error model in the likelihood is correct).

The resulting distributions of the $\Delta$ across the 100 simulations are shown in~\reffig{validation}, with the average value of $\Delta$ quoted in each panel. The biases are generally small compared to the statistical errors, and the scatter of the posterior means across simulations is roughly in line with expectations. An exception is the Poisson correlation coefficient, $r^{\text{ps}}_{545\times 857}$, which shows a large relative bias of $2.29\,\sigma$. However, the absolute bias is only around 1\,\% since the statistical error is very small. For this reason, we have not investigated the source of bias in this parameter in detail.

\section{Poisson Powers from Source Counts}
\label{app:ps}
\subsection{Data}
For 353\,GHz, our main analysis is based on source count measurements
from~\citet{Geach2016}. These are based on the SCUBA-2 Cosmology
Legacy Survey (S2CLS), the largest survey to date at a wavelength of
850\micron, which covers around $5\,\text{deg}^2$ of extragalactic
survey fields (an order of magnitude larger than any previous survey
at this wavelength).
\citet{Geach2016} present a catalogue of nearly 3\,000 submillimeter
sources, each detected with significance greater than $3.5\,\sigma$. Their
counts are in reasonable agreement with previous surveys (see below), but are higher than some previous
counts by almost a factor of two at flux-densities $S<10$\,mJy.

To illustrate the differences between earlier source counts at faint flux-densities we
also plot counts from~\citet{Weiss2009},~\citet{Simpson2015}, and~\citet{Karim2013} in 
\reffig{sourcecount}. Briefly, these data are as follows.
\begin{enumerate}
\item \citet{Weiss2009} estimate counts from the Large Apex Bolometer Camera (LABOCA) Extended Chandra Deep Field South (ECDFS) field covering $0.35\,\text{deg}^2$,
 using $P(D)$ analysis.\footnote{A statistical method used to estimate
   the number counts in a field from the pixel histogram of an
   extragalactic map.}
They present a catalogue of 126 sub-millimeter galaxies (SMGs) detected
with a significance level above $3.7\,\sigma$. \\
\item \citet{Simpson2015} analyse 30 ALMA maps centered on bright
  sub-millimetre sources selected from the UKIDSS UDS
  field~\citep{Lawrence2007} of the S2CLS [the median flux-density is
  $S=(8.7\pm 0.4)$\,mJy]. They find that these SCUBA bright sources are comprised of a blend of multiple sources brighter than 1\,mJy in the ALMA maps. The multiplicity of the sources means that the intrinsic number counts originally measured with SCUBA measurements (e.g.,~\citealt{Coppin2006}) are likely overestimated and that many of the sources detected by SCUBA
are in fact multiple fainter sources. \\
\item \citet{Karim2013} follow up sources from the LABOCA ECDFS
  sub-millimeter survey with high-resolution ALMA observations.
Their results are in broad agreement with those of the LABOCA survey, but show a deficit
at the bright end ($S\simgt 2$\,mJy) caused by multiplicity of bright sources . There is evidence from observations at other wavelengths
(e.g., K-band selected galaxies at $z>2$;~\citealt{vanDokkum:2006})
that the ECDFS is under-dense by a factor of around
two~\citep[see][]{Weiss2009}. We therefore follow the suggestion
by~\citet{Weiss2009} and Karim (private communication, 2015) to scale
up the source count measurements of \citet{Karim2013} by a factor of
two. In fact, making this correction brings their counts into  better agreement with SCUBA and  other observations with ALMA  (e.g.,~\citealt{Hatsukade2013}).
\end{enumerate}

\noindent For 545 and 857\,GHz, we use counts based on \Herschel/SPIRE
data at 500 and $350\,\mu\text{m}$, respectively. We take counts from
several works that use different analyses and sources selected from
different fields, as follows.
\begin{enumerate}
\item \citet{Glenn2010}: Number counts are reconstructed based on $P(D)$ analysis using the sources in the GOODS-N, Lockman-North, and Lockman-SWIRE field from the \Herschel\ Multi-tiered Extragalactic Survey (HerMES).\\
\item \citet{Bethermin2012b}: Field selection is similar to~\citet{Glenn2010}, but the counts are reconstructed based on stacking analysis.\\
\item \citet{Oliver2010}: Field selection is similar to~\citet{Glenn2010}, with additional fields FLS and A2218, but the counts are directly measured from the resolved sources.\\
\item \citet{Clements2010}: Sources are selected from the largest \Herschel\ survey, ATLAS, which covers $550\,\text{deg}^2$, and separate fields from the HerMES. The counts are directly measured from resolved sources.
\end{enumerate}

\noindent For high flux densities, $S>1$\,Jy, was also use the number
counts measured in the \Planck\ HFI bands from~\citet{planck2012-VII}. As noted in~\refsec{ps}, the contribution to the Poisson power from the bright counts are subdominant. 

\subsection{Infrared source count model}
To model the faint source counts we fit  a double power-law to the data:
\begin{equation}
\frac{dN}{dS} = A\left[ \left( \frac{S}{B} \right )^{n_1} +\left( \frac{S}{B} \right )^{n_2} \right ]^{-1} .
\label{eqn:dndsf}
\end{equation}
This model has four free parameters, $A$, $B$, $n_1$, and $n_2$, which we fit to the faint source counts 
using a Gaussian likelihood. Since $n_1$ and $n_2$ are degenerate with each other, we apply priors on both parameters:  $6\le n_1\le 10$ and $0\le n_2 \le 2.5$, at all three frequencies. We treat each dataset as independent and give them equal weight in the likelihood, neglecting any correlations
between the differential source counts in different flux-density bins.
 We apply correction factors of 1.016 and 0.805 at 857 and 545\,GHz,
 respectively, to the derived Poisson power in order to convert from \Herschel\ to \Planck\ frequencies. We additionally apply cross-calibration factors of $1/1.047^2$ and $1/1.003^2$ at 545 and 857\,GHz, respectively, that account for the SPIRE/HFI relative gains~\citep{Bertincourt2015}.

\begin{figure}
  \begin{center}
        \includegraphics[width=180mm]{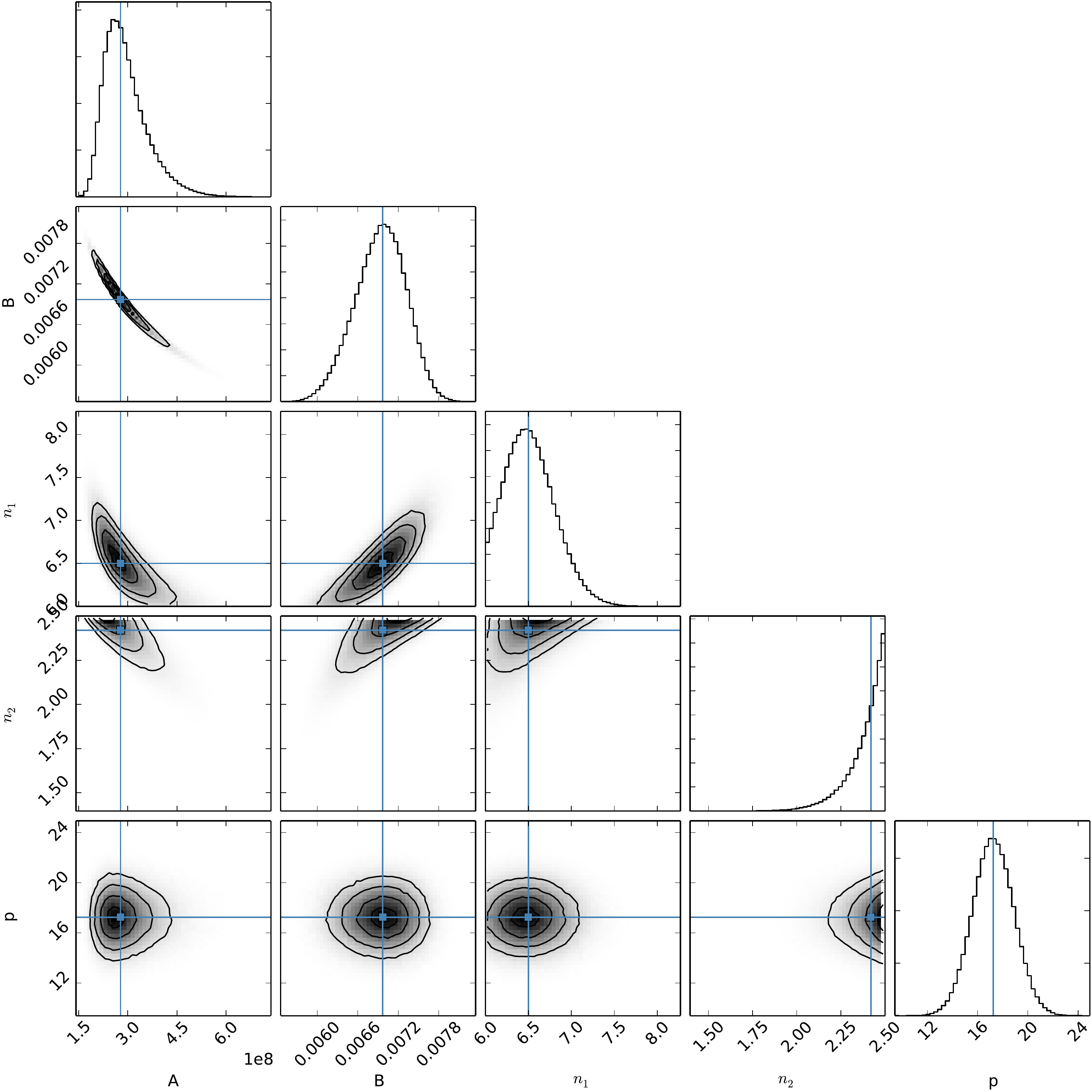}   
\caption{Constraints on the five-parameter source count model at 353\,GHz.
 The blue dots indicate the best-fit values to the MCMC chains.}
     \label{fig:tri_353}
  \end{center}
\end{figure}

\begin{figure}
  \begin{center}
        \includegraphics[width=180mm]{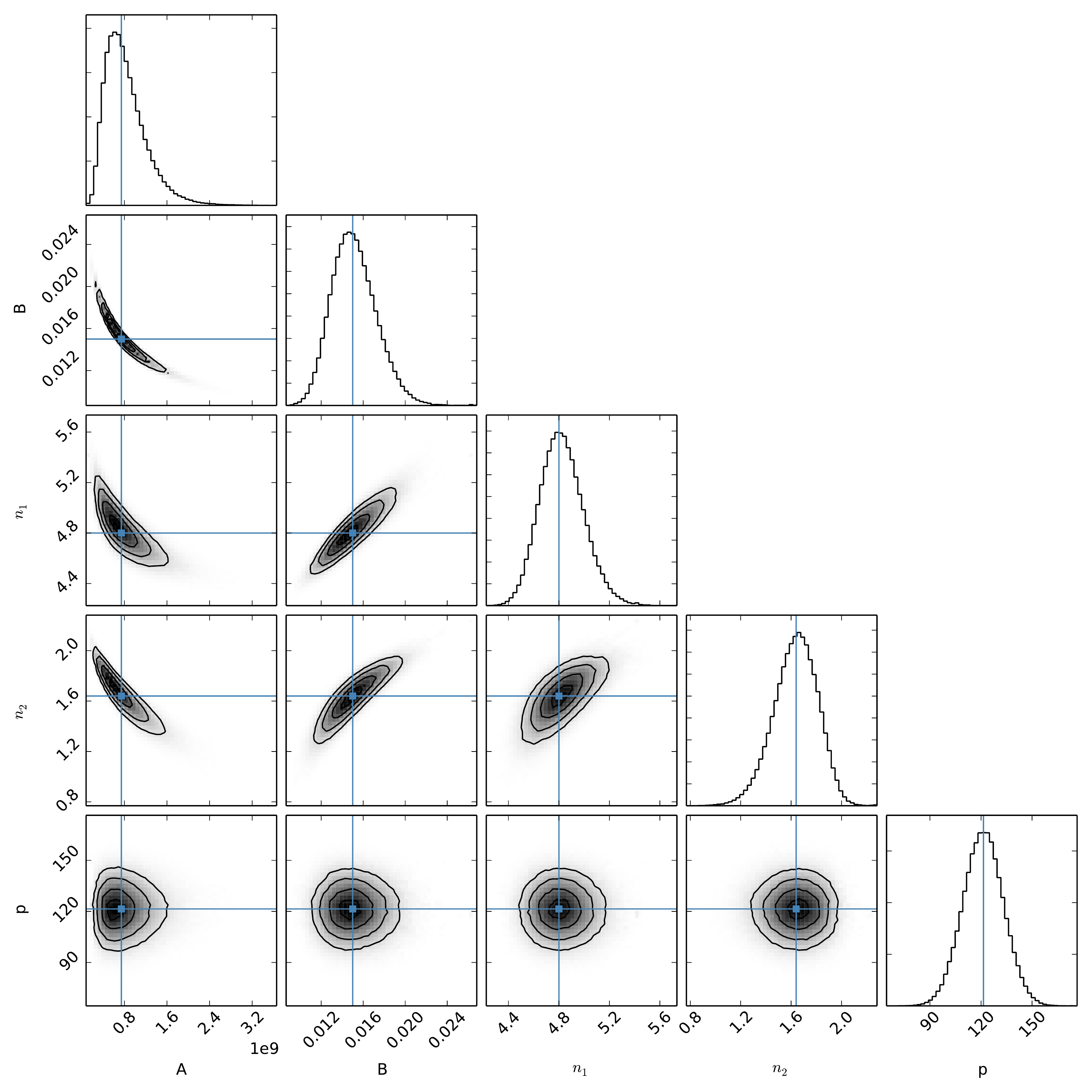}  
\caption{Same as~\reffig{tri_353}, but for 545\,GHz.}
     \label{fig:tri_545}
  \end{center}
\end{figure}

\begin{figure}
  \begin{center}
        \includegraphics[width=180mm]{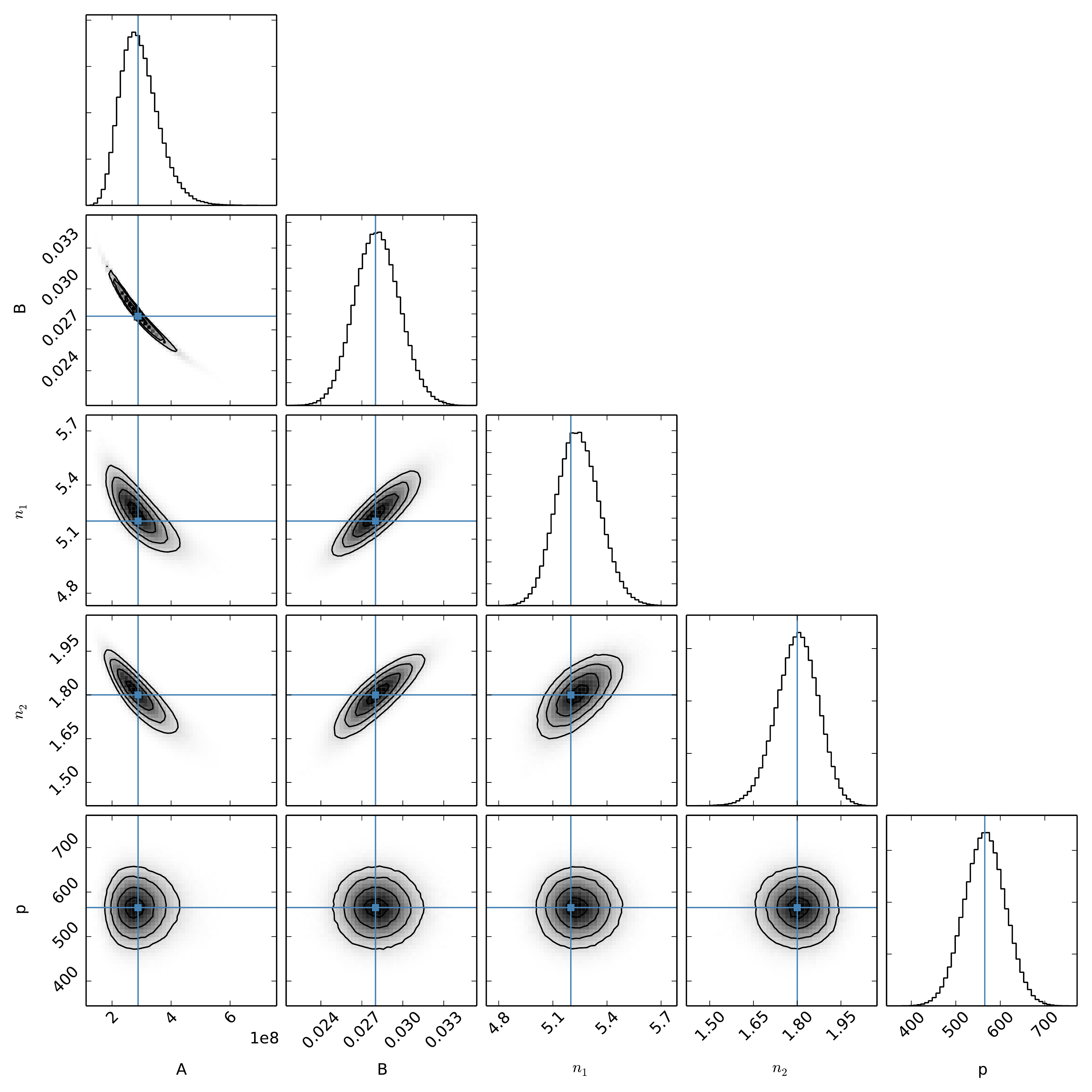}  
\caption{Same as~\reffig{tri_353}, but for 857\,GHz.}
     \label{fig:tri_857}
  \end{center}
\end{figure}

For the bright flux-densities, the differential source count distribution can be described by a single power-law:
\begin{equation}
\frac{dN}{dS} = pS^{-2.5}  .
\label{eqn:dndsfbright}
\end{equation}
We therefore add this contribution to Eq.~(\ref{eqn:dndsf}), adding an additional free parameter $p$ (the Euclidean plateau level). 
We summarise the best-fit parameters of this model
in~\reftab{pspriormodel}, and show in Figs.~\ref{fig:tri_353},~\ref{fig:tri_545}, and~\ref{fig:tri_857}, the constraints on the five parameters at 
each frequency.

\begin{table}
\begingroup
\newdimen\tblskip \tblskip=5pt
\caption{Best-fit parameters of the source count model.}
\label{t:pspriormodel}                            
\nointerlineskip
\vskip -3mm
\footnotesize

\setbox\tablebox=\vbox{
   \newdimen\digitwidth
   \setbox0=\hbox{\rm 0}
   \digitwidth=\wd0
   \catcode`*=\active
   \def*{\kern\digitwidth}

   \newdimen\signwidth
   \setbox0=\hbox{+}
   \signwidth=\wd0
   \catcode`!=\active
   \def!{\kern\signwidth}
\halign{\tabskip 0pt#\hfil\tabskip 1.5em&
\hfil#\hfil&
\hfil#\hfil&
\hfil#\hfil&
\hfil#\hfil&
\hfil#\hfil\tabskip 0pt\cr                            
\noalign{\doubleline}
Frequency&$A$ [$\text{Jy}^{-1}\,\text{sr}^{-1}$]&$B$ [$\text{Jy}$]&$n_1$&$n_2$&$p$ [$\text{Jy}^{1.5}\,\text{sr}^{-1}$]\hfill\cr
 \noalign{\vskip 3pt\hrule\vskip 5pt}
353&*$2.82^{+0.75}_{-0.50}\times10^{8}$&$0.007^{+0.004}_{-0.004}$&$6.5^{+0.5}_{-0.5}$&$2.42^{+0.06}_{-0.13}$&*$17.24^{+1.64}_{-1.63}$\cr
 545& $7.44^{+0.44}_{-0.28}\times10^{8}$&$0.015^{+0.002}_{-0.002}$&$4.80^{+0.2}_{-0.2}$&$1.64^{+0.16}_{-0.18}$&$121.44^{+11.68}_{-11.75}$\cr
 857&$2.88^{+0.70}_{-0.56}\times10^{8}$&$0.028^{+0.002}_{-0.002}$&$5.20^{+0.30}_{-0.50}$&$1.80^{+0.07}_{-0.07}$&$565.15^{+45.72}_{-45.78}$\cr
       \noalign{\vskip 3pt\hrule\vskip 5pt}
}}
\endPlancktablewide 
\endgroup
\end{table}


\section{Modelling the power spectrum covariance for statistically-anisotropic dust}
\label{app:dust}
\subsection{The modulated dust model and power spectrum covariance}

In~\refsec{ngnoise}, we introduced a simple model for statistically-anisotropic dust emission: $d(\hat{\mathbfit{n}}) = m(\hat{\mathbfit{n}})[1+g(\hat{\mathbfit{n}})]$, where $g(\hat{\mathbfit{n}})$ is a statistically-isotropic, zero-mean Gaussian field with power spectrum $C_\ell^g$, and $m(\hat{\mathbfit{n}})$ is a more slowly-varying modulation field. We begin by considering the sample variance of power spectrum estimates derived from masked versions of 
$d(\hat{\mathbfit{n}})$.

If the mask $w(\hat{\mathbfit{n}})$ and the modulation field $m(\hat{\mathbfit{n}})$ are sufficiently slowly varying, the expected value of the pseudo-spectrum is approximately
\begin{equation}
\langle \tilde{C}_\ell^d \rangle = \langle w^2 m^2 \rangle_\Omega C_\ell^g \, ,
\end{equation}
where $\langle \cdot \rangle_\Omega$ denotes an average over the sky. We remind the reader that the pseudo-spectrum is formed from the spherical multipoles $\tilde{d}_{\ell m}$ of the masked signal as $\tilde{C}_\ell^d = \sum_m |\tilde{d}_{\ell m}|^2 / (2\ell+1)$. The mask-deconvolved spectrum $\hat{C}_\ell^d$ is approximately a renormalised version of the pseudo-spectrum, $\hat{C}_\ell^d \approx \tilde{C}_\ell^d / \langle w^2 \rangle_\Omega$. In the mean,
\begin{equation}
\langle \hat{C}_\ell^d \rangle \approx \frac{\langle w^2 m^2 \rangle_\Omega}{\langle w^2 \rangle_\Omega} C_\ell^g \, ,
\end{equation}
which is mask-dependent because of the modulation field $m(\hat{\mathbfit{n}})$. It is this quantity that is modelled in the likelihood with the fitting function of Eq.~(\ref{eqn:dust_mask50}). In the limit of wide bandpowers (such that $\Delta \ell$ is large compared to the support of the power spectrum of the product of the mask and modulation fields), the bandpower variance of the pseudo-spectrum is~\citep{Hivon2002}
\begin{equation}
\text{var}(\tilde{C}_\ell^d) \approx \frac{2}{(2\ell+1)\Delta \ell} \langle w^4 m^4 \rangle_\Omega \left(C_\ell^g\right)^2 \, .
\end{equation}
Renormalising to give the variance of the $\hat{C}_\ell^d$, and expressing the result in terms of $\langle \hat{C}_\ell^d \rangle$, we find
\begin{equation}
\text{var}(\hat{C}_\ell^d) \approx \frac{2}{(2\ell+1)\Delta \ell} \frac{\langle w^4 m^4 \rangle_\Omega}{\langle w^2 m^2 \rangle_\Omega^2} \langle \hat{C}_\ell^d \rangle^2 \, .
\label{eq:moddustvar}
\end{equation}
It follows that the effective number of degrees of freedom is $\nu_\ell \approx (2\ell+1) \Delta \ell / H[m]$, where
\begin{equation}
H[m] \equiv \frac{\langle w^4 m^4 \rangle_\Omega} {\langle w^2 m^2 \rangle^2_\Omega}
\label{eq:moddustH}
\end{equation}
plays the role of an effective inverse sky fraction for the variance. $H[m]$ is larger than its equivalent, $\langle w^4\rangle_\Omega / \langle w^2 \rangle^2_\Omega$, in the absence of modulation, reducing $\nu_\ell$ and increasing the variance of the power spectrum. 

\begin{figure}
  \begin{center}
  \begin{tabular}{cc}
        \includegraphics[width=80mm, angle=180]{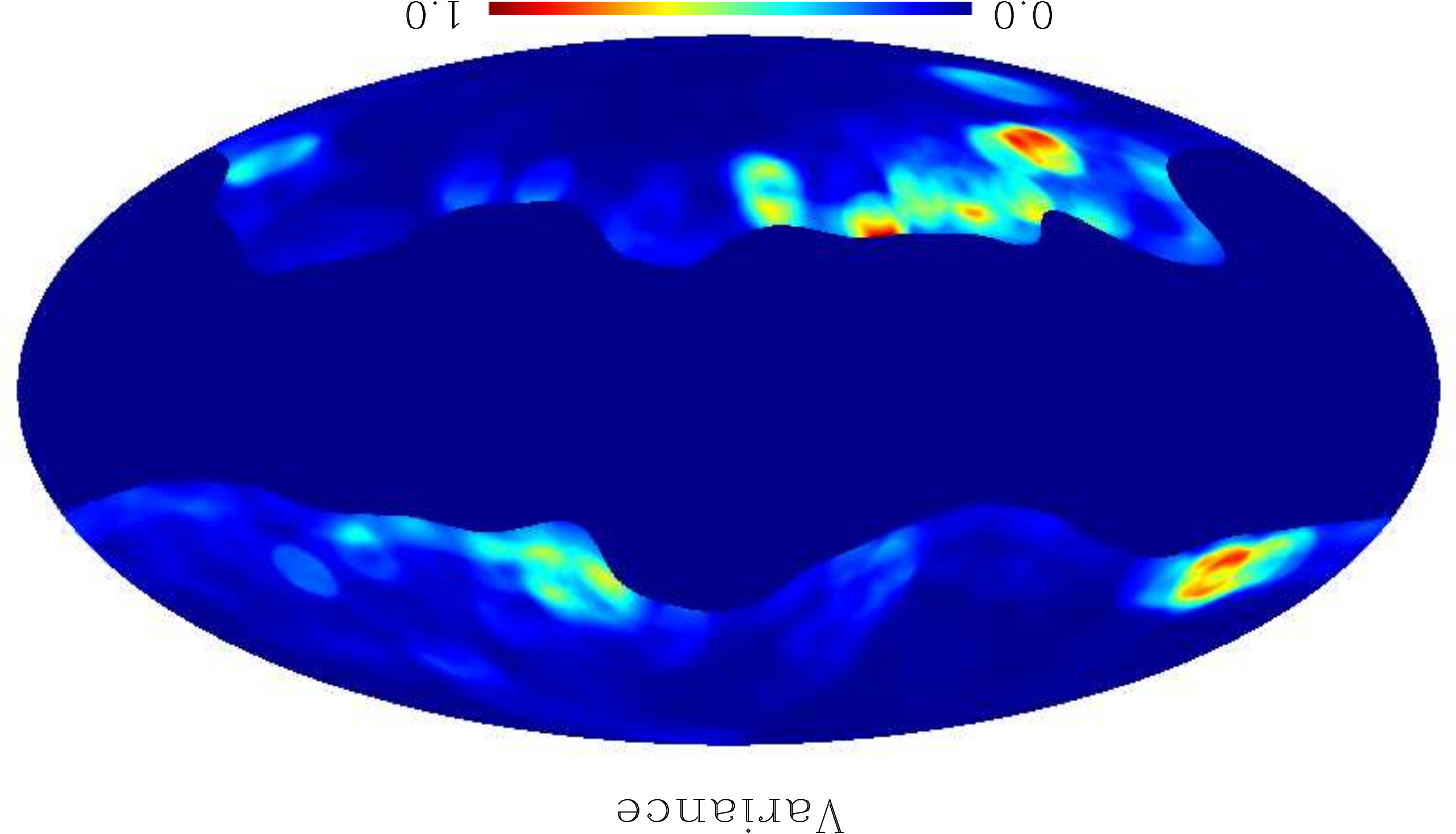} &
        \includegraphics[width=80mm,angle=180]{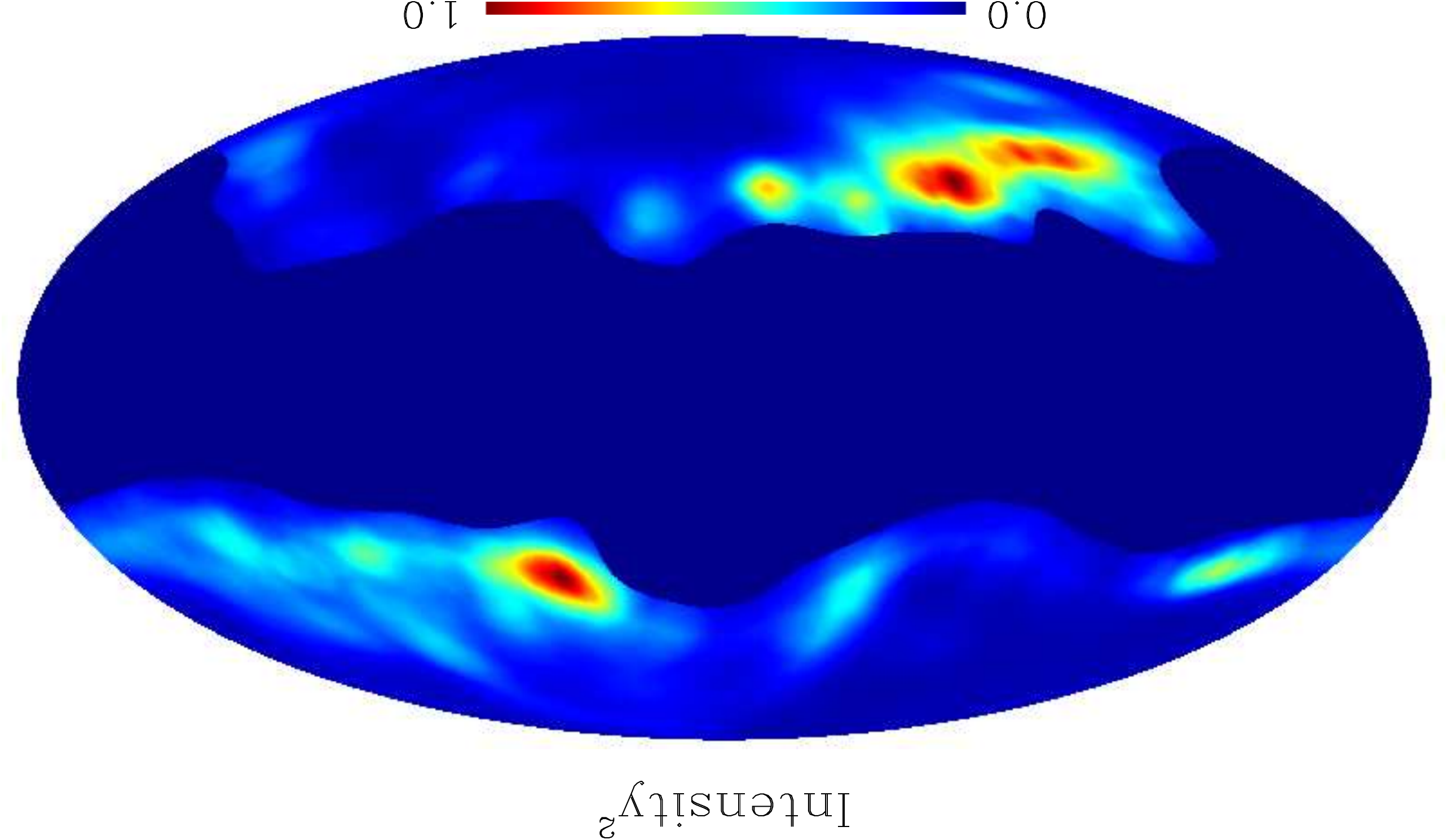} \\
   \end{tabular}
       \caption{{\it Left}: Normalised variance map of dust emission in circular appertures of $6^\circ$ radius constructed from smoothed 857\,GHz maps. This map traces the square of the modulation field. {\it Right}: Square of the mean intensity constructed from the same procedure as for the variance map. }
     \label{fig:mod}
  \end{center}
\end{figure}

\subsection{Approximating the modulation field}
\label{app:corr}

We estimate the modulation field $m(\hat{\mathbfit{n}})$ using the \Planck\ 857\,GHz data on intermediate and large scales as a tracer of dust emission. The modulation controls both the local mean level of dust emission, $\langle d \rangle_R$, and the local variance, $\sigma^2_R$. Here, the local mean and variance are calculated within circular apertures of radius $R$. For $R$ small compared to the scale of the modulation, the model predicts $\sigma_R^2 \propto \langle d \rangle_R^2$. This scaling is consistent with the results reported in~\citet{Miville2007} in regions of low emission. We estimate the local mean and variance of the 857\,GHz maps as follows.
\begin{enumerate}
\item We first smooth the masked, 857\,GHz Year-1 and Year-2 maps with a Gaussian of $\text{FWHM}=1^\circ$ in order to suppress the small-scale CIB and instrument noise.
\item Since we are interested in the large-scale modulation field, we estimate this at lower resolution (\healpix $N_{\rm side} = 256$). At the centre of each $N_{\rm side} = 256$ pixel that lies outside the Galactic mask, we compute the mean of each smoothed map and the cross-variance (to avoid noise bias) within a circular aperture of radius $R= 6^\circ$. We note that the smoothed maps are retained at their native resolution, and that pixels in these maps that are masked by the point source mask are discarded.
\end{enumerate}

Figure~\ref{fig:mod} shows the variance map and the map of the squared mean
calculated with this procedure. The square of the modulation field can be estimated from these maps up to an irrelevant normalisation. The variance and mean-squared maps both show significant variation across the sky and are clearly correlated. We note that the mean is sensitive to the estimation of the zero-point level of the maps, and the CIB monopole, while the variance is not. In the following, we use the variance map as a proxy for $m^2(\hat{\mathbfit{n}})$.

Evaluating $H[m]$ from Eq.~(\ref{eq:moddustH}), we find $H=5.77$ for mask 40. This is enhanced by a factor $1.94$ by the modulation, so we expect roughly a factor of 
$\sqrt{1.94}\approx 1.39$ increase in the bandpower errors on scales
where dust is dominant. For a crude estimate of how much the increased
errors might reduce the $\chi^2$ of the best-fit model, we simply
inflate the bandpower errors of the $857\times 857$ spectrum by
$40\,\%$ at multipoles $\ell <1000$. Keeping the best-fit model
unchanged, we find that $\hat{\chi}^2=1.77$ for 74 degrees of freedom,
with a PTE of $0.1\,\%$. The equivalent value without modelling the
dust modulation (from Table~\ref{t:chi2_binned}) is
$\hat{\chi}^2=2.28$ so the correction is very significant. Repeating
for mask 50, we find a similar boost of $1.42$ in the bandpower errors
on large scales. At this mask, we apply the correction to $\ell<1500$
since dust dominates out to higher multipoles, and obtain an improved
$\hat{\chi}^2=1.46$ (again, for 74 degrees of freedom) corresponding to a $\Delta \chi^2=56$. A bigger effect is seen for mask 50 since the dust emission is more intense. 

\begin{figure}
  \begin{center}
        \includegraphics[width=120mm]{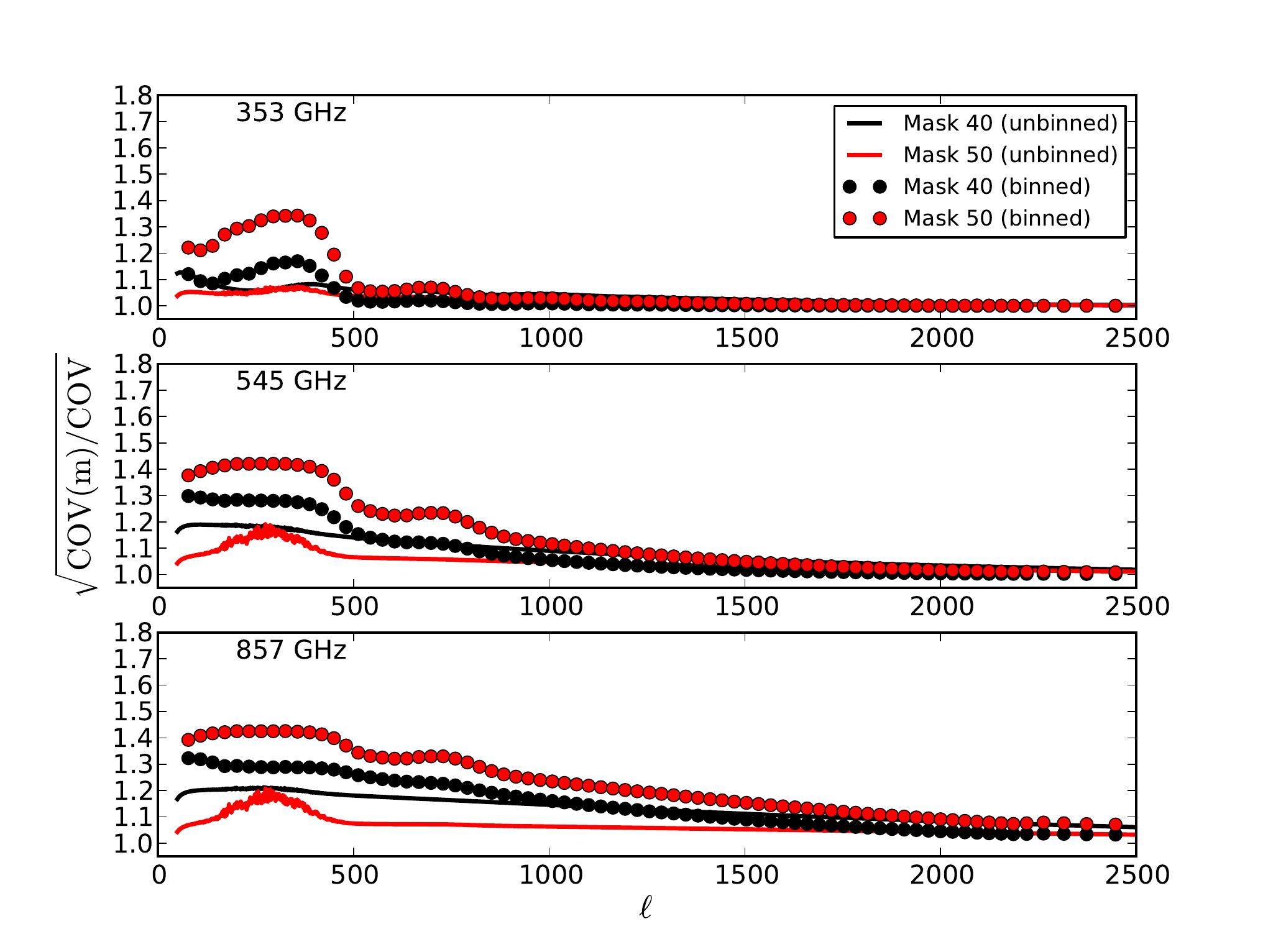}
       \caption{Ratio of the square roots of the diagonal elements of the covariance matrices with and without corrections for the statistical anisotropy of Galactic dust. The ratios are shown for the $353\times 353$ (top), $545\times 545$ (middle), and $857\times 857$ (bottom) spectra for mask 40 (black) and mask 50 (red). In all cases, results for both unbinned (lines) and binned spectra (points) are shown.}
     \label{fig:ratio}
  \end{center}
\end{figure}

\begin{figure}
  \begin{center}
        \includegraphics[width=150mm]{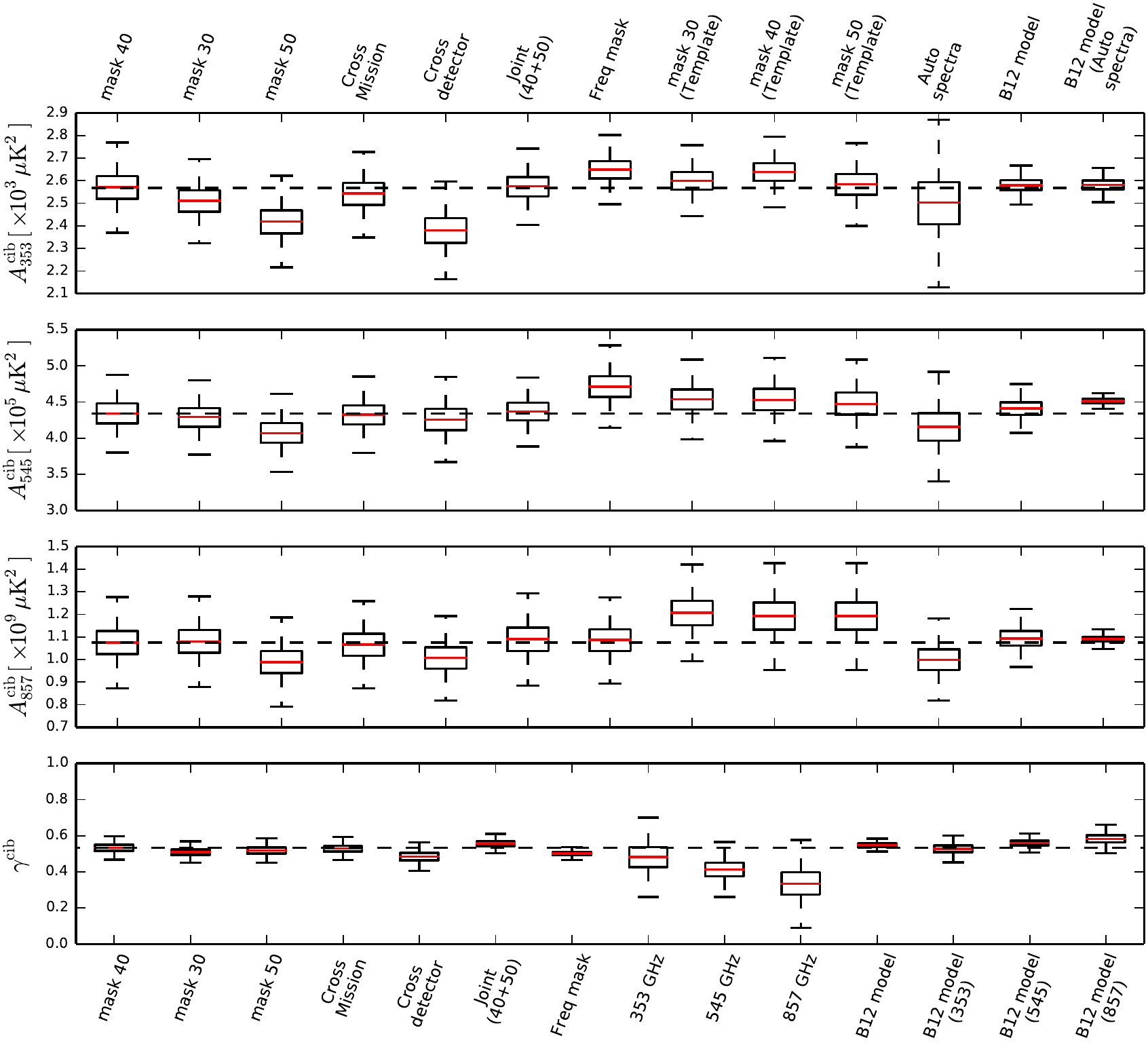}  
       \caption{Comparison of the distributions of $A^{\rm cib}$ and
         $\gamma^{\rm cib}$ in the reference case (first data point in
         each panel) with a set of validation test cases. The red
         lines indicate the best-fit values of the parameters. The
         boxes show the $68\,\%$ confidence interval and the outer
         lines show the $95\,\%$ interval. We use the results obtained
         with mask 40 (and the union point source mask),
         cross-correlating the Year-1 and Year-2 data, as our baseline
         (column 1). The other columns show variations with respect to
         this baseline. {\it Common to all four panels}: columns 2 and
         3 shows the results using masks 30 and 50, respectively;
         column 4 shows results obtained by cross-correlating the
         first and second halves of the \Planck\ mission data; column
         5 shows results from cross-correlating the detector-set maps;
         column 6 shows the results of the joint-mask analysis (using
         masks 40 and 50); column 7 shows results using
         frequency-dependent point source masks. {\it Common to top
           three panels ($A^{\rm cib}_\nu$)}: columns 8--10 show the
         results from replacing the power-law clustered CIB model with the
         halo model template for masks 30, 40, and 50, respectively;
         column 11 shows results when using auto-frequency spectra
         only; columns 12 and 13 show results with the Poisson amplitudes,
$A^{\rm ps}$, kept fixed to \bcount\ model predictions, using all six
auto- and cross-frequency spectra (column 12) and only auto-frequency
spectra (column 13). {\it Bottom panel $\gamma^{\rm cib}$}: columns
8--10 show results using only the auto-frequency spectrum at 353, 545,
and 857\,GHz, respectively; columns 11--14 show results with the
Poisson amplitudes, $A^{\rm ps}$, kept fixed to the \bcount\ model
predictions and using either all six auto- and cross-frequency spectra (column 11),
or only the auto-frequency spectrum at each of 353, 545, and 857\,GHz
(columns 12, 13, and 14, respectively).}
\label{fig:robustness}
  \end{center}
\end{figure}

\subsection{Corrections to covariance}
\label{app:corr_cov}
Equation~(\ref{eq:moddustvar}) gives the approximate sample variance of the anisotropic dust in broad bandpowers. In the likelihood results presented in Sec.~\ref{sec:ngnoise}, we use a more sophisticated approach whereby we include the dust modulation field directly in the covariance matrices of the pseudo-spectra. We split the signal into the statistically-anisotropic dust, $d$, and statistically-isotropic components (CIB and CMB) with total theory power spectra $C_\ell^{\text{iso},ij}$ for maps $i$ and $j$. We model the dust as a modulated Gaussian field, as above. For modulation that is smooth on the beam scale, the convolution with the instrument beam commutes with the modulation, and so the modulation can be treated as an additional mask that is applied to the beam-smoothed, statistically-isotropic Gaussian field $g^i(\hat{\mathbfit{n}})$. This modifies the covariance matrix of the pseudo-spectra; for example, the signal-signal part of the covariance matrix [i.e., the first term on the right of Eq.~(\ref{eq:cov})] becomes
\begin{multline}
\text{cov}(\tilde{C}_\ell^{ij},\tilde{C}_{\ell'}^{pq}) \supset \sqrt{C_{\ell}^{\text{iso},ip}
C_{\ell'}^{\text{iso},ip}}\sqrt{C_{\ell}^{\text{iso},jq} C_{\ell'}^{\text{iso},jq}} \Xi(\ell,\ell',W^{(ip)(jq)}) +
\sqrt{C_{\ell}^{\text{iso},ip}
C_{\ell'}^{\text{iso},ip}}\sqrt{C_{\ell}^{g,jq} C_{\ell'}^{g,jq}} \Xi(\ell,\ell',W_{\text{mod.}}^{(ip)(jq)}) \\ +
\sqrt{C_{\ell}^{\text{iso},jq}
C_{\ell'}^{\text{iso},jq}}\sqrt{C_{\ell}^{g,ip} C_{\ell'}^{g,ip}} \Xi(\ell,\ell',W_{\text{mod.}}^{(jq)(ip)})+
\sqrt{C_{\ell}^{g,ip}
C_{\ell'}^{g,ip}}\sqrt{C_{\ell}^{g,jq} C_{\ell'}^{g,jq}} \Xi(\ell,\ell',W_{\text{mod.}\,\text{mod.}}^{(jq)(ip)}) + p \leftrightarrow q\, ,
\end{multline}
where
\begin{align}
W^{(ip)(jq)}_{\text{mod.},\ell} &= \frac{1}{2\ell+1}\sum_m w_{\ell m}^{(ip)}\left(w^{(jq)}_{\text{mod.},\ell m} \right)^\ast \, , \\
W^{(ip)(jq)}_{\text{mod.}\,\text{mod.},\ell} &= \frac{1}{2\ell+1}\sum_m w_{\text{mod.},\ell m}^{(ip)}\left(w^{(jq)}_{\text{mod.},\ell m} \right)^\ast \, ,
\end{align}
with
\begin{equation}
w^{(ij)}_{\text{mod.},\ell m} = \int m^2(\hat{\mathbfit{n}}) w^i(\hat{\mathbfit{n}}) w^j(\hat{\mathbfit{n}}) \, d \hat{\mathbfit{n}} \, .
\end{equation}
Here, we have assumed that the modulation is the same for all maps, so that $g^i(\hat{\mathbfit{n}})$ carries the frequency dependence of the dust. We estimate the cross-power spectrum of $g^i$ and $g^j$, $C_\ell^{g,ij}$, from the best-fit dust model in the likelihood, $\hat{C}_\ell^{\text{dust},ij}$, as
\begin{equation}
\sum_{\ell'} M_{\ell\ell'}[w^i\times w^j] \hat{C}_{\ell'}^{\text{dust},ij} = \sum_{\ell'} M_{\ell\ell'}[mw^i \times mw^j] C_{\ell'}^{g,ij} \, ,
\end{equation}
where the mask-coupling matrix $M_{\ell\ell'}[mw^i \times mw^j]$ is calculated as in Eq.~\ref{eq:maskcouple} but with $W_\ell^{ij}$ replaced by the cross-spectrum of the modulated masks $m w^i$ and $m w^j$.

Figure~\ref{fig:ratio} shows the ratio of the square roots of the diagonal elements of the covariance matrices with and without the correction for dust modulation. We show results for both unbinned and binned spectra. The latter is sensitive to the enhanced mode-coupling across multipoles induced by the modulation. In this case, the diagonal errors are increased by about 20--40\,\%, depending on mask and frequency, at $\ell<1000$ and drop to less than $5\,\%$ at higher multipoles. As expected, the effect of the modulation is greatest in cases where the dust is more dominant, i.e., large scales, high frequencies, and masks that retain a larger fraction of the sky.
The amplitude of the corrections roughly corresponds to our initial estimates of around 40\,\%.

\section{Stability and robustness tests}
\label{app:con}
In this appendix we investigate the stability of the distribution of
the model parameters to technical choices that we make in the
analysis. These choices fall into two broad categories. The first
involves the choices of fields, i.e., masks. The second involves the
selection of data, such as number and type of spectra used. We performed a number of tests to investigate the impact of these choices 
on the parameters, and to compare with the results of the baseline
analysis used in most of this paper (six auto- and cross-frequency
spectra obtained by cross-correlating Year-1 and Year-2 maps outside
mask 40 and with the union point source mask). The results are summarised in \reffig{robustness}, which compares the CIB amplitudes  $A^{\rm cib}$ and CIB power
law index $\gamma^{\rm cib}$ for various analysis choices.

Increasing the sky fraction between 30\,\% and 50\,\% (i.e., using
masks 30, 40, and 50) gives consistent results for the CIB amplitudes,
as already illustrated in Figs.~\ref{fig:dx11-p2013-herschel} and
\ref{fig:dx11-p2013} and discussed in \refsec{cs}. In addition,
switching to frequency-dependent point source masks, or changing to
spectra constructed by cross-correlating the first and second halves
of the mission data, or detector-set maps, we see changes of less than
$1\,\sigma$ in the CIB parameters. Our best-fit Poisson amplitudes
are actually in very good agreement with the predictions of the \bcount\ models, so imposing the \bcount\ constraints on the point source amplitudes has
very little effect (apart from shrinking the errors on the clustered CIB parameters).

\end{document}